\documentstyle[12pt]{article}

\def\hybrid{\topmargin -20pt  \oddsidemargin 0pt
      \headheight 0pt   \headsep 0pt
      \textwidth 6.25in 
      \textheight 9.5in 
      \marginparwidth .875in
      \parskip 5pt plus 1pt   \jot = 1.5ex}

\hybrid

\def\x{\times}
\def\ox{\otimes}
\def\o+{\oplus}
\def\ra{\rightarrow}
\def\lra{\longrightarrow}
\def\Lra{\Longrightarrow}
\def\Lla{\Longleftarrow}
\def\Llra{\Longleftrightarrow}
\def\hra{\hookrightarrow}
\def\da{\downarrow}
\def\beqa{\begin{eqnarray}}
\def\eeqa{\end{eqnarray}}
                       
\newcommand{\un}{\underline}

\newcommand{\ti}{\tilde}
\newcommand{\al}{\alpha}
\newcommand{\be}{\beta}

\newcommand{\la}{\lambda}
\newcommand{\si}{\sigma}
\newcommand{\om}{\omega}

\newcommand{\D}{{\cal D}}
\newcommand{\E}{{\cal E}}
\newcommand{\cF}{{\cal F}}
\newcommand{\G}{{\cal G}}
\newcommand{\M}{{\cal M}}

\newcommand{\cL}{{\cal L}}

\newcommand{\cW}{{\cal W}}
\newcommand{\cO}{{\cal O}}
\def\Si{\Sigma}

\newcommand{\resetcounter}{\setcounter{equation}{0}}

\begin{document}

\thispagestyle{empty}
\rightline{LMU-ASC 51/08}
\vspace{2truecm}
\centerline{\LARGE \bf{World-sheet Instanton Superpotentials}}
\vspace{.5truecm}
\centerline{\LARGE \bf{in Heterotic String theory and their}}
\vspace{.5truecm}
\centerline{\LARGE \bf{Moduli Dependence}}

\vspace{1.5truecm}
\centerline{Gottfried Curio\footnote{gottfried.curio@physik.uni-muenchen.de}} 

\vspace{.6truecm}

\centerline{{\em Arnold-Sommerfeld-Center for Theoretical Physics}}
\centerline{{\em Department f\"ur Physik, 
Ludwig-Maximilians-Universit\"at M\"unchen}}
\centerline{{\em Theresienstr. 37, 80333 M\"unchen, Germany}}

\vspace{1.0truecm}

\begin{abstract}
To understand in detail the contribution 
of a world-sheet instanton to the superpotential
in a heterotic string compactification,
one has to understand the moduli dependence
(bundle and complex structure moduli)
of the one-loop determinants from the fluctuations,
which accompany the classical exponential contribution
(involving K\"ahler moduli) 
when evaluating the world-volume partition function.
Here we use techniques to describe geometrically these Pfaffians 
for spectral bundles over rational base curves in
elliptically fibered Calabi-Yau threefolds, 
and provide a (partially exhaustive) list of
cases involving {\em factorising} (or vanishing) superpotential.
This gives a conceptual explanation and generalisation
of the few previously known cases
which were obtained just experimentally by a numerical computation.
\end{abstract}

\newpage

\section{Introduction and description of results}

In heterotic $E_8\times E_8$ string theory
models of $N=1$ supersymmetry in 4D arise by compactification 
on a Calabi-Yau threefold $X$ with vector bundle $V$. 
Originally the case of $V$ the tangent bundle was considered
which led to an unbroken gauge group $E_6$ (times a hidden $E_8$).
The generalisation to an $SU(n)$ bundle $V$ gives unbroken GUT groups like 
$SO(10)$ and $SU(5)$ (we will in the following focus on the visible sector 
and may assume an $E_8$ bundle $V_2$ embedded in the second $E_8$). 
To be able to handle the $SU(n)$ bundle $V$ most explicitly we assume that $V$
arises by the spectral cover construction
for bundles on an $X$ which has an elliptic fibration $\pi:X\ra B$.
This description uses a surface $C\subset X$ given by
an $n$-fold (ramified) cover of the base $B$ and a line bundle $L$ on $C$.
Assuming $C$ ample the continuous moduli of $V$ 
come just from the deformations of $C$ in $X$; these 
are given by the polynomial
coefficients entering the defining equation of $C$.

For a holomorphic curve $b$, arising as support of a 
world-sheet instanton, to contribute to the superpotential 
one assumes that $b$ is isolated and rational.
In the following we want to bring to bear the explicit information about the 
bundle $V$ provided by the spectral cover construction; we will 
therefore restrict us to the case of horizontal curves, i.e.~curves $b$ 
lying in $B$.
If the world-sheet instanton
contribution $W_b$ supported on $b$ is generically
nonzero one wants to have the finer information how this contribution
depends on the bundle (and possibly complex structure)
moduli. This is described by the Pfaffian prefactor 
$Pfaff$ of the classical instanton contribution $e^{i\int_b J}$. 
$Pfaff$ is given by a generally complicated determinantal
expression in the bundle moduli.

Because of holomorphic dependence crucial is here 
the sublocus in the moduli space where $Pfaff$ vanishes.
This leads to an identification of $Pfaff$ with a geometrical
determinantal expression whose vanishing controls the vanishing of $Pfaff$.
In some cases $Pfaff$ vanishes 
identically in the moduli for well-understood reasons. 
More interesting is the case where $Pfaff$ is generically nonvanishing 
in the moduli. Especially interesting, and the raison d'\^{e}tre of the
present paper, is the case where the Pfaffian shows some structure, 
i.e.~is not as complicated as generically under the given circumstances.
This means more precisely that one has a nontrivial factorisation 
like $Pfaff=f^k$ with $k>1$, or $Pfaff=fg$ or even $Pfaff=f^kg$.
This can simplify the search for zeroes of the superpotential,
and in a case with a multiplicity $k>1$ also of its derivative.

A small set of three examples of such behavior was found [\ref{BDO}]
by using computer calculation of some large determinants.
This had the character (like a fourth example of vanishing Pfaffian)
of a surprising simplification arising by an intransparent 
(purely numerical) brute-force computation. 
Our goal here is to get a conceptual understanding 
(i.e.~beyond doing algebra for concrete matrix expressions)
of the way such a simplifying structure arises. 
In the present paper we explain the case of the vanishing Pfaffian 
and the occurrence of the factor $f$; here explaining means we give
a conceptual, non-computational reason 
for the examples and generalise them to further cases;
the question of multiplicity $k$ of $f$ will be dealt with in a separate paper
as will be the treatment of the somewhat differently behaved
second factor $g$ (the Giant $Q_{11}$ in Example 1, c.f.~below).

In the rest of this {\em Introduction} we will first recall more precisely
the numerical results of [\ref{BDO}]; then we will describe what
is derived conceptually in the present paper and state the generalisations.
In {\em sect.~\ref{spectralcovermethod}} we recall some facts about
the spectral cover construction of bundles.
In {\em sect.~\ref{restriction to E}} we apply this to our case by restricting
the construction to the elliptic surface $\E$ 
lying above the instanton curve $b\subset B$ in the elliptic fibration,
providing a spectral curve $c$ and a line bundle $l$ on it.
We investigate some special loci in moduli space 
for one of our main example classes, the $SU(3)$ bundles.
In {\em sect.~\ref{WS instantons}} 
we recall from [\ref{Wittenhet}] and [\ref{BDO}]
some general conditions for world-sheet instantons to contribute to the
superpotential and make these conditions explicit in the parameters of the
construction.

In {\em sect.~\ref{section reduction}}
we explain the main idea of the paper: how a factor in the Pfaffian
can be explained by reduction to a different line bundle $\bar{l}$
which is 'simpler' than $l$.
In {\em sect.~\ref{Equality condition}} we describe how this
idea can be practically implemented if one can imposes
a certain ('equality') condition such that the resulting question for 
$\bar{l}$ is again controlled by a determinantal expression.
Here we give an (in a certain sense) exhaustive list of cases
where the reasoning described applies.
In {\em sect.~\ref{The vanishing condition}} we describe how the case of 
a generically vanishing Pfaffian fits into the set-up outlined so far.
In {\em sect.~\ref{difference subsection}}, which has the character of an
insert, we show how the argument described so far 
has to be supplemented if some assumptions are weakened; 
this will be relevant for Example 1.
In {\em sect.~\ref{Examples}} we discuss in detail the Examples of [\ref{BDO}]
and corresponding generalisations to which they give rise.

The {\em Appendices} collect auxiliary investigations.
App.~\ref{Polynomial factors} gives the polynomial factors of $Pfaff$ 
in the Examples. 
App.~\ref{rational curves} collects facts about
(horizontal) rational curves in the Calabi-Yau threefold $X$.
App.~\ref{Lemmata} gives Lemmata
conc.~the cohomological contribution criteria for instantons.
App.~\ref{augmentation} points to a different cohomological method,
alternative to the procedure in sect.~\ref{section reduction}.
App.~\ref{negative lambda} shows how the case of a negative bundle parameter
$\lambda$ is related to the usual procedure for $-\lambda$
(cf.~Example 1). App.~\ref{Reduction cases} 
shows how one obtains the exhaustive lists
for cases of reduction or vanishing Pfaffian.
App.~\ref{concrete matrix representations} gives
details connecting the structural investigations and concrete
matrix representations.
App.~\ref{chi = 0 app} illustrates the special case $\chi=0$ and shows a 
case of $Pfaff\equiv 0$ algebraically.
App.~\ref{sym} establishes needed facts and notation for symmetrized
tensor products and resultants.

\subsection{\label{Some experimental results}Some experimental results}
Some Examples were found experimentally via
computer evaluations of the occurring large determinants [\ref{BDO}].
The cases concerned spectral $SU(3)$ bundles 
of bundle parameters $\la\in \frac{1}{2}{\bf Z}$ and
$r=\eta b\in {\bf Z}$ (they are defined in 
sect.~\ref{spectralcovermethod} and \ref{coordinates for curve case} 
and are needed here only to display the results in an overview)
with $b$ the base of the ${\bf P^1}$-fibration on $B={\bf F_k}$
\begin{center}
\begin{tabular}{|c||c|c|c||c|}
\hline
Example & $\la$ & $r$ & ${\bf F_k}$ & $Pfaff$\\
\hline\hline
$1$ & $-5/2$ & $4$ & ${\bf F_1}$ & $f^{11}g$\\
\hline
$2$ & $3/2$  & $5$ & ${\bf F_1}$ & $f^4$\\
\hline
$3$ & $3/2$  & $1$ & ${\bf F_2}$ & $0$\\
\hline
$4$ & $3/2$  & $2$ & ${\bf F_2}$ & $f^4$\\
\hline
\end{tabular}
\end{center}

The detailed polynomial expressions of the factors of $Pfaff$ are 
given in app.~\ref{Polynomial factors}.

\subsection{Conceptual explanations and generalisations}

When we make 
in sect.~\ref{restriction to E} the transition from
a description of $V$ over $X$ via the spectral surface $C$ and the line
bundle $L$ on it to the corresponding notions on $\E$ we will,
besides the spectral curve $c:=C|_{\E}$, also introduce the corresponding
line bundle $L|_c$. And as we are able, under our assumptions, to describe
the line bundle $L$ on $C$ as a restriction $L=\un{L}|_C$
from a line bundle $\un{L}$ on $X$, we will similarly be able to describe 
the line bundle $L|_c$ on $c$ as a restriction $L|_c=l|_c$
from a line bundle $l$ on $\E$; actually, of course, $l=\un{L}|_{\E}$.
For all these relations cf.~sect.~\ref{twist class}.

We note that when we wish to emphasize the dependence on a modulus $t$
we write $V_t$, $C_t$ and so on for the corresponding objects 
(cf.~sect.~\ref{Overview and summary}). The
abstract modulus $t$ will actually turn out to be a modulus of the
surface $C$, respectively the curve $c$, and will be given concretely 
by polynomial coefficients of its defining equation
(this will be studied very explicitly for the case of $SU(3)$ bundles,
cf.~(\ref{equ spec curve}) and (\ref{coefficient expressions}), for example).

We usually take $\chi=c_1(B)\cdot b$ to be $0$ or $1$, 
with $1$ the relevant case, cf.~sect.~\ref{chi subsection}; the class
in $\E=\pi^{-1}(b)$
of the spectral cover curve $c$ over $b$ is $ns+rF$ with $s=\sigma|_{\E}$ 
the section of the elliptic surface $\E$ and $F$ the class 
of the elliptic fibre.

We will explain and generalise the experimental results 
in sect.~\ref{Some experimental results}.
For the degenerate case of Example 3 a conceptual interpretation 
from the existence of nontrivial sections of $l(-F)|_c$, 
cf.~sect.~\ref{The vanishing condition} and below, will be given
(besides an algebraic explanation, cf.~sect.~\ref{chi = 0 app}). 
Example 1 has a somewhat exceptional status, cf.~sect.~\ref{Exceptional} and 
\ref{Ex 1 sect}.
Examples 2 and 4 are covered below
(full details for Examples 2,3,4 are given in the sections indicated
below).

\newpage

{\em \underline{Vanishing Pfaffian:} $Pfaff\equiv 0$}

Let us begin with the case where the Pfaffian vanishes identically.
This happens in the following cases (in sect.~\ref{red cases for bar al > n},
\ref{The vanishing condition} and app.~\ref{cases al > n} it is described 
under which assumptions these cases give an exhaustive description; a
corresponding remark applies below)
\begin{itemize}
\item $SU(n)$ bundles with $\chi=0, r=1, \la \in \frac{1}{2}+{\bf Z}^{\geq 1}$ 
\item $SU(3)$ bundles with $r\not\equiv \chi (2)$ and $\la =3/2$
\end{itemize}

This constitutes a vast generalisation of Example 3.

{\em \underline{Factorising Pfaffian:} $Pfaff=fg$} 
\hspace{2cm}(including the case $g=f^m$)

In the main case the Pfaffian factorises such that $f|Pfaff$. 
More precisely we will identify a concrete factor 
$f=\det \bar{\iota}_1$ in $Pfaff=\det \iota_1$ 
where\footnote{with $\beta = \frac{r+\chi}{2}-\la(r-n\chi)-1$ and 
$\bar{\beta}=(r+\chi)-\frac{1}{2}\frac{n+1}{n-1}r-\frac{1}{2}(r-n\chi)-1$}
(cf.~sect.~\ref{red cases for bar al = n} and app.~\ref{list subsection}) 
\beqa
\iota_1: H^1\Big(\E, l(-F-c)\Big)\lra H^1\Big(\E, l(-F)\Big) ,&& \;\;
l(-F)=\cO_{\E}\Big(n(\la + \frac{1}{2})s+\beta F\Big)\;\;\\
\bar{\iota}_1: H^1\Big(\E, \bar{l}(-F-c)\Big)\lra H^1\Big(\E, \bar{l}(-F)\Big),
&&\;\;
\bar{l}(-F)=\cO_{\E}\Big(ns+\bar{\beta}F\Big)\;\;
\eeqa
\begin{itemize}
\item $SU(3)$ bundles 
\begin{center}
\begin{tabular}{|c||c|c|c||c|c||c|c|}
\hline
$\sharp$ &$\chi$ & $r$ & $\la $ & $l(-F)=\cO_{\E}(\cdot)$ 
& $\bar{l}(-F)=\cO_{\E}(\cdot)$ & $\deg Pfaff$ & $\deg f$ \\
\hline
\hline
1 & $0$ & $2$& $ \frac{3}{2}, \frac{5}{2}, ...$ 
& $3(\la + \frac{1}{2})s-2\la F$ & $3s-2F$ & $6(\la^2-\frac{1}{4})$ & $3$ \\
\hline
2 & $\geq 1$ & $5\chi$ & $\frac{3}{2}$ 
& $6s-F$ & $3s-F$ & $20\chi$ & $5\chi$ \\
\hline
3 & $\chi$ & $\geq 5\chi, \equiv \chi(2)$ & $\frac{3}{2}$ 
&$6s-(r-5\chi+1)F$ &$3s-(\frac{r-5\chi}{2}+1)F$ & $6r-10\chi$ & 
$\frac{3r-5\chi}{2}$\\
\hline
4 & $0$ & $4$ & $\frac{5}{2}$ & $9s-9F$ & $3s-3F$& $72$ & $6$\\
\hline
5 & $1$ & $5$ & $\frac{5}{2}$ & $9s-3F$ & $3s-F$ & $62$ & $5$\\
\hline
\end{tabular}
\end{center}
Case 3 constitutes a vast generalisation of Example 2 and 4
(why case 2 is listed separately besides case 3 will only become clear
later and is of no concern here). 
\item $SU(4)$ bundles
\begin{center}
\begin{tabular}{|c||c|c|c||c|c||c|c|}
\hline
$\sharp$ & $\chi$ & $r$ & $\la $ & $l(-F)=\cO_{\E}(\cdot)$ 
& $\bar{l}(-F)=\cO_{\E}(\cdot)$ & $\deg Pfaff$ & $\deg f$ \\
\hline
\hline
1 & $\geq 1$  & $9\chi$ & $1$ & $6s-F$ & $4s-F$ & $20\chi$ & $9\chi$ \\
\hline
2 & $0$ & $3$ & $\frac{3}{2}$ & $8s-4F$ & $4s-2F$ & $24$ & $4$\\
\hline
\end{tabular}
\end{center}
\end{itemize}
So case 5 of $SU(3)$ and case 1 of $SU(4)$ have a second factor
(there certainly $Pfaff\neq f^k$).

\subsection{\label{Overview and summary}Overview and summary}

As the issue in question - the vanishing behaviour of the Pfaffian prefactor
(of a world-sheet instanton superpotential) in dependence 
on the vector bundle moduli - 
necessarily uses a heavy amount of algebraic-geometric notions it may be 
useful to provide here also a nontechnical overview of the more detailled
investigations which follow in the later sections
(herein we allow ourselves to give an only approximate description of various
issues whose more detailed aspects are dealt with in the main text).

According to the main contribution criterion which will be recalled in
sect.~\ref{contribution criteria} one has as precise vanishing criterion 
for the Pfaffian
\beqa
Pfaff(t)=0& \Llra & \Gamma\Big(b, V_t|_b\ox \cO_b(-1)\Big)
\; \cong \; \Gamma\Big(c_t, l(-F)|_{c_t}\Big)\; \neq \; 0 
\eeqa

Clearly, if the line bundle\footnote{the notations for this 
and the following line bundles are used - 
with their here indicated meaning - only in this subsection}
$L=\cO_c(D)$ over $c$ 
whose possible sections are concerned 
here is a tensor product $L_1\ox L_2=\cO_c(D_1)\ox \cO_c(D_2)$ 
of line bundles which themselves {\em have} 
a nontrivial section,
then also their product bundle, in question here, would have such a section.
For example, for one - auxiliary - factor $L_2$ the existence of such a section
might be assured by a general argument, like that the line bundle is
associated to an effective divisor; thereby the problem of existence of a 
nontrivial section would have been reduced from the original problem 
for $L$ to the other factor $L_1$. Because $D-D_1$ is, as assumed, effective
one might say that the problem is reduced to a smaller line bundle
\beqa
D-D_1 \;\; \mbox{effective}\;\;\;\;\;\; & \lra & 
\;\;\;\;\;\; \mbox{reduction from $L$ to $L_1$}
\eeqa
Actually
there will be an even more important sense of such a 'reduction in size'
when going from $L$ to $L_1$ described later (the relevant vanishing condition
will be expressed by a determinant of a matrix of smaller size; we start
to develop this technically in sect.~\ref{section reduction}).

Now, one of the easiest possibilities for an accessible criterion for the
existence of a section would arise if for $L_1$ a similar determinantal 
expression could be found (whose vanishing controls the existence of a 
nontrivial section) as for $L$.
For this one considers for $L_1$ similar cohomological sequences as for $L$
and has to see whether again, under certain conditions, one can relate
the space of sections in question to a map between $H^1$-cohomologies
of line bundles over $\E=\pi^{-1}(b)$ where the latter spaces have
{\em equal dimension}; if that {\em equality condition} (which we study
in sect.~\ref{Equality condition}) 
is satisfied one gets indeed again a determinantal 
expression which controls the issue in question here, the existence 
of a nontrivial section of $L_1$
\beqa
\mbox{equality condition}\;\;\;&\lra & 
\;\;\; \mbox{determinantal expression controlling $h^0(c, L_1)\neq 0$}
\;\;\;\;\;\;
\eeqa

Occasionally, if another special condition is fulfilled, it may also 
happen that the line bundle $L_1$ does always\footnote{independently of any
special choice of moduli (such as making a determinantal expression vanish)}
have a section; then, of course, the same holds for $L$ and the Pfaffian
will vanish identically on the moduli space (the corresponding {\em vanishing
condition} is studied in sect.~\ref{The vanishing condition})
\beqa
\mbox{vanishing condition}\;\;\;\;\;\;&\lra &
\;\;\;\;\;\; Pfaff \equiv 0
\eeqa

In carrying out this program there occurs a difficulty which deserves to be
mentioned. In the reduction step one needed to show that a certain line bundle
(the bundle $L_2$) above is related to an {\em effective} divisor.
Now in many cases when line bundles on spectral surfaces $C$ in $X$
or on spectral curves $c$ in $\E$ are considered it is enough
to restrict attention to objects induced by restriction 
from the ambient space, be it $X$ or $\E$. For example, in our case 
where line bundles over the curve $c\subset \E$ are relevant,
by restricting generality in the manner indicated and looking only
to objects like $\cO_{\E}(\alpha s + \beta F)$, where $b$ and $F$ are
the rational base and elliptic fibre of the elliptic surface $\E$, 
the question of effectiveness reduces simply to the question whether 
integral coefficients are non-negative. This way of doing things suffices
for some major examples, like the Examples 2 and 4 which are repeatedly
studied in this paper like in [\ref{BDO}]. However there are other examples
which show quite interesting behaviour where employing this restriction
is not sufficient, like in the Example 1. More precisely 
what happens is this: although $L_2$ arises as a restriction from $\E$ to $c$
of a line bundle $L_2'=\cO_{\E}(D_2')$ on $\E$ 
it is not necessary that $D_2'$ is effective as a divisor on $\E$; this is
only sufficient 
\beqa
D_2'\;\;\; \mbox{effective} \;\;\;
& \stackrel{\mbox{\normalsize{$\not\Lla$}}}{\Lra} &
\;\;\; D_2=D_2'|_c \;\;\; \mbox{effective} 
\eeqa
(the difference between the easily applicable, but too strong 
condition of effectiveness of $D_2'$ and the precise condition of 
effectiveness of $D_2$ is studied in sect.~\ref{difference subsection}).
What one needs to study actually is whether $D_2$ is effective
as a divisor on $c$ which clearly is more difficult. Nevertheless
cases like the example 1 mentioned need for their 'factorisation reduction'
the employment of such more subtle line bundles. 

After having gone in the required technical detail 
through the steps described above
in sections \ref{section reduction} to \ref{difference subsection}
we will apply these methods to our main Examples 2 or 4 and 1
in sect.~\ref{n=3, la =3/2} and \ref{Ex 1 sect}, respectively.
All other cases are listed in sect.~\ref{red cases for bar al = n} 
and app.~\ref{list subsection}.

\newpage

\section{\label{spectralcovermethod}Spectral $SU(n)$ bundles over the elliptic
Calabi-Yau threefold $X$}

\resetcounter

In case $X$ admits an elliptic fibration $\pi: X\rightarrow B$ with 
a section $\sigma$ one can describe the bundle $V$ 
explicitly via the spectral cover $C$ of $B$: 
the data of an $SU(n)$ bundle are encoded by 
an $n$-fold ramified cover surface $C$
of the base $B$, which datum comes down to a class $\eta$ in $H^{1,1}(B)$,
and a line bundle $L$ over $C$; the latter,
under the standing assumption $h^{1,0}(C)=0$, reduces to the datum given 
by a class $\gamma$ in $H^{1,1}(C)$ (whose non-triviality is crucial to get 
chiral matter [\ref{C}]).

\subsection{\label{Elliptic CY}
The elliptic Calabi-Yau space $X$ over the surface $B$}

Before describing the bundle in greater detail let us first elucidate
the structure of the space $X$. Thre threefold is actually given as a 
hypersurface in an ambient four-fold ${\cal W}_B$ 
which itself is defined as a ${\bf P^2}$-bundle over the base $B$
\beqa
{\cal W}_B&=& {\bf P}(\cL^2\oplus \cL^3\oplus \cO)
\eeqa
where $\cL=K_B^{-1}$.
The homogeneous coordinates in the fibre ${\bf P^2}$ are denoted by $x,y,z$
and $X$ is the divisor given by vanishing of the defining equation
$zy^2=4x^3-g_2xz^2-g_3z^3$ where $g_2$ and $g_3$ are sections of 
$\cL^4$ and $\cL^6$. 

The base $B$ is actually either a del Pezzo surface
(including ${\bf P^2}$ and some blow-ups), a Hirzebruch surface
(plus some blow-ups) or an Enriques surface.
As we will have to consider rational curves, as support of the 
world-sheet instanton, in $B$ we recall in app.~\ref{rational curves}
the relevant facts in this regard and list the possible cases.

\subsection{The spectral cover surface $C$ over $B$}

The idea of the spectral cover description of an $SU(n)$ bundle $V$ is to 
consider first the bundle over an elliptic fibre $F$, 
and then to paste together these descriptions allowing global twisting data. 
$V$ (assumed to be fibrewise semistable)
decomposes fibrewise as a direct sum of line bundles of degree zero,
encoded by a set of $n$ points summing up to 
zero in the group law:
this is the point $p_0=(0,1,0)$, the point  
at infinity $x=y=\infty$ in affine coordinates with $z=1$;
it is globalized by the section $\si$.
Letting 
this vary over the base $B$ one gets a hypersurface $C \subset X$,
a ramified $n$-fold cover of $B$. Denoting the cohomology
class of $\si(B)\subset X$ by $\sigma$ one finds, allowing the twist by a 
line bundle $\M=\cO_B(M)$ over $B$ with $c_1(\M)=[M]=\eta \in H^{1,1}(B)$ 
\beqa
C=n\sigma + \eta
\eeqa
For the $n$-tuple of points $\{ p_i \; | \; i=1, \dots , n\}$ (on a fibre)
there exists a unique (up to a factor in ${\bf C^*}$) meromorphic function
$w$ of divisor $(w)=\sum_{i=1}^n (p_i - p_0)$: this means
in the standard convention for divisors
of meromorphic functions that $w$ has zeroes at the $p_i$
(i.e.~$f$ would be holomorphic for $(f)$ effective). This $w$ is, 
given in inhomogeneous form, a polynomial in $x$ and $y$
(which have a double and triple pole at $p_0$, respectively)
\beqa
\label{locus}
w & = & a_0+a_2x+a_3y+\dots \, a_nx^{(n-3)/2}y\; =\; 0
\eeqa
(this is for $n$ odd\footnote{for $n=3$ in sect.~\ref{Ri section}
we call $a_0=C, a_2=B, a_3=A$ and write
$D_m=\sum_{i=0}^m d_i u^i v^{m-i}$ for $D=C,B,A$ when restricting the 
consideration to the projective line $b\subset B$ 
with its homogeneous coordinates $(u,v)$; the subscript $m$ denotes
the degree (identified as $r, r-2\chi, r-3\chi$ 
in sect.~\ref{coordinates for curve case}) of the homogeneous polynomial $D$ 
(the reader will not confuse the
$a_i$ in (\ref{locus}) with the coefficients of $A$)}; 
for $n$ even the last term reads $a_nx^{\frac{n}{2}}$).
Globally $C$ is given as the locus (\ref{locus}) 
with $w$ a section of ${\cal O}(\sigma)^n \ox \M$ 
(here $\M$ being understood as pulled back to $X$)
and $a_i\in H^0(B, \M \ox \cL^{-i})$.

$\eta$ has to fulfill some conditions. As the spectral cover 
is an actual surface one needs
\beqa
C \;\;\;\; \mbox{effective} \;\; \;\; \;\; \;\; \;\; \;\; 
(\; \Llra \;\;\; \eta \geq 0)
\eeqa
(i.e. $\eta$ effective).
Second, to guarantee [\ref{FMWIII}]
that $V$ is a {\em stable} vector bundle, one needs [\ref{irred}]
\beqa
\label{irreducibility}
C \;\; \mbox{irreducible} & \Llra &\;\;\;\;\;\;\;
\{a_0=0\} \; \mbox{irreducible} \;\;\;\;\;\;\; \;\;\;
\;\;\; (\Llra \;  \eta \cdot b \geq 0)\\
\label{irreducibility effectivity}
&&\mbox{and} \,\; \{a_n=0\}=C\cdot \si \;\; 
\mbox{effective}\;\;\; (\Llra \; \eta - n c_1 \geq 0)
\eeqa

\subsubsection{The moduli of $V$}

The isomorphism class of $V$ in this set-up will be determined by
$C$ and a certain line bundle $L$ over it, specified in the next subsection;
its cohomology class will have to take a specific form (to get $c_1(V)=0$). 
Therefore an important subclass of cases 
(and the one to which we restrict ourselves throughout)
is given by spectral cover surfaces with
the divisor $C$ not just effective but even ample (positive)
\beqa
C \;\;\;\; \mbox{ample} \;\; \;\; \;\; \;\; \;\; \;\; 
(\; \Lra \;\;\; \eta -n c_1 >0)
\eeqa
$C$ will then have the property $h^{1,0}(C)=0$ (inherited from $X$)
and so line bundles on $C$ are characterised by their Chern classes. 
That is, the (continuous) bundle moduli of $(X,V)$ 
are then given just by ${\bf P}H^0(X, \cO_X(C))$, 
i.e.~the different choices of $C$ which in turn are parametrised 
by the polynomial coefficients of its
defining equation (in addition there is a discrete parameter $\lambda$
described in the next subsection).


\subsection{The line bundle $L=\un{L}|_C$ over the spectral cover surface $C$}

One describes the $SU(n)$ bundle $V$ over $X$ by a line bundle 
$L$ over $C$ 
\beqa
V=p_*(p_C^*L\otimes {\cal P})
\eeqa
with $p:X\times_B C\ra X$ and $p_C: X\times_B C\ra C$ the projections and 
${\cal P}$ the global variant of (a symmetrized version of) the Poincare line 
bundle over $F_1\times F_2$, i.e. the universal bundle which realizes $F_2$ as
moduli space of degree zero line bundles over $F_1$.
$L$ is specified by a half-integral number $\lambda$. This occurs as
$c_1(V)=\pi_*\big(c_1(L)+\frac{c_1(C)-c_1}{2}\big)=0$ implies
\beqa
\label{c_1(L)}
c_1(L) \; = \; -\frac{1}{2}\Big(c_1(C)-\pi_{C*}c_1\Big)+\gamma \;
= \; \frac{n\si + \eta + c_1}{2}|_C+\gamma
\eeqa
(we will omit usually the obvious pullbacks).
Here $\gamma$ denotes the only generally given class in the kernel of
$\pi_{C*}:H^{1,1}(C)\ra H^{1,1}(B)$, i.e. 
($\underline{\gamma} \in H^{1,1}(X)$)
\beqa
\label{gamma definition}
\gamma \; =  \; \underline{\gamma}|_C \;\;\;\; \mbox{with} \;\;\;\;
\underline{\gamma}  \; =  \; \lambda \, \Big(n\sigma-(\eta-nc_1)\Big)
\eeqa
This gives precise integrality conditions for $\la$:
if $n$ is odd, then one needs actually $\la\in \frac{1}{2}+{\bf Z}$;
if $n$ is even, then $\la \in \frac{1}{2}+{\bf Z}$ needs 
$c_1 \equiv 0 \; \mbox{mod}\, 2$
and $\la \in {\bf Z}$ needs $\eta \equiv c_1 \; \mbox{mod}\, 2$.

Assuming $h^{1,0}(C)=0$ line bundles on $C$ are characterised by their 
Chern classes. Therefore one can define line bundles $\underline{\G}$ and $\G$ 
on  $X$ and $C$, respectively, by
\beqa
c_1(\underline{\G}) \; = \; \underline{\gamma} \;\; ,  
\;\; c_1(\G) \; = \; \gamma
\eeqa
(when we want to make the $\lambda$-dependence explicit we denote these by 
$\underline{\G}_{\lambda}$ and $\G_{\lambda}$). They are related to
corresponding divisor classes (modulo linear equivalence) 
$\underline{G}$ and $G$ with
\beqa
\underline{\G} \; = \; \cO_X\big( \underline{G}\big)\;\; , \;\;
\G \; = \; \cO_C\big( G \big)
\eeqa
i.e. one has  $\underline{G}=\lambda (n\si - \pi^*(M + n K_B))$
and, for example, $G|_c=\la(ns-(r-n\chi)F)|_c$.

Note that all these considerations of $\underline{\G}$ and $\G$ apply strictly
only formally as the corresponding Chern classes will, taken alone for 
themselves, be only half-integral in general; only the full combination in 
(\ref{c_1(L)}) will be integral and define a proper line bundle. 
Similar remarks apply to the formal decompositions written below
($K_B$ denotes here a line bundle and also the corresponding divisor class).

Explicitly one finds for the various incarnations of the spectral line bundle
\beqa
\label{cal L def}
\underline{L} & = & \Big( \cO_X(\si)^n \ox \pi^* \M \ox 
\pi^* K_B^{-1}\Big)^{1/2}\ox 
\underline{\G}\; = \; 
\cO_X\Big( \frac{C+\pi^*\, K_B^{-1}}{2}+\underline{G}\Big)\\
L & = & K_C^{1/2} \otimes\pi_C^*K_B^{-1/2}\otimes \G
\eeqa
Explicitly one has for the Chern class 
\beqa
c_1(\underline{L}) & = & 
n\Big(\la + \frac{1}{2}\Big)\si + \Big(\frac{1}{2}-\la\Big)\eta 
+ \Big(\frac{1}{2}+n\la\Big)c_1 
\eeqa


\section{\label{restriction to E}
Spectral $SU(n)$ bundles over the elliptic surface $\E$}

\subsection{\label{chi subsection}
The elliptic surface $\E$ over the instanton curve $b$}

\resetcounter

A horizontal curve lies in two surfaces in $X$: in the base $B$
and in $\E= \pi^{-1}b$, the elliptic surface over $b$.
Let us define the following expressions related to the restriction to $\E$
\beqa
s\; :=\; \si|_{\E}\;\;\;\;\; , \;\;\;\;\; \chi\; := \; c_1 \cdot b \in {\bf Z}
\eeqa
(with $c_1:=c_1(B)$).
With the adjunction relation 
$\si^2=-c_1\si$ and (\ref{irreducibility}) 
one finds
\beqa
\label{chi k relation}
s^2\; =\; -\chi \leq 0
\eeqa
with $\chi \geq 0$ from our assumptions (\ref{resBdef}), (\ref{defE}).
The tangent bundle decomposes over $b$ 
\beqa
TX|_b & = & \cO_b(2)\; \oplus \cO_b(a_h)\; \oplus \cO_b(a_v)
\eeqa
with the latter two terms comprising the normal bundle where $a_h+a_v=-2$.
If $b$ is {\em isolated} then $a_h=a_v=-1$.
Clearly for our horizontal curve $b$ one has
\beqa
a_h \; = \; (b|_B)^2\, , \;\;\;\;\;\; a_v \; = \; (b|_{\E})^2:= s^2=-\chi
\eeqa

The canonical class of the elliptic surface $\E$ is
\beqa
\label{K relation}
K_{\E}=\pi_{\E}^* \Big(K_b+{\cal O}_b(\chi)\Big)=\cO_{\E}\big( (\chi-2)F\Big)
\; 
\Lra \; c_1(\E)=(2-\chi)F
\eeqa
Equivalently the relative dualizing sheaf is
\beqa
\omega_{\E/b}\; & = & \; K_{\E}\ox \pi_{\E}^*\, K_b^{-1}\; 
= \; \pi_{\E}^* \, \cO_b(\chi)
\eeqa
Here 
(cf. for example [\ref{AChor}]; note that the discriminant of $X$ over $B$ has
class $\Delta = 12 c_1$) 
\beqa
\chi=\chi(\E, {\cal O}_{\E})=\frac{1}{12}e(\E)=c_1\cdot b
\eeqa

So, all one needs to know about the position of $b$ in the base $B$ 
is encoded by the number $\chi$.
Beyond that we will need the rank $n$ of the vector bundle $V$ 
and its spectral data $\eta, \lambda$, or rather $r, \lambda$ after the
restriction to $\E$, cf. (\ref{def r}). One upshot of the discussion above
is that $\chi =1$ is the relevant case to consider; for contrast we 
also consider $\chi=0$ (where the whole interpretation changes)
and so keep the parameter $\chi$ manifest throughout.

For $b$ the base in the ${\bf P^1}$-fibered Hirzebruch surface ${\bf F_k}=B$
one has the following cases (where $\chi=2-k\geq 0$ and $c_1(\E)=kF$)
\begin{center}
\begin{tabular}{|c||c|c|c|}
\hline
$B$ & ${\bf F_0}$ & ${\bf F_1}$ & ${\bf F_2}$\\
\hline
$\E$ & $K3$  & $dP_9$ & $b\x F$ \\
\hline
\end{tabular}
\end{center}
The case $B={\bf P^2}$ 
of $c_1=3l$ and $c_1\cdot l=3$ 
or $ c_1\cdot 2l=6$ gives an $\E$ of Euler number $36$ or $72$, respectively.

\subsection{\label{coordinates for curve case}
The spectral cover curve $c$ over $b$}

The spectral surface $C$ of our bundle $V$ being an $n$-fold 
cover of the base one has 
\beqa
\label{big proj}
C \;\;\;\;\;\;\;\; & \hra & \;\;\; X\nonumber \\
\;\;\; \pi_C\,\da \; n:1& &\pi \,\da \, F \;\;\; \;\;\; \\
B \;\;\;\;\;\;\;\; & = & \;\;\; B \nonumber
\eeqa
The support of the world-sheet instanton we 
consider is a rational curve $b$ inside the base $B$
(in turn embedded in $X$ by the zero section $\si$); specifically
one may think of $b$ as given by the base ${\bf P^1_b}$
inside the ${\bf P^1_f}$-fibered surface $B={\bf F_k}$.
Let $\E:= \pi^{-1}b$ be the elliptic surface over $b$
and $c:=C|_{\E}$ the corresponding spectral curve of $V|_{\E}$
\beqa
\label{small proj}
c \;\;\;\;\;\;\;\;\; & \hra & \;\;\;\;\; \E\nonumber \\
\pi_c \,\da \; n:1& & \pi_{\E} \,\da \, F  \;\;\;\; \\
b \;\;\;\;\;\;\;\;\; & = &\;\;\; \;\; b\nonumber 
\eeqa
(which we assume irreducible as we did for $C$).
If the whole description is restricted from the 
elliptic threefold $X\subset \cW_B$ over $B$
(where the fourfold $\cW_B$ is the ${\bf P^2}$-bundle 
of Weierstrass coordinates over $B$)
to the elliptic surface $\E$ over $b$ one gets again the 
equation (\ref{locus}) for $c\subset \E \subset \cW_{b}$
(in the threefold given by the ${\bf P^2}$ bundle 
of Weierstrass coordinates over $b\cong {\bf P^1}$); 
what was $\cL=K_B^{-1}$ for the situation over $X$, 
becomes here $\cL|_b=\cO_b(\chi)$ such that now
$a_i|_b\in H^0(b, \cO_b(r-i\chi))$
(where $r:=\eta\cdot b$ such that $\M|_b=\cO_b(r)$, cf.~below).
The section $s$, i.e.~concretely the (group-)zero point
$p_0=(x_0, y_0, z_0)=(0,1,0)=(z)\cap F_t$ 
in each fibre over a point $t=u/v\in {\bf P^1}=b$,
consists, when restricted to $c$, 
out of $s\cdot c=r-n\chi$ points (here $(z)$ is the locus where $z=0$). 

Let us define the following expression related to the restriction to $\E$
\beqa
\label{def r}
r & := & \eta \cdot b \in {\bf Z}
\eeqa
So $C=n\si + \eta$ gives $c=ns+rF$. With $\eta = xb+yf$ on ${\bf F_k}$
one finds $r\geq 0$
with
$r=0 \Leftrightarrow \eta = x \, b_{\infty}$ where 
$b_{\infty}=b+kf$. This case can occur only over ${\bf F_2}$ as 
by (\ref{irreducibility effectivity}) 
\beqa
\label{r chi relation}
r\geq n\, \chi
\eeqa
For $C$ ample (as we will assume) one gets even (such that 
in particular always $r>0$)
\beqa
\label{r chi relation sharp}
r> n\, \chi
\eeqa

The integer $r$ has an interpretation as an 
instanton number\footnote{The class $\gamma$ would not occur 
in the specification 
of a spectral bundle $V_{\E}$ over the one-dimensional
base $b$ where $\pi_{c\, *}:H^{1,1}(c)\ra H^{1,1}(b)$ 
is injective, cf.~[\ref{FMW}],
in accord with the fact that by (\ref{r instnr})
$c_2(V|_{\E})$ sees only the $\gamma$-free part of
$c_2(V)$ and not
$\omega =  -\frac{n^3-n}{24}c_1^2-\frac{n}{8}\eta(\eta-nc_1)
-\frac{1}{2}\pi_*(\gamma^2)\; = \; 
-\frac{n^3-n}{24}c_1^2+(\la^2-\frac{1}{4})\frac{n}{2}\eta(\eta-nc_1)$.}
\beqa
\label{r instnr}
c_2(V)=\eta \si + \omega & \Lra & c_2(V|_{\E})=r 
\eeqa

The canonical bundle of $c$ is given by (cf.~(\ref{K relation}))
\beqa
\label{can bund of c}
K_c&=&\Big( \cO_{\E}(c)\ox K_{\E}\Big)|_c\; =
\; \cO_{\E}\Big( ns+(r+\chi-2)F\Big)|_c
\eeqa
For later use, cf.~the comments after (\ref{specialisation}) below, 
we remark that (which is $\geq 0$ under our assumptions)
\beqa
\label{canonical degree}
\deg K_c^{1/2} =g_c-1& = & n (r - \frac{n-1}{2}\chi -1)
\eeqa

The cohomologically nontrivial line bundles on $\E$ 
which come from $X$, i.e.~are of the form $\cO_{\E}(xs+yF)$, and
which become flat on $c$ are powers of\footnote{for $gcd(n, r-n\chi)=1$,
which is fulfilled automatically in the cases of application 
(where $\chi=1, n\leq 5$)}
\beqa
\label{Lambda}
\Lambda&:=& \cO_{\E}\Big( ns - (r-n\chi)F\Big)
\eeqa

\subsubsection{\label{specialisation subsection}Moduli of the spectral curve}

The coefficients of the homogeneous polynomials $a_i$ 
are (after one overall scaling) 
moduli $m\in {\cal M}_{\E}(c)={\bf P}H^0(\E, \cO_{\E}(c))$ 
of external motions of $c$ in $\E$, that is of those part of the moduli 
in ${\cal M}_{bun}(X,V)$ which is relevant in our consideration over $\E$.

{\em Behaviour over the special sublocus 
$\Sigma_{\Lambda}=\{f_{\Lambda}=0\}$ of the moduli space ${\cal M}_{\E}(c)$}

Consider in the moduli space ${\cal M}_{\E}(c) (\ni m)$ the 
specialisation locus $\Sigma_{\Lambda}$ where
\beqa
\label{Lambda triviality on c}
\Lambda|_c &\cong & \cO_c
\eeqa
and let us assume that this locus is characterised by
a (set of) condition(s) $f_{\Lambda}(m)=0$, 
say\footnote{\label{vector-valued fct}This locus will 
in general (cf.~later Example $1$) have codimension higher
than one, so $f$ should be interpreted as a vector-valued 'function';
i.e.~two or more conditions have to be posed at the same time.}.
Now, one of the principal objects of our study (cf.~(\ref{prec criterion})), 
the line bundle $l(-F)|_c=\cO_{\E}\Big(n(\la + \frac{1}{2})s+\beta F\Big)|_c\; 
\cong  K_c^{1/2}\ox \Lambda|_c^{\la}$, 
cf.~(\ref{degree equalities}) and (\ref{l(-F)}), 
becomes along $\Sigma_{\Lambda}$
\beqa
\label{specialisation}
l(-F)|_c & \buildrel \mbox{\scriptsize on} \;\Sigma_{\Lambda} \over \lra & 
\cO_{\E}\Big([(\la + \frac{1}{2})(r-n\chi)+\beta]F\Big)|_c=
\cO_{\E}\Big((r-\frac{n-1}{2}\chi-1)F\Big)|_c\, 
\buildrel \Sigma_{\Lambda} \over 
\cong \, K_c^{1/2}
\;\;\;\;\;\;\;\;\;\;\;
\eeqa
Note that (\ref{specialisation}) is a bundle of integral degree on $c$.
We did exclude here the case $n$ even, $\chi$ odd, $\la\in {\bf Z}$
where the alternative evaluation $\cO_{\E}(\frac{n}{2}s+\frac{r-1}{2}F)|_c$ 
applies.
So (with the mentioned alternative evaluation understood) we see 
{\em that $l(-F)|_c$ becomes along $\Sigma_{\Lambda}$ effective}:
$l(-F) \cong  \cO_{\E}(mF)|_c$ where $m=r-\frac{n-1}{2}\chi-1$
(and similarly for $n$ even).

\subsubsection{\label{Ri section}The case of $SU(3)$ bundles}

Here ($C=a_0, B=a_2, A=a_3$; $m\in  H^0\Big( \E, \cO_{\E}(3s+[m]\chi F)\Big)$
for $m=z, x, y; [m]= 0, 2, 3$)
\beqa
\label{equ spec curve}
w\; = \; C_r\, z + B_{r-2\chi} \, x + A_{r-3\chi}\, y\; 
& \in & H^0\Big( \E, \cO_{\E}(3s+rF)\Big) \\
& \cong &
zH^0\Big(b, \cO(r)\Big)\oplus xH^0\Big(b, \cO(r-2\chi)\Big)
\oplus yH^0\Big(b, \cO(r-3\chi)\Big) \nonumber
\eeqa

The defining equation for $c$ is
$w=C_r(t)z+B_{r-2\chi}(t)x+A_{r-3\chi}(t)y=0$
where $t=(u,v)\in b$ with homogeneous coordinates $u,v$ on $b$, 
i.e.~(suitable pull-backs understood) 
$u,v\in H^0\Big( \E, \cO_{\E}(0s+1F)\Big)$. So 
for the special case $n=3$ the number $r-n\chi$ of points of $s\cdot c$
coincides with the degree of $A_{r-3\chi}$,
and the points are just the $r-3\chi$ points $\{p_0, t_i\}\in \E$
in the fibres  $c_{t_i}:=F_{t_i}|_c$ 
over the zeroes $t_i$ of $A_{r-n\chi}=\prod_{i=1}^{r-n\chi}A_1^{(i)}$
\beqa
\label{s|_c generically}
s|_c&=&\sum_{A_{r-3\chi}(t_i)=0}\{p_0, t_i\}
\eeqa
Generically in $c$-moduli (for $B(t_i)\not = 0$)
these are {\em simple}\footnote{though a three-fold touching point of the line 
$(z)_t$ and the elliptic curve $F_t$ in the
${\bf P}^{\bf 2}_t$ over $t\in b$}
points in the fibres $F_{t_i}$ w.r.t.~the intersection of $s=p_0$
with the defining line $(w)_{t_i}=\{C(t_i)z+B(t_i)x=0\}$ of $c$ 
(unequal to the line $(z)_{t_i}=\{z=0\}\subset {\bf P}^{\bf 2}_{t_i}$): 
here $C(t_i)z+B(t_i)x=0$ determines $x$ from $z$, and $y$ from
the Weierstrass equation, giving the other two points $p_i^{\pm}$ of
$c_{t_i}=\{p_0+p_i^++p_i^-, t_i\}$.

\noindent
{\em The codimension one sublocus $Res(A,B)=0$ in ${\cal M}_c$ }

The generic result (\ref{s|_c generically})
changes at the specialisation locus $Res(A,B)=0$ 
(where $B_3(t_1)=0$, cf.~app.~\ref{resultant}): 
the $(w)$ line (where $w=0$ in the ${\bf P}^{\bf 2}_{t_1}$-fibre in $\cW_b$)
becomes  just the $(z)$ line and $p_0$ becomes a three-fold point of $c_{t_1}$ 
(where $\sim$ means linear equivalence)
\beqa
(z)|_c=3s|_c=\sum_i \{3p_0, t_i\} = F_{t_1}|_c + \sum_{i\geq 2}\{3p_0, t_i\}
\neq  F_{t_1}|_c +\sum_{i\geq 2}F_{t_i}|_c \sim (r-n\chi)F|_c
\eeqa
{\em The codimension $r-n\chi$ sublocus $A|B$ or $R_i=0, i=1, ..., r-n\chi$ 
in ${\cal M}_c$ }

Here we demand that {\em not one but all} $r-n\chi$ roots
of $A$ are in common with $B$, i.e.~$A|B$.
Now $f_{\Lambda}=0$ is (implied by) a set of $r-n\chi$ conditions
$R_i=0$ which are just the $r-n\chi$ conditions $B(t_i)=0$,
i.e.~$R_i=Res(B, A_1^{(i)})$: demanding that all $R_i=0$ 
the previous argument gives now in each $c_{t_i}$ a threefold $\{p_0, t_i\}$,
so $\cO_{\E}(3s)|_c  \cong  \cO_{\E}((r-3\chi)F)|_c$ or
$\Lambda|_c \cong  \cO_c$ (cf.~sect.~\ref{specialisation subsection}), 
so (here no converse, cf.~Example 4 in sect.~\ref{Ex 4 case})
\beqa
\label{Ri = 0 implication}
A|B & \Llra & R_i=0, \forall i \;\;\; \Longrightarrow \Lambda|_c \cong \cO_c
\eeqa
So one has $(R_i)_i=0 \Lra f_{\Lambda}=0$.
Note further that the codim $r-n\chi$ locus 
$\{R_i(m)=0, \forall i\} (\subset \Sigma_{\Lambda})$ 
is a subset
of the codim 1 locus $\{Res(B, A)(m)=0\}$, 
i.e.~$(R_i)_i=0\Lra Res(B, A)=0$, so
(a power of) $Res$ is a 
combination\footnote{\label{multiple resultant zero}From
$R_i=0, \forall i$ one expects 
an $(r-3\chi)$-fold zero of the resultant; for a representation
$Res=P_{r-n\chi}(R_1, ..., R_{r-n\chi})$, with $P_{r-n\chi}$ a homogeneous 
polynomial of degree $r-n\chi$ in the $R_i$, 
cf.~(\ref{f in the R_i}),(\ref{Ex 4 resultant}).}
(in the ideal sense) of the $R_i$.

\subsection{\label{twist class}The line bundle $l|_c$
over the spectral cover curve $c$}

As a further datum describing $V$ beyond the 
surface $C$, which encodes $V$ just fiberwise,
one has a line bundle $L$ over $C$ with  $V=p_*(p_C^*L\otimes {\cal P})$.
$L$ arises in the simplest case as a restriction $L=\underline{L}|_C$ 
to $C$ of a line bundle $\underline{L}$ on $X$.
We define also a corresponding restriction 
$l:=\underline{L}|_{\E}$ to $\E$ (with $l|_c=L|_c$).
So one has the inclusions of line bundles
\beqa
L \;\;\;\;\;\; \hra \;\;\; \underline{L} \;\;\;\;\;
&\;\;\;\;\;\;\;\;\;\;\;\;\;\;\;&\;\;\;\;\;
\; l|_c \;\;\;\;\;\;  \hra  \;\;\;\; l
\nonumber \\
\da \;\;\;\;\; \;\;\;\;\;\;\;\;  \;\;\; \da \;\;\;\;\;
& \;\; \mbox{and} \;\;& \;\;\;\;\;
\; \da \;\;\;\;\;\;\;\;  \;\;\; \;\;\; \;\;\; \da 
\\
C \;\;\;\;\;\; \hra  \;\;\; X\;\;\;\;\;
& \;\;\;\;\; &\;\;\;\;\;
\; c \;\;\;\;\;\;\;  \hra  \;\;\;\; \E 
\nonumber
\eeqa

The crucial fact is that one has for spectral bundles
\beqa
\label{bundle projection}
V|_{B}=\pi_{C*} L  \;\;\;\;\;\;\;\;\;\;\;\;\;\;\; 
\mbox{such that}\;\;\;\;\;\;\;\; V|_b=\pi_{c*}\; l|_c
\eeqa
Similarly to (\ref{cal L def}) 
one has with $\underline{\gamma}|_{\E}=\lambda\big(ns-(r-n\chi)F\big)$ 
and $\cL|_b=K_B^{-1}|_b=\cO_b(\chi)$
that\footnote{\label{K_c footnote}The equality to the second line
in (\ref{degree equalities}) follows from
$l^2|_c \ox \cF^{-2}\cong
\cO_{\E}\Big( (ns+rF) + \chi F\Big)|_c=\cO_{\E}\Big( (ns+rF-kF)+2F\Big)|_c
\cong K_c \otimes \pi_c^*K_b^{-1}$
using (\ref{can bund of c}).} 
\beqa
\label{l def}
l&=&\underline{L}|_{\E}\; \cong \; 
\cO_{\E}\Bigg( \frac{c+\pi_{\E}^{*}\cO_b(\chi)}{2}+\underline{G}|_{\E}\Bigg)
\; = \; 
\cO_{\E}\Bigg( \frac{ns+(r+\chi)F}{2}\Bigg)\ox \underline{\G}|_{\E} 
\;\;\;\;\;\;\;\;\; \nonumber\\
&& \Lra \;\;\;\;\;\; l(-F) \; \cong \; \Big( K_{\E}\ox \cO_{\E}(c)\Big)^{1/2}
\ox \Lambda^{\la}\\
\label{degree equalities}
l|_c&=&\cO_c\Bigg( \frac{ns + (r+\chi)F}{2}|_c \Bigg)\otimes \cF = L|_c \cong
\Big(K_C^{1/2} \otimes\pi_C^*K_B^{-1/2}\Big)|_c \otimes \cF
\nonumber\\
&\cong& K_c^{1/2}\otimes\pi_c^*K_b^{-1/2}\otimes \cF
\;\;\; \Lra \;\; l(-F)|_c\; \cong  K_c^{1/2}\ox \Lambda^{\la}|_c
=\; K_c^{1/2}\otimes \cF
\eeqa 
Here we have introduced the flat (cf.~below) line bundle on $c$
given by the restriction
\beqa
\cF & = & \G|_c\; = \; \cO_c\big( G|_c\big)
\eeqa
Explicitly one has for the Chern class of the line bundle $l$ on $\E$
\beqa
c_1(l) & = & n\Big(\la+\frac{1}{2}\Big)s
+\Big( (\frac{1}{2}-\la)r+(\frac{1}{2}+n\la)\chi\Big) F 
\eeqa
One notes 
that the line bundle $\cF$ over $c$ is flat as
\beqa
\label{ineffectivity}
\Big(ns-(r-n\chi)F\Big)(ns+rF)=0& \Lra & \deg \, G|_c=0
\eeqa
Note that the flat bundle $\cF$ has continuous moduli corresponding 
to $Jac(c)$ as $h^{1,0}(c)\neq 0$ whereas we did assume $h^{1,0}(C)= 0$
so that we had only the discrete twist datum $\gamma$
(we will not have to treat the ambiguities of $K_c^{1/2}$ 
in the present paper).

\section{\label{WS instantons}World-sheet instantons and 
superpotential contribution}

\resetcounter

\subsection{General case: $SU(n)$ bundles over an instanton curve $b$}

To set the stage we recall first some versions of the criterion
for a world-sheet instanton to contribute to the superpotential $W$. 
Let us assume that $V$ is an $SU(n)$ bundle, embedded in the first $E_8$ group.
A world-sheet instanton, supported on an isolated rational curve $b$, 
contributes according to the following criterion [\ref{Wittenhet}] 
(with $V|_b(-1)$ denoting $V|_b \ox \cO_b(-1)$)
\beqa
\label{main criterion}
W_b\neq 0 & \Llra & h^0\Big(b, V|_b(-1)\Big)=0
\eeqa
Considered with respect to the structure group $SO(2n) \supset SU(n)$ 
(with $n\leq 8$) one has 
\beqa
V|_b&=&\bigoplus_{i=1}^n\, \cO_b(\kappa_i)\oplus \cO_b(-\kappa_i)
\;\;\;\;\;\;\;\;\;\;\;\;\;\; (\kappa_i\geq 0)
\eeqa
such that 
$V|_b(-1)=\bigoplus_{i=1}^n\, \cO_b(\kappa_i -1)\oplus \cO_b(-\kappa_i -1)$
gives 
\beqa
h^0(b, V|_b(-1))=\sum_{\kappa_i -1 \geq 0} \kappa_i=\sum \kappa_i
\eeqa
Therefore $b$ contributes precisely if $V|_b$ is trivial, i.e.
\beqa
\label{triviality criterion}
W_b\neq 0 & \Longleftrightarrow & V|_b = \bigoplus_{i=1}^n\, \cO_b
\eeqa
A corresponding framing would give $n$ linearly independent 
global sections such that
\beqa
\label{n global sections}
W_b\neq 0 & \Longrightarrow & h^0(b, V|_b)=n
\eeqa
(This is, of course, also directly a consequence of (\ref{main criterion}), 
cf. app.~\ref{Lemmata}, Lemma 1). A counterexample to the converse
of (\ref{n global sections})
is given by $V|_b = \cO_b\oplus \cO_b(1)\oplus \cO_b(-1)$.

Considered in $SU(n)$ (as we will henceforth do) one has 
\beqa
\label{decomp over b}
V|_b&=&\bigoplus_{i=1}^n\, \cO_b(k_i)\;\;\;\;\;\;\;\;\;\;\;\; 
\mbox{with} \;\;\;\;
\sum k_i=0 \;\;\;\;\;\; (k_i\in {\bf Z})
\eeqa
where now 
\beqa
h^0(b, V|_b(-1))=\sum_{k_i -1 \geq 0} k_i=\sum_{k_i\geq 0}k_i
\eeqa
such that one gets (noting $\sum k_i=0$) again (\ref{triviality criterion})
\beqa
h^0(b, V|_b(-1))=0&\Longleftrightarrow & k_i=0\;\;\;\;
\mbox{for all $i$ with $k_i\geq 0$}\nonumber\\
&\Longleftrightarrow & k_i=0 \;\;\;\;
\mbox{for all $i$}  
\eeqa

\subsection{\label{contribution criteria}Elliptic case with
spectral bundles and a base curve $b$}

\noindent
{\bf \underline{Precise criterion for $W_b\neq 0$}}
\beqa
\label{prec criterion}
W_b\neq 0 \Llra
0=h^0\Big(b, V|_b(-1)\Big)= 
h^0\Big(b, \pi_{c*}l|_c\otimes {\cal O}_b(-1)\Big)=h^0\Big(c, l(-F)|_c\Big)
\eeqa

Note that $l(-F)|_c$ occurs in the short exact sequence of sheaves on $\E$ 
\beqa
0 \lra l(-F-c) \buildrel \iota \over \lra l(-F) \lra l(-F)|_c \lra 0
\eeqa 
(cf.~for the following also [\ref{BDO}])
which gives a long exact sequence of cohomology groups
\beqa
\label{long exact sequence}
0 \lra H^0\Big(\E, \, l(-F-c)\Big) \buildrel \iota_0 \over 
\lra H^0\Big(\E, \, l(-F)\Big) \lra H^0\Big(c, \, l(-F)|_c\Big) \lra ...
\eeqa

\noindent
\underline{{\bf First, necessary criterion for $W_b\neq 0$}}  
\beqa
\label{criterion}
W_b\neq 0 \;\;\;\; \Lra \;\;\;\; 
dim \; \iota_0 \; H^0\Big(\E, \, l(-F-c)\Big) \; 
= \; dim \; H^0\Big(\E, \, l(-F)\Big)
\eeqa

As $\iota_0$ is an embedding this amounts just to 
the condition $h^0 \Big(\E, \, l(-F-c)\Big) 
= h^0\Big(\E, \, l(-F)\Big)$. 
From (\ref{criterion}) one gets as {\em sufficient criterion 
for non-contribution of $b$} 
\beqa
\label{further criterion}
h^0\Big(\E, \, l(-F-c)\Big)= 0  \;\;\; , \;\;\;
h^0\Big(\E, \, l(-F)\Big) >  0  
\;\;\;\; \Lra  \;\;\;\; W_b=0
\eeqa

In the following we make a technical assumption
on the bundle parameter $\la$ (cf. below)
\beqa
\label{lambda assumption}
\la > 1/2
\eeqa
Note that then, as recalled after (\ref{technical assumption}), 
the second cohomology groups 
in the long exact sequence (\ref{long exact sequence}) vanish,
leaving only the three $H^0$- and the three $H^1$-terms 
(cf.~Ex. after (\ref{H^1 reduction conditions})).
By $c_1(V|_b)=0$ and (\ref{index}) the two terms over 
$c$ have equal dimension 
such that
\beqa
\label{dimensions}
h^0\Big( \E, l(-F)\Big) - h^0\Big( \E, l(-F-c)\Big)=
h^1\Big( \E, l(-F)\Big) - h^1\Big( \E, l(-F-c)\Big)
\eeqa
If criterion (\ref{criterion}) is 
fulfilled there is a further, more precise assertion.

\noindent
\underline{{\bf Second, (conditional,) precise criterion for $W_b\neq 0$}}\\
Let the necessary condition for contribution (\ref{criterion}) be
fulfilled and $\la > 1/2$. Then
\beqa
\label{criterion 2}
W_b\neq 0 \;\;\;\; \Llra \;\;\;\;
dim \; \iota_1 \; H^1\Big(\E, l(-F-c)\Big) \; 
= \; dim \; H^1\Big(\E, l(-F)\Big)
\eeqa
This follows by noting that in the long exact sequence 
(which can be written here as)
\beqa
0 \lra H^0\Big(c, l(-F)|_c\Big)\lra H^1\Big(\E, l(-F-c)\Big) 
\buildrel \iota_1 \over \lra 
H^1\Big(\E, l(-F)\Big) \lra H^1\Big(c, l(-F)|_c\Big)\lra 0 \nonumber
\eeqa
the outer (by $c_1(V|_b)=0$, cf.~(\ref{index})), 
and so the inner, two spaces have equal dimension.
So (\ref{criterion 2}) amounts to $\iota_1$ being an isomorphism
(generically in the moduli).

Concretely the map $\iota_1$ is induced from multiplication 
with a moduli-dependent element
\beqa
\label{moduli-dependent 1}
\tilde{\iota} & \in & H^0\Big(\E, \cO_{\E}(c)\Big)
\eeqa

\subsection{\label{Criteria Evaluation}Evaluation 
of the extrinsic contribution criteria}

To evaluate concretely the contribution criteria note first
(by $c=ns+rF$ and (\ref{l def}))
\beqa
\label{l(-F)}
l(-F)&=&{\cal O}_{\E}\Big( \al s + \beta F\Big)\\
\label{l(-F-c)}
l(-F-c)&=&{\cal O}_{\E}\Big( (\al-n) s + (\beta - r) F\Big)
\eeqa
We use the numerical parameters $\al, \beta$ 
given by\footnote{The numerical specification of $\beta$ by (\ref{beta eval}) 
(which in the end goes back just to (\ref{c_1(L)}),
restricted to $\E\subset X$) expresses just that
we tuned parameters to get $c_1(V)=0$ (resp.~the
corresponding version restricted to $\E$). This was crucial to have
$h^1(\E, l(-F-c))=h^1(\E, l(-F))$, cf.~the proof after (\ref{criterion 2})
(working within the assumption $\al>n$, 
so that the $H^2$-terms vanished; 
for a converse  cf.~sect.~\ref{equ cond al > n}).}
(where $n\chi-r\leq 0$ by (\ref{r chi relation}))
\beqa
\label{al eval}
\al & = & n(\la+\frac{1}{2})\\
\label{beta eval}
\beta &=& \beta^{(\chi)}_{n,r}(\lambda) :=\frac{r+\chi}{2}-\la(r-n\chi)-1
\; = \; \Big(\frac{1}{2}-\la\Big)\, r + \Big(\frac{1}{2}+n\la\Big)\, \chi -1
\eeqa
It will be useful to keep on record the following rewriting 
relating $\al$ and $\beta$
\beqa
\label{beta rewriting}
-(\beta+1)& = & \frac{\al -n}{n} \, (r-n\chi) - \frac{n+1}{2}\, \chi
\eeqa

Now $h^0\Big(\E, {\cal O}_{\E}(p s + q F)\Big)$ 
vanishes, if $p > 0$, just for negative $q$ 
(cf. app.~\ref{Lemmata}, Lemma 2)
\beqa
\label{vanishing crit}
\underline{p >0:}\;\;\;\;\;\;\;\;
h^0\Big(\E, {\cal O}_{\E}(p s + q F)\Big)=
h^0\Big(b, \pi_{\E *}{\cal O}_{\E}(p s + q F)\Big)
=0 \;\; \Llra \;\; q < 0 \;\;\;\;\;
\eeqa

In view of (\ref{l(-F-c)}) and (\ref{vanishing crit}) 
we will in the following assume 
\beqa
\label{technical assumption}
\al - n > 0 \;\;\;\;\;\;\;\; \;\;\;\;\; 
\mbox{, i.e.} \;\;\;\;\;\;\;\; \la > 1/2
\eeqa
(for $\la < -1/2$ cf.~app.~(\ref{negative lambda})).
Then sheaf cohomology groups on $\E$ reduce, 
by (\ref{vanishing R^1}), (\ref{reduction}),
to the corresponding push-forwards
(direct images) on $b$, cf.~remark after (\ref{lambda assumption}).

Following criterion (\ref{further criterion}) 
a sufficient condition for $W_b=0$ is by (\ref{vanishing crit}) 
\beqa
\label{beta crit}
\be \geq 0\;\;\; , \;\;\; \be - r < 0
\;\;\;\; \Lra  \;\;\;\; W_b=0
\eeqa
whereas an equivalent formulation of the
condition (\ref{criterion}) (necessary for $W_b\neq 0$) is $\beta < 0$
\beqa
\label{dim equality}
h^0\Big(\E, \, l(-F-c)\Big) \; = h^0\Big(\E, \, l(-F)\Big)
\;\;\;\; \stackrel{\mbox{\normalsize{$\Lla$}}}{\Lra} \;\;\;\; \beta < 0 
\eeqa
$\beta <0$ is sufficient as then both $h^0$ vanish by (\ref{vanishing crit}).
Remarkably, the converse holds: the $h^0$ in (\ref{dim equality}) 
can be equal {\em only} if both are zero (cf.~app.~\ref{Lemmata}, Lemma 3).
So one gets

\noindent
\underline{{\bf Third, necessary criterion for $W_b\neq 0$} 
(case $\la > 1/2$)}  
\beqa
\label{third criterion}
W_b\neq 0 \;\;\;\; \Lra \;\;\;\; \beta < 0 \;\;\;\; \Llra \;\;\;\; 
H^0\Big(\E, \, l(-F)\Big) = 0
\eeqa

\noindent
{\em Remarks:} 1) Note that $\beta <0$ is automatically 
fulfilled for $\chi=0$.\\
2) (\ref{third criterion}) can be formulated as a stronger bound on $\eta|_b$
((\ref{irreducibility}) gave already $r \geq n \chi$)
\beqa
\label{sharper bound}
W_b\neq 0 \Lra 
(\la - \frac{1}{2})\, r \geq (\frac{1}{2}+n\la)\, \chi
\eeqa
This is indeed sharper:
$\frac{1/2 + n \la }{\la - 1/2} = \frac{ (1+n)/2}{\la - 1/2} + n > n$.
So one gets as necessary condition 
\beqa
\label{sharper bound made explicit}
r-n\chi & \geq & \frac{ (1+n)/2}{\la - 1/2}\chi
\eeqa

Let us now come back to the process of making the criteria 
(\ref{criterion}) and (\ref{criterion 2}) more explicit.
If one has $\beta <0$, as we will assume, then
(\ref{criterion}) is fulfilled by (\ref{dim equality})
and one has, by (\ref{criterion 2}), to consider the map
(which can be further explicated using 
(\ref{l(-F)})-(\ref{beta eval}))
\beqa
\label{the critical map}
H^1\Big(\E, l(-F-c) \Big) \stackrel{\iota_1}{\lra} H^1\Big(\E, l(-F)\Big) 
\eeqa
This map is induced from multiplication with an element
$\tilde{\iota}  \in  H^0\Big(\E, \cO_{\E}(ns+rF)\Big)$
and so depends $m\in {\cal M}_{\E}(c)$ (so actually $c=c_m$). 
What (\ref{criterion 2}) says is that 
\beqa
\label{extrinsic determinant criterion}
Pfaff\, (m)\; = \; 0 & \Llra & \det \; \iota_1\; (m) \; = \; 0
\eeqa
(such that
$Pfaff$ equals, up to a constant factor, 
$(\det \iota_1)^m$; actually $m=1$ [\ref{BDO}]).
The moduli space ${\cal M}_{\E}(c)$ has dimension
$h^0\big(\E, \cO_{\E}(c)\big)-1$ which is by the index theorem 
\beqa
\dim \M_{\E}(c)\, = \,
h^0\big(\E, \cO_{\E}(c)\big)-1& = & n(r+1)-\Big(\frac{n(n+1)}{2}-1\Big)\chi
-1\\
&=&n\Big(r-\frac{n}{2}\chi \Big)-(\frac{n}{2}-1)\chi+n-1
\eeqa
in case $c$ is positive (for $b$ isolated we had $\chi=1$).
Finally the degree of the determinant in 
(\ref{extrinsic determinant criterion}) is given 
(using (\ref{h^1 reduction})) by 
(where explicitly $\frac{\al(\al-n)}{n}=(\la^2-\frac{1}{4})n$)
\beqa
\label{h^1 rewriting}
h^1\Big(\E, l(-F)\Big) & = & 
\al\Big(-(\beta+1)\Big)+\frac{\al(\al+1)}{2}\chi -\chi
\; =\; \frac{\al(\al-n)}{n}(r-\frac{n}{2}\chi)-\chi\;\;\;\;\;\;
\eeqa

Another way to express the precise criterion (\ref{prec criterion}) is
(with $\sim$ linear equivalence)
\beqa
\label{effectivity equivalence}
Pfaff(m)=0 & \Llra & h^0\Big(c, l(-F)|_c\Big) > 0 \;
\Llra \;\al s + \beta F|_c \sim \mbox{effective}
\eeqa
So one has not only 
$R_i=0, \forall i \Lra f_{\Lambda}=0$ by (\ref{Ri = 0 implication}) 
but also 
$f_{\Lambda}=0 \Lra  Pfaff =0$ by the remarks after (\ref{specialisation});
if actually codim $\Sigma_{\Lambda}=1$ 
(cf.~the remark in footn.~\ref{vector-valued fct})
one gets\footnote{\label{reduced factor}generally $f=0\Rightarrow g=0$ 
gives immediately only $f^{red}|g$ where $f=\prod f_i^{k_i}$
and $f^{red}=\prod f_i$,
but in our actual
cases (cf.~sect.~\ref{Examples} where this remark applies
in various places) $f$ will be irreducible}
\beqa
\label{general Pfaff factor}
f_{\Lambda}\, | \, Pfaff 
\eeqa


\section{\label{section reduction}The idea of reduction}

\resetcounter

The question of contribution of the world-sheet instanton 
supported on $b$ to the superpotential amounts
to decide whether $h^0(c, l(-F)|_c)=0$ or not
in dependence on the moduli of $c$ (concretely, in the case of $SU(3)$, 
the coefficients of the polynomials $C, B, A$ in the equation of $c$).
The idea of reduction is to translate the
question about $h^0(c, l(-F)|_c)$ to a simpler case. For this one 
introduces another ('smaller') line bundle $\bar{l}$ 
\begin{itemize}
\item leading to a map $\bar{\iota}_1$ between spaces, 
now of lower dimension (arguing as for $l(-F)$);
now one has two interesting cases to consider
\begin{itemize}
\item under a certain
{\em equality condition} these spaces have equal dimension and one is 
led again to the consideration of a (moduli dependent)
determinantal function $f=\det \bar{\iota}_1$ 
\item under a {\em vanishing condition} one has (universally in the moduli)
$\ker \bar{\iota}_1\not =0$, i.e.~$f\equiv 0$
\end{itemize}
\item under a {\em reduction condition}
$\bar{l}(-F)$ is related to the original $l(-F)$ such that 
if one has $\ker \bar{\iota}_1\not =0$ 
one gets $\ker \iota_1\not =0$
(both either for certain moduli or universally), 
i.e.~the vanishing of $Pfaff$ (at a special locus or universally); 
that is one gets
\begin{itemize}
\item $f|Pfaff$
in case the equality condition applies (as $f=\det \bar{\iota}_1$ 
has lower degree this gives a
'reduction' in the problem of finding zeroes of $Pfaff$)
\item $Pfaff\equiv 0$ 
in case the vanishing condition applies
\end{itemize}
\end{itemize}
(properly taking into account footn.~\ref{reduced factor}).
In the present section we will treat the reduction condition, 
in section \ref{Equality condition} the equality condition 
and in section \ref{The vanishing condition} the vanishing condition.

\subsection{The reduction condition: precise version (on $c$)}

Concerning $l(-F)=\cO_{\E}(\al s+\beta F)$ 
we will assume in the following 
$\al >0$ (actually even $\al > n$ by (\ref{technical assumption}))
and the necessary condition (for contribution to the superpotential)
$\beta <0$, cf.~(\ref{third criterion}). 
We consider a second line bundle
$\bar{l}(-F):=\cO_{\E}(\bar{\al} s+\bar{\beta} F)$ 
and denote the restrictions to $c$ by
$l(-F)|_c=\cO_c(D)$ and $\bar{l}(-F)|_c=\cO_c(\bar{D})$.
Now assume
\beqa
\label{precise reduction condition}
\Big(\bar{l}(-F)\Big)^p|_c & \hra & l(-F)|_c
\;\;\;\;\;\;\;\;\;\;\;\;\;\;\;\;\;\;\;\;
\hspace{2cm}
\mbox{(precise reduction condition)}
\eeqa
for a positive integer $p$. So the relevant condition is 
\beqa
\label{divisor condition}
p\bar{D}  & \leq & D \;\;\;\;\;\; , \;\;\;\;\;\; 
\mbox{i.e.}\;\; \tilde{D}=D-p\bar{D} \;\; \mbox{is effective}
\eeqa
such that $t\in H^0(c, \cO_c(D-p\bar{D})), t \neq 0$ exists.
This gives then
\beqa
\label{reduction phenomenon}
s\in H^0\Big(c, \bar{l}(-F)|_c\Big) & \Lra &
s^p\in H^0\Big(c, \cO_{\E}(p\bar{\al} s+p\bar{\beta} F)|_c\Big)
\nonumber\\
& \Lra &
s^p t\in H^0\Big(c, \cO_{\E}(\al s+\beta F)|_c\Big)
\eeqa 
as implication for the existence of nontrivial sections. Therefore one has
\beqa
\label{implication}
h^0\Big(c, \bar{l}(-F)|_c\Big)>0 & \Lra & h^0\Big(c, l(-F)|_c\Big)>0 
\eeqa
Here the section $t$ has just an auxiliary status:
as the difference of $D$ and $p\bar{D}$ is effective
(this is just the assumption of reduction)
a nontrivial section like $t$ will exist in any case
and the problem of existence of a nontrivial section of $l(-F)|_c$
is just reduced to the corresponding problem for $\bar{l}(-F)$
(note that neither $D$ nor $\bar{D}$ will be effective).\\
\noindent
{\em \underline{Remark:}} One can rephrase the procedure as follows. If one has
$i)$ the condition for a nontrivial section 
(on the lhs of (\ref{implication})) fulfilled
and $ii)$ that the reduction condition holds
\beqa
i) \;\;\; h^0(c, \bar{l}(-F)|_c)>0 &\Llra& \bar{D}\sim \bar{D}' \geq 0 
\; \Llra \; \bar{D}+(f)\geq 0\\
ii) \;\;\; \;\;\; \;\;\; \;\;\; \;\;\; reduction &\Llra& D\geq p\bar{D}
\eeqa
(here $\sim$ is linear equivalence)
then one gets a nontrivial section for the original bundle
\beqa
\label{reduction in remark}
h^0(c, l(-F)|_c)>0 &\Llra& D\sim D' \geq 0 
\; \Lla \; D+(f^p)\geq p(\bar{D}+(f))\geq 0
\eeqa

Precise reduction amounts according to (\ref{divisor condition})
to the effectivity of $\tilde{D}$; this implies that the 
complete linear system $|\tilde{D}|$ 
(of effective divisors, linearly equivalent to $\tilde{D}$)
is nonempty, what just signals the existence of 
a nonzero section $t$ of $\tilde{\cF}$
where $\tilde{\cF}=\cO_c(\tilde{D})$ is the line 
bundle on $c$ such that $\bar{l}(-F)^p|_c\ox \tilde{\cF}\cong l(-F)|_c$.
So reduction implies $h^0(c, \tilde{\cF})>0$.

\subsection{Strong version of the reduction condition (on $\E$)}

Actually we will usually assume the following sharper condition on $\E$
\beqa
\label{sufficient reduction condition}
\Big(\bar{l}(-F)\Big)^p & \hra & l(-F)\;\;\;\;\;\;\;\;\;\;\;\;\;\;\;\;\;\;\;\;
\hspace{2cm}
\mbox{(strong reduction condition)}
\eeqa
This condition will imply (\ref{precise reduction condition})
and is easy to check; however is is unnecessarily sharp, 
i.e. it is only a sufficient condition.
(\ref{sufficient reduction condition}) amounts to the effectiveness of
$(\al - p\bar{\al})s$ and $(\beta - p\bar{\beta} )F$, in other words
\beqa
\label{nec cond for suff reduc}
p\bar{\al} \leq  \al \;\; , \;\;\;\; p\bar{\beta}  \leq  \beta 
\eeqa
As $\beta < 0$ one therefore needs to have $\bar{\beta} <0$; 
we will also assume $\bar{\al}>0$.


\section{\label{Equality condition}The equality condition}

\resetcounter

The equality condition is the condition which will connect
the existence of a nontrivial section of $\bar{l}(-F)|_c$ with
the vanishing of a corresponding determinantal expression, in precise analogy
to the corresponding phenomenon for $l(-F)|_c$, 
cf.~(\ref{extrinsic determinant criterion}).

If one finds a line bundle $\bar{l}(-F)$ fulfilling 
(\ref{precise reduction condition}) 
and wants to use this to find a factor of $Pfaff$
one still has to make sure some things.
The first is the possibility to control $h^0(c, \bar{l}(-F)|_c)>0$ in
(\ref{implication}) again by a determinantal function like in 
(\ref{extrinsic determinant criterion}).
For this one needs an equality condition
(\ref{equality of h^1 dim's}) and when this holds one will have 
$\det \bar{\iota}_1 \, | \, \det \iota_1$
(modulo the remark before (\ref{general Pfaff factor}))
which reduces the problem of finding a zero of the polynomial $\det \iota_1$
of degree $h^1(\E, l(-F))$ to the polynomial $\det \bar{\iota}_1$ of 
degree $h^1(\E, \bar{l}(-F))$ (the degree is the sum of all degrees 
in the individual moduli, cf.~for example (\ref{degree 3 pol})).

In this connection we note that one has to check that the lhs of 
(\ref{equality of h^1 dim's}) should be indeed nonvanishing. In this question 
one has {\em with the assumptions}
$\bar{\al}>0, \bar{\beta}<0$ and (\ref{h^1 reduction}) that
$h^1(\E, \bar{l}(-F))>0$ if not either i) $\bar{l}(-F)=\cO_{\E}(s-F)$ 
or ii) $\bar{l}(-F)=\cO_{\E}(\bar{\al}s-F)$ with $\chi=0$;
i) is excluded by $\bar{\al}\geq n$
(cf.~remark after (\ref{equality of h^1 dim's})), 
ii) by (\ref{betabar fixing}) and (\ref{last expression}).

\subsection{\label{Numerical evaluation}
Numerical evaluation of the equality condition}

{}For $\bar{l}$ most of the arguments 
in subsect.~\ref{contribution criteria} are not applicable: one has neither
$h^0(c, \bar{\l}(-F)|_c)=h^1(c, \bar{\l}(-F)|_c)$, because now
$c_1(\bar{V}|_b)\neq 0$, 
nor\footnote{as $\al-n>0$ from $\al=n(\la+1/2), \la > 1/2$ 
does not give necessarily $\bar{\al}-n>0$} 
the vanishing of the $H^2$-terms in the long exact
sequence (\ref{long exact sequence}). 
Nevertheless, from the assumption $\bar{\al}>0$ 
and from $\bar{\beta}<0$, one has the 
vanishing of the first two terms such that still 
$H^0(c, \bar{\l}(-F)|_c) \cong \mbox{Ker}\;\bar{\iota}_1$:
\beqa
0 \ra H^0\Big(c, \bar{\l}(-F)|_c\Big)
\ra H^1\Big(\E, \bar{l}(-F-c)\Big) \stackrel{\bar{\iota}_1}{\ra} 
H^1\Big(\E, \bar{l}(-F)\Big) \ra 
H^1\Big(c, \bar{\l}(-F)|_c\Big) \ra \dots \;\;\;
\eeqa
One now wants to find cases where $\bar{\iota}_1$ is a map between spaces of 
equal dimension\footnote{Note that the evaluation by (\ref{reduction}) 
also of the lhs of (\ref{equality of h^1 dim's}), 
which has with $\bar{\al}-n$ a smaller $s$-coefficient in 
$\bar{l}(-F)=\cO_{\E}((\bar{\al}-n)s+(\bar{\beta}-r)F)$
which could be potentially $\leq 0$,
will be appropriate as the condition  
$\bar{\beta}\leq (\bar{\al}-n+1)\chi+r$ from (\ref{H^1 reduction conditions}) 
is fulfilled 
(the rhs is $> 0$ by (\ref{r chi relation}) and $\bar{\al}\geq 0$).}
\beqa 
\label{equality of h^1 dim's}
h^1\Big(\E, \bar{l}(-F-c)\Big)&=&h^1\Big(\E, \bar{l}(-F)\Big)
\;\;\;\;\;\;\;\;\;\;\;\;\;\;\;\; \mbox{(equality condition)}
\eeqa
We proceed now by distinguishing 
the cases\footnote{\label{al geq n footn}note that $\bar{\al}<n$ where 
$h^1(\E, \bar{l}(-F-c))=0$
is not interesting for us as $h^0(c, \bar{l}(-F)|_c)=0$ would mean that
one has no nontrivial section to start with 
in the procedure (\ref{reduction phenomenon}).} 
$\bar{\al}>n$ and $\bar{\al}=n$.

\subsubsection{\label{equ cond al > n}
The equality condition in the case $\bar{\al}>n$}

To compute the difference of the sides of 
(\ref{equality of h^1 dim's}), one can use formula (\ref{h^1 reduction})
for $\bar{\al}, \bar{\beta}$ {\em and}, in this case of $\bar{\al}>n$, 
{\em also for} $\bar{\al}-n, \bar{\beta}-r$
and gets
\beqa
h^1\Big(\E, \bar{l}(-F-c)\Big)-h^1\Big(\E, \bar{l}(-F)\Big)
&=& (\bar{\al}-\frac{n}{2})(r-n\chi)+(\beta+1-\frac{r+\chi}{2})n
\eeqa
Equivalently, with $h^2(\E, \bar{l}(-F-c))=0$, one can
use the index formula to compute
\beqa
\label{index computation}
h^1\Big(\E, \bar{l}(-F-c)\Big)-h^1\Big(\E, \bar{l}(-F)\Big)
&=&
h^0(c, \bar{l}(-F)|_c)-h^1(c, \bar{l}(-F)|_c)\nonumber\\
&=&\deg \bar{l}(-F)|_c - \deg K_c^{1/2}\nonumber\\
&=&\Big(\bar{\al}s+\bar{\beta}F-\frac{1}{2}(ns+rF+(\chi-2)F)\Big)(ns+rF)
\nonumber\\
\label{final expression}
&=&(\bar{\al}-\frac{n}{2})(r-n\chi)+(\bar{\beta}+1-\frac{r+\chi}{2})n\;\;\;
\eeqa
Thus one gets vanishing just for
\beqa
\label{betabar fixing}
-(\bar{\beta}+1)&=&\frac{\bar{\al}-n}{n}(r-n\chi)-\frac{n+1}{2}\chi
\eeqa
Note that this is just again the condition (\ref{beta rewriting})
(i.e., one has (\ref{equality of h^1 dim's}) if {\em and only if}
$c_1(\bar{V}|_{\E})=0$, expressed by the relation (\ref{beta rewriting})
between $\bar{\al}$ and $\bar{\beta}$, in this case).
The integrality requirement means for $n$ odd 
(or equally for $n$ even and $\chi$ even) that $n|\bar{\al} r$ 
while for $n=2m$ and $\chi$ odd that $\bar{\al}r/m$ must be an odd integer.

\subsubsection{The equality condition in the case $\bar{\al}=n$}

Here one computes $h^1(\E, \cO(0s+(\bar{\beta}-r)F))=-(\bar{\beta}+1)+r$ 
and gets
\beqa
h^1\Big(\E, \bar{l}(-F-c)\Big)-h^1\Big(\E, \bar{l}(-F)\Big)
&=& -(\bar{\beta}+1)+r+\bar{\al}(\bar{\beta}+1)
-\Big(\frac{\bar{\al}(\bar{\al}+1)}{2}-1\Big)\chi\;\;\;\;\;\;\;\;\;
\eeqa
Alternatively one gets this by computing the index on $c$
(using (\ref{final expression})) and adding 
\beqa
h^2\Big(\E, \bar{l}(-F-c)\Big)
&=&h^0\Big(\E, K_{\E}\ox \cO_{\E}(c)\ox (\bar{l}(-F))^{-1}\Big)
=h^0\Big(b, \cO_b(\chi-2+r-\bar{\beta})\Big)\nonumber\\
&=&-(\bar{\beta}+1)+r+\chi
\eeqa

Thus one gets vanishing just for 
\beqa
\label{last expression}
-(\bar{\beta}+1) & = & \frac{r}{n-1} - (\frac{n}{2}+1)\chi
\eeqa
As this expression has to be integral one notes
that for $n$ or $\chi$ even one needs to have $r=(n-1)\rho$ with 
$\rho$ a positive integer; if $n=2m+1$ and $\chi$ are odd 
one needs to have $\frac{r}{2m}-\frac{1}{2}\in {\bf Z}$, i.e.~that
$r/m$ is an odd integer 
(for example $r$ must be odd for $n=3$ with $\chi$ odd). 

\subsection{\label{interpretation}
Interpretation of the equality and reduction condition}

We rephrase the condition (\ref{equality of h^1 dim's}) for reduction, 
given in the numerical parameters above, 
in a more geometric form by restricting the discussion of the line bundles
from $\E$ to $c$.

\subsubsection{\label{interpret al > n}
The conditions in the case $\bar{\al}>n$}

The index computation (\ref{index computation}) shows that
the {\em equality condition} amounts to 
\beqa
\label{interpretation equality condition}
\bar{l}(-F)|_c & = & K_c^{1/2}\ox \bar{\cF}
\eeqa
with $\bar{\cF}$ a flat bundle on $c$ playing the same role for 
$\bar{l}(-F)|_c$ as does $\cF$ for $l(-F)|_c$, cf.~(\ref{degree equalities}).
$\bar{\cF}$ being a flat bundle this is just an equation of degrees:
$\deg \bar{l}(-F)=\deg K_c^{1/2}=\deg l(-F)$.

Now the precise {\em reduction condition} (\ref{precise reduction condition})
implies, together with (\ref{interpretation equality condition}) 
and (\ref{degree equalities}), 
the necessary condition $p\, \deg K_c^{1/2}\leq \deg K_c^{1/2}$. 
So for $p\geq 2$ one gets the condition $\deg K_c \leq 0$
which in view of (\ref{canonical degree})
comes down to $r - \frac{n-1}{2}\chi -1\leq 0$ (compare the reasoning 
in app.~\ref{Reduction cases}). One gets, in view of 
(\ref{r chi relation sharp}), that $r=1, \chi=0$, i.e. the spectral
curve $c=C\cap \E$ is an elliptic curve.

\subsubsection{The conditions in the case $\bar{\al}=n$}

Here the {\em equality condition} amounts to
\beqa
\label{interpretation2 equality condition}
\bar{l}(-F)|_c & = & K_c^{1/2}\ox ( \omega_{c/b}^{1/2})^{-\frac{1}{n-1}}
\ox \bar{\cF}
\eeqa
where $\omega_{c/b}=K_c\ox \pi_c^* K_b^{-1}$ is the relative dualizing sheaf
with $\deg \omega_{c/b}^{1/2}=n(r - \frac{n-1}{2}\chi)$.
Although written, for analogy with (\ref{interpretation equality condition}), 
with the $(n-1)^{th}$ root of the line bundle $\omega_{c/b}^{1/2}$ 
this is essentially just meant as an equation of degrees
(cf.~also the remarks after (\ref{last expression})).

Now the precise {\em reduction condition} (\ref{precise reduction condition})
gives, together with (\ref{interpretation2 equality condition}) 
and (\ref{degree equalities}), 
the necessary condition 
$p \, \deg K_c^{1/2}-\frac{p}{n-1}\deg \om_{c/b}^{1/2}\leq 
\deg K_c^{1/2}$ and so
\beqa
(p-1)(r - \frac{n-1}{2}\chi -1)& \leq &\frac{p}{n-1}(r - \frac{n-1}{2}\chi)
\eeqa
or
\beqa
\label{final form}
\Big(p(n-2)-(n-1)\Big) (r - \frac{n-1}{2}\chi -1)& \leq & p
\eeqa
This is usefully applicable for $n=3, p>2$ and $n=4, p\geq 2$,
giving all cases in app.~\ref{list subsection}. 
Actually we will proceed slightly differently in 
the concrete derivation in app.~\ref{list subsection}
(using $\la \geq p - \frac{1}{2}$ in (\ref{first bound})
one would get back (\ref{final form})).

\subsection{All (strong) reduction cases with equality condition}

One has always $\la > \frac{1}{2}$ and
$\la \in \frac{1}{2}+{\bf Z}$ for $n$ odd, while for $n$ even
the case $\la \in \frac{1}{2}+{\bf Z}$ needs $\chi$ even
and $\la \in {\bf Z}$ needs $r-\chi$ even.
The condition $\beta <0$ can, according to (\ref{sharper bound}), 
be rephrased in the parameters as
$r-n\chi\geq \frac{(1+n)/2}{\la - 1/2}\chi$; furthermore $r-n\chi>0$
even for $\chi =0$ by (\ref{r chi relation sharp}).
We are usually interested mainly (physically) in $n=2,3,4,5$ and 
$\chi =0$ (for illustration) or $1$ ($b$ isolated).
Now the detailed study in app.~\ref{Reduction cases} gives the following.

\subsubsection{\label{red cases for bar al > n}Cases with $\bar{\al}>n$}

For $SU(n)$ bundles one has for
$\chi=0, r=1, \la +\frac{1}{2}\in p{\bf Z}^{>1}$ 
just the $p$-case 
\beqa
\l(-F)&=&\cO_{\E}\Big(n(\la+\frac{1}{2})s-(\la + \frac{1}{2})F\Big)\\
\label{r=1 general case}
\bar{l}(-F)&=&
\cO_{\E}\Big(n\frac{\la+\frac{1}{2}}{p}s-\frac{\la + \frac{1}{2}}{p}F\Big)
\eeqa
Now $\chi=0, r=1$ were just the conditions for
an elliptic $c$, cf.~(\ref{canonical degree}). One gets
\beqa
K_c&=& \cO_{\E}(ns+(r+\chi-2)F)|_c=\cO_{\E}(ns-F)|_c\\
l(-F)&=&\cO_{\E}(ns-F)^{\ox (\la + \frac{1}{2})}|_c
\eeqa
So here $K_c\cong \cO_c\cong l(-F)|_c$, 
$h^0(c, l(-F)|_c)=1>0$ and $Pfaff \equiv 0$, cf.~Ex.~3 below.

\subsubsection{\label{red cases for bar al = n}Cases with $\bar{\al}=n$}

Here one has 
\beqa
\l(-F)&=&\cO_{\E}\Big(n(\la+\frac{1}{2})s+\beta F\Big)
\; , \;\;\;\;\;\; 
\beta =  -\Bigg( (\la - \frac{1}{2})r-(n\la+\frac{1}{2})\chi+1\Bigg)\\
\label{bar l(-F)}
\bar{l}(-F)&=&\cO_{\E}\Big(ns + \bar{\beta}F\Big)\; , \;\;\;\;\;\;\;\;\;\;
\;\;\;\;\;\;\;\,
\bar{\beta}=-\Bigg( \frac{1}{n-1}r-(\frac{n}{2}+1)\chi+1\Bigg)
\eeqa
The $(\chi, r, \lambda, p)$-list 
of occurring cases for $SU(n)$ bundles is given in app.~\ref{list subsection}.
Furthermore 
\beqa
\deg \det \iota_1 & = & h^1(\E, l(-F))=-\al(\beta+1)
+\Big(\frac{\al(\al+1)}{2}-1\Big)\chi\nonumber\\
&=&n(\la^2-\frac{1}{4})(r-\frac{n}{2}\chi)
-\chi\\
\deg \det \bar{\iota}_1&=&h^1(\E, \bar{l}(-F))=
-(\bar{\beta}+1)+r = \frac{n}{n-1}r-(\frac{n}{2}+1)\chi\nonumber\\
&=&\frac{r}{n-1}+(r-\frac{n}{2}\chi)-\chi
\eeqa

\section{\label{The vanishing condition}The vanishing condition
(including Example 3)}

\resetcounter

When considering in sect.~\ref{red cases for bar al > n} the reduction case 
with $\bar{\al}>n$ we saw that for $\chi=0, r=1, \la \in \frac{1}{2}+{\bf Z}$ 
with $c$ elliptic one gets $Pfaff \equiv 0$ by directly computing 
$h^0(c, l(-F)|_c)=1>0$. In that case both, $l(-F)|_c$ and $\bar{l}(-F)|_c$, 
were powers of the trivial bundle $K_c$.

This type of argument can be applied more widely.
Note that always $\dim \ker \bar{\iota}_1\leq h^0(c, \bar{l}(-F)|_c)$
by the analog of the long exact sequence (\ref{long exact sequence})
for $\bar{l}$ (with equality if $h^0(\E, \bar{l}(-F))=0$). Therefore, 
even is one is not in an equality case where (\ref{equality of h^1 dim's}) 
holds (when vanishing of $h^0(c, \bar{l}(-F)|_c)=\dim \ker \bar{\iota}_1$ 
will be controlled by $\det \bar{\iota}_1$) one gets 
\beqa
\mbox{(vanishing condition)}\;\;
h^1\Big(\E, \bar{l}(-F-c)\Big)-h^1\Big(\E, \bar{l}(-F)\Big)>0& \Lra & 
h^0\Big(c, \bar{l}(-F)|_c\Big)>0\;\;\;\;\;\;\;\;\;
\eeqa
If the reduction condition holds for $p$ then the argument will be completed
in the usual way by noting that the existence of $s\in H^0(c, \bar{l}(-F)|_c),
s\not = 0$ implies $s^p t\in H^0(c, l(-F)|_c)$; therefore if the 
(moduli-independent) vanishing condition holds then $Pfaff \equiv 0$.

Let us see how this condition can be applied. If one is in the case 
$\bar{\al}>n$ one has $h^1(\E, \bar{l}(-F-c))-h^1(\E, \bar{l}(-F))
=\deg \bar{l}(-F)|_c-\deg K_c^{1/2}$ which can not be positive if we want at 
the same time the necessary condition for reduction 
$\deg \bar{l}(-F)|_c\leq \deg l(-F)|_c=\deg K_c^{1/2}$ to be fulfilled. 
Therefore 
$\bar{\al}=n$, i.e.~$\bar{l}(-F)=\cO_{\E}(ns+\bar{\beta}F)$ and we have
\beqa
\label{positivity condition}
h^1\Big(\E, \bar{l}(-F-c)\Big)-h^1\Big(\E, \bar{l}(-F)\Big)&=&
(n-1)(\bar{\beta}+1)+r-\Big(\frac{n(n+1)}{2}-1\Big)\chi\nonumber\\
&=&(n-1)\Big( \bar{\beta}+1-\frac{n+2}{2}\chi\Big)+r
\eeqa

For $SU(3)$ or $SU(4)$ bundles, the only new case (besides the above mentioned 
$\chi=0, r=1$ case) where this is strictly positive is $n=3, \la = 3/2$ with 
$r\not\equiv \chi(2)$, cf.~app.~\ref{vanishing n=3 cases}.

Let us take a closer look on the case of $SU(3)$ bundles with $\la = 3/2$ as 
this will be an important case in sect.~\ref{Examples}. Here one has $l(-F)=
\cO_{\E}(6s-(r-5\chi+1)F)$ where $\beta<0$ means $r\geq 5\chi$. Let us take
\beqa
\bar{l}(-F)&=&\left\{ \begin{array}{lll}
\cO_{\E}\Big(3s-(\frac{r-5\chi}{2}+1)F\Big) & \;\;\;\mbox{if} & 
r\equiv \chi(2)\\
\cO_{\E}\Big(3s-\frac{r-5\chi+1}{2}F\Big) & \;\;\;\mbox{if}&r\not\equiv\chi(2)
\end{array} \right. 
\eeqa

If $r\equiv \chi(2)$ one gets a moduli-dependent statement, 
cf.~sect.~\ref{n=3, la =3/2}.

By contrast, if $r\not\equiv\chi(2)$ one gets the moduli-independent
statement $Pfaff\equiv 0$ as here (\ref{positivity condition}) becomes
just $1$ (alternatively one can also argue directly
that $h^0(c, \bar{l}(-F)|_c)=\frac{3}{2}(r-\chi-1)-3(r-\chi-1)
+h^0(c, \cO_{\E}((r+\chi-2+\frac{r-5\chi+1}{2})F)|_c)\geq 1$).
For $\chi=0, r=1$ cf.~Example $3$.
For an algebraic vanishing argument 
in the somewhat special $\chi=0$ case cf.~app.~\ref{chi = 0 app}.

\section{\label{difference subsection}On the difference between
the strong and the precise reduction condition}

\resetcounter

Let us now come back to the precise version of the reduction condition
on $c$ instead of the strong version on $\E$. 
The latter had with (\ref{nec cond for suff reduc}) 
a precise numerical expession whereas (\ref{precise reduction condition}) 
implies only the necessary condition for the degrees
\beqa
\label{nec cond for prec reduc}
p(\bar{\al}s+\bar{\beta}F)(ns+rF) & \leq & (\al s+\beta F)(ns+rF)
\eeqa
or explicitly\footnote{\label{barbeta geq 0}
Note that we always work under the assumption 
$\bar{\al}\geq n$, cf.~footn.~\ref{al geq n footn}; here the case
$\bar{\al}>n$, which implies that the rhs of 
(\ref{nec cond for prec reduc explic}) is $\leq 0$ 
(cf.~sect.~\ref{interpret al > n}), 
still gives $\bar{\beta}\leq 0$ and even $<0$ by (\ref{r chi relation sharp});
however a case $\bar{\al}=n, \bar{\beta}\geq 0$ is possible and will 
become relevant in sect.~\ref{Exceptional} 
and in the Example 1 in sect.~\ref{Ex 1 sect}.}
\beqa
\label{nec cond for prec reduc explic}
p\Big( \bar{\al}(r-n\chi)+n\bar{\beta}\Big) & \leq & 
n \Big(r-\frac{n-1}{2}\chi -1\Big)
\eeqa
So the logical relations are
\beqa
strong\; reduction \;(equ.~(\ref{nec cond for suff reduc}))\;
\buildrel i) \over \Lra \;
precise\; reduction \;(equ.~(\ref{precise reduction condition}))\;
\buildrel ii) \over \Lra \;equ.~(\ref{nec cond for prec reduc})\;\;\;\;\;
\eeqa
From the onesided implications two questions arise. 

i) How much widened is the domain of possible $\bar{l}(-F)$'s 
by considering the precise condition (\ref{precise reduction condition})
instead of the sharpened condition (\ref{nec cond for suff reduc}) ?

ii) The question about sufficiency of the condition 
(\ref{nec cond for prec reduc}) for the precise reduction
(\ref{precise reduction condition}):
when is $D-p\bar{D}$ on $c$ (cf.~(\ref{divisor condition})), 
which has degree $\geq 0$ by (\ref{nec cond for prec reduc}), 
actually even effective ? 

Also in connection with these questions we notice the following.

ad i) Note that if the strong reduction condition is violated
we {\em need actually to check explicitely}
that one has $h^0(c, \bar{l}(-F)|_c)>0$ 
for which we argued only under certain assumptions
in the remark before sect.~\ref{Numerical evaluation}; for example
$\bar{\beta}<0$ followed only because 
(\ref{nec cond for suff reduc}).$^{\ref{barbeta geq 0}}$

ad ii) 
{\em If the necessary condition
(\ref{nec cond for prec reduc}) for precise reduction is actually saturated} 
then $t\in H^0(c, \tilde{\cF})$, cf.~remark after (\ref{reduction in remark}), 
is constant ($\tilde{\cF}$ is flat and then even trivial; 
a degree zero effective divisor on $c$ is zero);
so, then reduction holds just if $l(-F)|_c\cong \bar{l}(-F)^{\ox p}|_c$.

\subsection{\label{Exceptional}The exceptional case $\bar{\beta}=0$}

We remarked above that $\bar{\beta}<0$ is necessary to fulfill
(\ref{sufficient reduction condition}), 
i.e. concretely $p\bar{\beta}\leq \beta$.
However, as pointed out, this condition is unnecessarily sharp.
Actually one wants only (\ref{precise reduction condition}). 
For this (\ref{nec cond for suff reduc}) is not necessary (so $\bar{\beta}$
might be $\geq 0$; cf.~footn.~\ref{barbeta geq 0});
necessary is (\ref{nec cond for prec reduc}).

Let us try to allow $\bar{\beta}=0$ 
(this will be relevant for Example 1 in sect.~\ref{Ex 1 sect});
consider the first two terms\footnote{Concerning the $H^2$-terms 
the second one vanishes
by (\ref{H^2 vanishing conditions}) as $\bar{\al}>0$
but $h^2(\E, \cO_{\E}((\bar{\al}-n)s-rF))=0$ only if $\bar{\al}>n$
whereas one gets for $\bar{\al}=n$ that $h^0(b, \cO_b(r+\chi-2))=r+\chi-1$
if not $\chi=0, r=1$.} 
in the long exact sequence
(\ref{long exact sequence}) for $\bar{l}(-F)$:
although still $h^0(\E, \cO_{\E}((\bar{\al}-n)s-rF))=0$ for the 
first term, one has $h^0(\E, \cO_{\E}(\bar{\al}s))=1$ for the 
second one if $\chi >0$ and $=\bar{\al}$ if $\chi=0$.
Thus $\dim \ker \bar{\iota}_1$ equals $h^0(c, \bar{l}(-F))$ {\em minus}
$1$ or {\em minus} $\bar{\al}$ if $\chi>0$ or $=0$, resp.; so the vanishing 
of $\det \bar{\iota}_1$ (at a point in the moduli space $\M_{\E}(c)$)
still indicates the existence of a nontrivial
section of $\bar{l}(-F)|_c$ and one still has 
$\det \bar{\iota}_1|\det \iota_1$.

Let us give the equality condition for the case $\bar{\beta}=0$. Then one has 
$h^1(\E, \bar{l}(-F))=-(\bar{\al}-1)(\bar{\beta}+1)
+(\frac{\bar{\al}(\bar{\al}+1)}{2}-1)\chi$ 
with the remark after (\ref{h^1 reduction}) and gets 
$-(\bar{\beta}+1)=\frac{r}{n-2}-\frac{(n+2)(n-1)}{2(n-2)}\chi$
which determines $r$ 
(where (\ref{r chi relation sharp}) excludes $n=2$ 
and (\ref{r chi relation}) $n>2, \chi=0$)
\beqa
\label{exceptional r fix}
r-n\chi & = & (n-2)\Big(\frac{n+1}{2}\chi-1\Big)
\eeqa
Here, with $r$ assumed fixed by (\ref{exceptional r fix}), 
the interpretation as in sect.~\ref{interpretation}
amounts to
\beqa
\bar{l}(-F)|_c & = & K_c^{1/2}\ox (K_c^{1/2})^{-\frac{1}{n-1}}\ox \bar{\cF}
\eeqa
For $n=3$, where previously $\beta<0\Leftrightarrow r\geq 5\chi$,
we get $\bar{\beta}=0\Leftrightarrow r=5\chi-1$, cf.~Example 1.

This exceptional bundle has the degree on $c$
(using in both cases the $r$-fix (\ref{exceptional r fix}))
\beqa
\deg (\bar{l}(-F))|_c=\deg \cO_{\E}(ns-0F)|_c&=&n(r-n\chi)
=n(n-2)\Big(\frac{n+1}{2}\chi-1\Big)\\
\deg K_c^{1/2}&=&n(n-1)\Big(\frac{n+1}{2}\chi-1\Big)
\eeqa
Therefore,
although the conditions (\ref{nec cond for suff reduc}) for 
strong reduction (\ref{sufficient reduction condition}) 
are violated here as $p0\not\leq \beta <0$, 
the necessary condition (\ref{nec cond for prec reduc}) 
for the precise reduction condition (\ref{precise reduction condition})
is fulfilled if $p(n-2)\leq n-1$, i.e.~for $n\leq 1+ \frac{p}{p-1}$. This
holds for $n=2$ or for $n= 3, \, p=2$ 
where (\ref{nec cond for prec reduc}) is even saturated
(and of course trivially for $p=1$).
When is not only the necessary condition 
(\ref{nec cond for prec reduc}) for precise reduction but even the 
precise reduction condition (\ref{precise reduction condition}) itself 
fulfilled ? Consider the case $n=3, p=2$ where
\beqa
l(-F)&=&\cO_{\E}\Big(3(\la + \frac{1}{2})s-(\la -\frac{3}{2})(2\chi -1)F\Big)\\
\label{n=3 exceptional}
\bar{l}(-F)&=&\cO_{\E}(3s-0F)
\eeqa
Note that $l_{\la = 3/2}(-F)=\bar{l}(-F)^{\ox 2}$,
and so a fortiori $l_{\la = 3/2}(-F)|_c=\bar{l}(-F)^{\ox 2}|_c$,
but the necessary condition $\beta<0$ from (\ref{third criterion})
requires here $\la > 3/2$, cf.~Example 1 in sect.~\ref{Examples}. 
But according to the remark 
before sect.~\ref{Exceptional} here reduction does hold
precisely if $l_{\la}(-F)|_c=\bar{l}(-F)^{\ox 2}|_c$; 
so reduction holds just if $l_{\la}(-F)|_c=l_{\la = 3/2}(-F)|_c$
and one needs to know the sublocus in the moduli space $\M_{\E}(c)$
where (the isomorphism class of) $l(-F)|_c$ is actually independent of $\la$,
i.e.~where $\Lambda=\cO_{\E}(ns-(r-n\chi)F)|_c$ becomes trivial;
now cf.~sect.~\ref{Ex 1 sect}.

\section{\label{Examples}Some examples of $SU(3)$ bundles}

\resetcounter

\subsection{Overview over the three main examples}

We illustrate the theory with some examples
(for Example 3 of $Pfaff=0$ cf.~sect.\ref{The vanishing condition})
\begin{center}
\begin{tabular}{|c||c|c|c|}
\hline
                & Ex $1$      & Ex $2$        & Ex $4$ \\
\hline\hline
$\chi$          & $1$         & $1$           & $0$ \\
\hline
$\lambda$       & $5/2$       & $3/2$         & $3/2$ \\
\hline
$c$             & $3s+4F$     & $3s+5F$       & $3s+2F$ \\
\hline
$K_c$           &$\cO_{\E}(3s+3F)|_c$&$\cO_{\E}(3s+4F)|_c$&$\cO_{\E}(3s)|_c$ \\
\hline
$\Lambda$       & $\cO(3s-F)$ & $\cO(3s-2F)$  & $\cO(3s-2F)$  \\
\hline
codim $\Sigma_{\Lambda}$
                &      1      &      2        &    1          \\
\hline 
$l(-F)$         & $\cO(9s-F)$ & $\cO(6s-F)$   & $\cO(6s-3F)$\\
\hline
$Pfaff=\det \iota_1$         
                &$f^{11}Q_{11}$    & $f^4$         & $f^4$ \\
\hline
$\bar{l}(-F)$   & $\cO(3s)$   & $\cO(3s-F)$   & $\cO(3s-2F)$ \\
\hline 
$f=\det \bar{\iota}_1$
                &$f_{\Lambda}$&   $Res(B, A)$ & $f_{\Lambda}$\\
\hline
\end{tabular}
\end{center}
For these Examples we find that 
$f:=\det \bar{\iota}_1$ equals $f_{\Lambda}$ 
if that relation is possible at all, that is if the 
codimension of the $f_{\Lambda}=0$ locus $\Sigma_{\Lambda}$ is $1$, 
as in Examples $1$ and $4$:
in Example 4 the naive codimension $r-n\chi$ (cf.~sect.~\ref{Ri section})
of $\Sigma_{\Lambda}$ is $2$ 
but $\bar{l}(-F)=\Lambda$ leads to $f=f_{\Lambda}$; 
in Example 2 the codimension is $2$:
one has to demand $R_1=0$ and $R_2=0$, 
i.e.~$(R_1, R_2)=f_{\Lambda}=(Res(B, A_i))_i$ 
where $A=\prod_{i=1}^{r-n\chi}A_i$ 
(cf.~footn.~\ref{vector-valued fct})
and one has then that codim $\Sigma_{\Lambda}>1$ (the general case).
Although $\Sigma_{\Lambda}$ is a sublocus of $Pfaff=0$
(cf.~the remark after (\ref{effectivity equivalence}))
the 'scalar function' dividing $Pfaff$ is $Res(B, A)$, 
cf.~footn.~\ref{multiple resultant zero}.

\subsection{\label{n=3, la =3/2}Bundles with $\la =3/2$}

Here one has $l(-F)=\cO_{\E}(6s-(r-5\chi+1)F)$ 
where $\beta<0$ means $r\geq 5\chi$. Let us take
\beqa
\bar{l}(-F)&=&\left\{ \begin{array}{lll}
\cO_{\E}\Big(3s-(\frac{r-5\chi}{2}+1)F\Big) & \;\;\;\mbox{if} & 
r\equiv \chi(2)\\
\cO_{\E}\Big(3s-\frac{r-5\chi+1}{2}F\Big) & \;\;\;\mbox{if}&r\not\equiv\chi(2)
\end{array} \right. 
\eeqa

For $r\not\equiv\chi(2)$ we found $Pfaff\equiv 0$ 
in sect.~\ref{The vanishing condition}. 
For $r\equiv \chi(2)$ one gets a moduli-dependent statement: 
if an $\rho\in H^0(c, \bar{l}(-F)|_c), \rho\not =0$ exists 
(what is described by $\det \bar{\iota}_1=0$)
then $\rho^2u$ and $\rho^2v$ are non-trivial elements of $H^0(c, l(-F)|_c)$, 
what suggests for this case that $Pfaff=f^2g$. 
Note that at most $Pfaff=f^4$ as
$\deg Pfaff=6r-10\chi, \deg f = \frac{3r-5\chi}{2}$.

\subsubsection{\label{Exa 2}The case $\chi\geq 1$
(including Example 2 as minimal $r$-value)}

For $r=5\chi$, where
$\deg Pfaff=4r, \deg f = r$ and $\chi\geq 1$ by (\ref{r chi relation sharp}),
one gets
\beqa
l(-F)&=&\cO_{\E}(6s-F)\\ 
\bar{l}(-F)&=&\cO_{\E}(3s-F)
\eeqa
One then has also $\rho z\in H^0(c, l(-F)|_c)$
such that possibly $Pfaff=f^3h$ in this case.

For $\chi=1$, where $c=3s+5F$, this is realized in Example $2$. Note 
that there the matrix induced by $\bar{\iota}$ is $m_5=\D_5$ in (\ref{D_5}), 
such that $f=\det \bar{\iota}_1=Res(A_2,B_3)$, cf.~sect.~\ref{resultant}. 

This is case $3^{\chi=1}$ of app.~\ref{list subsection}.
According to (\ref{the critical map}) one has then to consider 
the map $\iota_1: H^1\Big(\E, \cO_{\E}(3s+(\be -r)F)\Big) 
\lra H^1\Big(\E, \cO_{\E}(6s+\be F)\Big)$ between spaces 
of dimension $6r-10\chi$
(for concrete matrix representations 
cf.~app.~\ref{concrete matrix representations})
This map is induced by multiplication with an element 
$\tilde{\iota}\in H^0\big( \E, \cO_{\E}(3s+rF)\big)$.
The determinant of the $\iota_1$ for $r\equiv \chi (2)$ has as factor
the determinant of a corresponding
map $i_r: i_r:  H^1\Big(\E, \cO_{\E}\big(0s+(\bar{\be}-r)F\big)\Big) \lra  
H^1\Big(\E, \cO_{\E}\big(3s+ \bar{\be}F\big)\Big)$ 
(again induced by multiplication with 
$\tilde{\iota}\in H^0\Big(\E, \cO_{\E}(3s+rF)\Big)$, 
so depending on the same moduli)
between two spaces of dimension $\frac{1}{4}(6r-10\chi)$.

Now, what was generically 
$s|_c=\{p_0, t_1\}+\{p_0, t_2\}\, , \;
F_{t_i}|_c=c_{t_i}=\{p_0+p_i^++p_i^-, t_i\}$
(with $A_2(t_i)=0$, $i=1,2$, cf.~(\ref{s|_c generically}))
changes at the specialisation locus 
$f=\det \bar{\iota}_1=\det \D_5 =Res(A_2,B_3)=0$ (where $B_3(t_1)=0$; 
the non-generic case for (\ref{s|_c generically})): 
the $(w)$ line becomes, in the ${\bf P}^{\bf 2}_{t_1}$-fibre of $\cW_b$, 
the $(z)$ line and $p_0$ becomes a three-fold point of $c_{t_1}$ 
\beqa
(z)|_c=3s|_c=\{3p_0, t_1\}+\{3p_0, t_2\} = F_{t_1}|_c + \{3p_0, t_2\}
\neq  F_{t_1}|_c +F_{t_2}|_c 
\eeqa
In other words, in the specialisation locus $f=0$
one gets $(3s-F)|_c \sim$ effective
or equivalently $h^0(c, \cO(3s-F)|_c)\geq 1$ (the existence of $\rho\neq 0$); 
therefore also $(6s-F)|_c \sim$ effective or $h^0(c, l(-F)|_c)\geq 1$; in 
other words $f\, | \, Pfaff$. 

To demand that even $(3s-2F)|_c \sim$ effective 
is a more restrictive condition,
i.e.~$R_1=0=R_2$ is a proper sublocus of $f=0$
(so here $f_{\Lambda}=(R_1, R_2)$, cf.~footn.~\ref{vector-valued fct}).
Note that $(3s-2F)|_c\sim $ effective $\Lra (3s-F)|_c\sim $ effective;
so, if\footnote{cf.~sect.~\ref{Ri section}, we call here $R_2=R_0$;
cf.~also app.~\ref{more than one root}} 
$R_1=0=R_0$  
at certain points in moduli space, then $f=0$ there, as
becomes manifest in the representation 
\beqa
\label{f in the R_i}
f&=&\frac{1}{a_2^2}\det \left( {\begin{array}{ccc}
a_0  & a_1  & a_2 \\
R_1  & R_0  &   0 \\
  0  & R_1  & R_0 
\end{array}} \right) 
\eeqa

\subsubsection{\label{Ex 4 case}The case $\chi=0$
(including Example 4 as minimal $r$-value)}

For $\chi=0$ one has $r$ even and 
$\deg Pfaff=12\frac{r}{2}$ and $ \deg f = 3\frac{r}{2}$.
The minimal $r$-value (which must be $> 0$)
is $2$, cf.~Example $4$, where $c=3s+2F$ and
\beqa
l(-F)&=&\cO_{\E}(6s-3F)\\
\bar{l}(-F)&=&\cO_{\E}(3s-2F)=\Lambda
\eeqa
Precisely on the specialisation locus $f=\det \bar{\iota}_1
=\det \D_3^{\chi=0}=0$, cf.~(\ref{D_3 chi=0}),
the degree $0$ line bundle $\bar{l}(-F)|_c$ 
gets a nonzero section and so becomes trivial,
in other words
\beqa
f_{\Lambda}&=&f
\eeqa
{\em \un{Remark:} i)} Following strictly the procedure of example $2$ one
would expect codim $\Sigma_{\Lambda}=2$ 
and $f_{\Lambda}$ a two-component expression: 
the condition $A_2|B_2$ is here\footnote{with minors 
w.r.t.~a development of (\ref{D_3 chi=0}) w.r.t.~the $c$-row, 
and in (\ref{f in the M_i}) also w.r.t.~the $b$-row} 
$M^{c}_0=0=M^{c}_1$ while
\beqa
\label{Ex 4 resultant}
Res(B_2, A_2)&=&\det \left( {\begin{array}{cccc}
b_2  &  b_1 & b_0 & 0  \\
0    &  b_2 & b_1 & b_0\\
a_2  &  a_1 & a_0 & 0  \\
0    &  a_2 & a_1 & a_0
\end{array}} \right) 
= \det \left( {\begin{array}{cc}
M_1^c  &  M_0^c  \\
M_2^c  &  M_0^c
\end{array}} \right) 
\eeqa
(note that $0=a_0M_0^c-a_1M_1^c+a_2M_2^c$ allows to eliminate $M_2^c$).
Now one has 
\beqa
\label{f in the M_i}
f&=&\frac{1}{a_2}\det \left( {\begin{array}{cc}
M_1^c  &  M_0^c  \\
M_1^b  &  M_0^b
\end{array}} \right) 
\eeqa
such that the locus where $M^{c}_0=0=M^{c}_1$ is certainly a subset
of the $f=0$ locus, cf.~the remark after (\ref{Ri = 0 implication}).
But $A_2|B_2$ is only a sufficient, not a necessary 
condition for $\Lambda|_c\cong \cO_c$, 
cf.~(\ref{Ri = 0 implication}).
Here, in this case of symmetry between $C, B$ and $A$ because of $\chi=0$,
the precise condition for $\Lambda|_c\cong \cO_c$ turns out to be 
just linear dependence among $C, B$ and $A$ 
which represents the {\em single} condition $f=0$; 
this comes as here $\Lambda = \bar{l}(-F)$ and
demanding a nontrivial section over $c$ leads to the 
determinantal condition $f=0$.\\
{\em ii)} That 
$3s|_c\sim 2F|_c$ on the locus $f=0$ can also be seen
directly from $\det \bar{\iota}_1=\det \D_3^{\chi=0}=0$, 
cf.~(\ref{D_3 chi=0}), instead of arguing via the long exact sequence. 
The equation for the fibre of $c$ over $t_1=u/v\in b$, say, is
$C_2(t_1)z +B_2(t_1)x+A_2(t_1)y=0$ and these
three points lie also in the divisor $\alpha z + \beta x +\gamma y=0$, 
understood as $3s|_c$, 
where $\alpha = C_2(t_1), \beta=B_2(t_1), \gamma=A_2(t_1)$; 
this divisor of degree $6$ contains three further points; we want
to see that these can arise as a further fibre triplet over $t_2$, say.
So we ask whether $t_1, t_2(\neq t_1) \in b$ 
exist such that the 3-vectors $(D_2(t_1))$ and $(D_2(t_2))$ 
where $D_2=C_2, B_2, A_2$ are linearly dependent, i.e.~whether
$k\in {\bf C^*}$ exists such that $D_2(t_2)=kD_2(t_1)$ 
for $D_2=C_2, B_2, A_2$ which indeed just amounts to the nontrivial solvability
of $ \D_3^{\chi=0}\cdot (kt_1^2-t_2^2, kt_1-t_2, k-1)^t=0$.

\subsection{\label{Ex 1 sect}Example 1 with $\la=5/2$}

An $SU(3)$ bundle, $\chi=1$ with $c=3s+4F$, 
so\footnote{this is reflected in the sense of 
app.~\ref{negative lambda}
from an Example $1$ with $\la=-5/2$}
$l(-F)=\cO_{\E}\big( 9s-F\big)$, gives a map
\beqa
\label{ex 1tilde map}
\iota_1: \; 
H^1\Big( \E, \cO_{\E}\big( 6s-5F \big) \Big)
\lra
H^1\Big( \E, \cO_{\E}\big( 9s-F\big) \Big)
\eeqa
between $44$-dimensional spaces by (\ref{h^1 reduction}).
$\iota_1$ is induced from multiplication with
$\tilde{\iota}  =  Cz+Bx+Ay$ where
(for the notation cf.~app.~\ref{concrete matrix representations})
\beqa
\tilde{\iota} & \in & H^0\Big( \E, \cO_{\E}(3s+rF)\Big)
=\bigoplus_{w'\in StB(\Si)}\, w' \, H^0\Big(b, r- [w'] \, \chi\Big)
=\bigoplus_{w'\in StB(\Si)}\, w'\, S^{r- [w'] \, \chi}\, V \;\;\;\;\;\;\;\;\;
\eeqa
(i.e. $w'\in H^0(\E, \cO_{\E}(3s+[w']\chi F))$, 
cf.~(\ref{standard base elements}))
with the accompanying coefficients
\beqa
\label{coefficient expressions}
C=c_4u^4+c_3u^3v+c_2u^2v^2+c_1uv^3+c_0v^4
&\in &S^r\, V \;\;\;\;\, = S^4\, V = Hom(S^4\, V^*, {\bf C})\;\;\;\;\;\;\\
B=b_2u^2+b_1uv+b_0v^2 \;\;\;\;\;\;\;\;\;\;\;
&\in &S^{r-2\chi}\, V 
= S^2\, V = Hom(S^2\, V^*, {\bf C})\;\;\;\;\;\;\;\;\\ 
A=a_1u+a_0v \;\;\;\;\;\;\;\;\;\;\;\;\;\;\;\;
& \in &S^{r-3\chi}\, V= \;\;\;\;\, V= Hom(V^*, {\bf C})\;\;\;\;
\eeqa
With $l(-F)$ being not among the strong reduction cases we
adopt the more general approach of sect.~\ref{Exceptional}: 
we use $\bar{l}(-F)=\cO_{\E}(3s-0F)$ and get the map
\beqa
\bar{\iota}_1: 
H^1\Big(\E, \cO_{\E}(0s-4F)\Big)\lra H^1\Big(\E, \cO_{\E}(3s-0F)\Big)
\eeqa
between $3$-dimensinal spaces.
Or, in matrix form, 
(having $\bar{\beta}=0$ the first line is absent)
\beqa
m_r\, : \;\;\;\;\;\;\;\;\;\;\;\;\;\;
\left( \begin{array}{c}
\, C \odot S^{-\bar{\beta}-2}\;\;\;\;\;\, V^*    \\
B \odot S^{-\bar{\beta}-2+2\chi}\, V^*     \\
A \odot S^{-\bar{\beta}-2+3\chi}\, V^*     
\end{array} \right)
\eeqa

Concretely this means one builds the map
\beqa
\D_3 := B \; \oplus \; A\; \underline{\odot}\; V^*  = 
\left( {\begin{array}{ccc}
b_2 & b_1  & b_0  \\
a_1 & a_0  & 0   \\
0   & a_1  & a_0 
\end{array}} \right) \; \in \; Hom(S^2\, V^*, {\bf C} \oplus V^*)
\eeqa
One finds (where $\det \, \D = Res(A_1, B_2)$, cf.~sect.~\ref{intro Ex 1} and
\ref{resultant})
\beqa
\det\, \iota_1 = -4^{10}\; \Big( \det \, \D_3 \Big)^{11} \; Q_{11}
\eeqa

Let us discuss the pair
\beqa
l(-F) & = & \cO_{\E}(9s-F)\\
\bar{l}(-F)&=&\cO_{\E}(3s-0F)
\eeqa
from the point of view of the reduction philosophy described in
sect.~\ref{Exceptional}. Concerning $\bar{l}(-F)$ we are here in the 
exceptional case $\bar{\al}=n, \bar{\beta}=0$; so the equality condition leads 
to the $r$-fix $r=4$ from (\ref{exceptional r fix}) and the discussion after
(\ref{n=3 exceptional}) shows that, although the $\la=3/2$ analogue of $l(-F)$ 
would have given $l(-F)=\bar{l}(-F)^{\ox 2}$, here $\la \geq 5/2$ is required.

Now the long exact sequence for $\bar{l}$
\beqa
 0\;\;\;\;\;\;\;\;\;\;\;\;\; &\lra &\; 
H^0\Big(\E, \cO_{\E}(3s)\Big)\; \buildrel \rho \over \lra \; 
H^0\Big(c, \cO_{\E}(3s)|_c\Big)\nonumber\\
\buildrel \delta \over \lra 
H^1\Big(\E, \cO_{\E}(-4F)\Big)\; &\buildrel \bar{\iota}_1 \over \lra &
\; H^1\Big(\E, \cO_{\E}(3s)\Big)
\eeqa
gives $\ker \bar{\iota}_1\cong 
H^0(c, \cO_{\E}(3s)|_c)/\rho H^0(\E, \cO_{\E}(3s))$.
So generically, for $f:=\det \bar{\iota}_1\neq 0$,
one has $h^0(c, \cO_{\E}(3s)|_c)=1$ from the section given by $z$.
The necessary condition  (\ref{nec cond for prec reduc})
for precise reduction $p\deg \bar{l}(-F)|_c\leq \deg l(-F)|_c$
gives $p=1$ or $2$; we consider here the case $p=2$. 
Then the bundle $\tilde{\cF}=\cO_c(D-2\bar{D})$ 
from sect.~\ref{difference subsection}, 
which needs to have a nontrivial section to carry through 
the reduction procedure, 
is the bundle $\Lambda|_c=\cO_{\E}(3s-F)|_c$, cf.~(\ref{Lambda})
(this is also in line with the remarks after (\ref{n=3 exceptional}));
this flat bundle has $h^0(c, \Lambda|_c)$ either $0$ or $1$.
So, if both, 
$\bar{l}(-F)|_c=\cO_{\E}(3s)|_c$ and $\Lambda|_c=\cO_{\E}(3s-F)|_c$,
have a nontrivial section (the first line bundle
has $z$) then so does $l(-F)|_c$.
So (cf.~(\ref{general Pfaff factor}))
\beqa
\label{f_L Pfaff relation}
f_{\Lambda}\, | \, Pfaff
\eeqa
To find the result $f|Pfaff$ we now show that here actually $f=f_{\Lambda}$.

First we argue for $f_{\Lambda} | f$.
Now, with (\ref{can bund of c}), one has 
\beqa
\label{P equ}
h^0(c, \cO_{\E}(3s)|_c)&=&3-6+h^0(c, \cO_{\E}(3F)|_c)=: -3+P\\
\label{Q equ}
h^0(c, \cO_{\E}(3s-F)|_c)&=&0-6+h^0(c, \cO_{\E}(4F)|_c)=: -6+Q
\eeqa
such that always $P\geq 3, Q\geq 6$.
Generically one has $P=4$, from the section $z$; note that
$P=h^0(c, \cO_{\E}(3F)|_c)\geq h^0(\E, \cO_{\E}(3F))=h^0(b, \cO_b(3))=4$. 
As remarked above actually $Q=6$ or $7$. 
By (\ref{general Pfaff factor}) the latter is
sufficient for reduction (giving (\ref{f_L Pfaff relation})):
$Q=7\Llra h^0(c,\Lambda|_c)\geq 1\Llra 
\Lambda|_c\cong \cO_c \Llra f_{\Lambda}=0 \Lra l(-f)|_c\cong \cO_{\E}(2F)|_c$.
Note that $Q\geq 7\Lra P\geq 5$ as then $\cO(3s)|_c\cong \cO(F)|_c$ 
with the two sections $\pi^*_cu$ and $\pi^*_cv$; so one gets
another (linearly independent) section $z'$ of $\cO_{\E}(3s)|_c$, such that
\beqa
\label{f_Lambda to f relation}
f_{\Lambda}\, | \, f
\eeqa
\hspace{.3cm}Now we argue for $f| f_{\Lambda}$: the generic result
$s|_c=\{p_0, t_0\}\, , \; F_{t_0}|_c=c_{t_0}=\{p_0+p_++p_-, t_0\}$
(with $A_1(t_0)=0$, cf.~(\ref{s|_c generically}))
changes at the locus 
$f=\det \bar{\iota}_1=\det \D_3 =Res(A_1, B_2)=0$
(where $B_2(t_0)=0$): $p_0$ becomes a three-fold point of $c_{t_0}$. 
So one has in ${\cal M}_{\E}(c)$
\beqa
f=0 & \Llra &
(z)|_c=3s|_c=\{3p_0, t_0\} = F_{t_0}|_c
\eeqa
Therefore $f\, | \, f_{\Lambda}$ and one gets here that actually even
$f=f_{\Lambda}$ and so $f|Pfaff$ by (\ref{f_L Pfaff relation}).

\appendix

\resetcounter

\section{\label{Polynomial factors}
The polynomial factors of the Examples in detail}

After restricting from the twodimensional base $B$ 
to the rational curve $b\subset B$ (with its own homogeneous
coordinates $u$ and $v$)
the defining equation for the spectral curve $c$ 
(of class $ns+rF$, cf.~sect.~\ref{coordinates for curve case})
in the elliptic surface $\E$ over $b$ is
\beqa
w:=C_r(u,v)z+B_{r-2\chi}(u,v)x+A_{r-3\chi}(u,v)y=0
\eeqa
Here the subscripts denote the degrees of the homogeneous polynomials and
for the cases of $B={\bf F_k}$ (as for the Examples in the table of
sect.~\ref{Some experimental results})
one has $\chi=k-2$, cf.~sect.~\ref{chi subsection}; $x,y,z$ are, of course,
the usual Weierstrass coordinates for the elliptic fibre, 
cf.~sect.~\ref{Elliptic CY}.

\subsection{\label{intro Ex 1}Detailed consideration of Example 1}

Here the spectral curve has the equation 
(with $D_m=\sum_{i=0}^m d_i u^i v^{m-i}$ for $D=A,B,C$)
\beqa
w=C_4z+B_2x+A_1y=0
\eeqa
and the experimental result is, in more detail, the following
\beqa
\label{exp result Example 1}
Pfaff&=&-4^{10}\, \D_3^{11}\, Q_{11}
\eeqa
where the factors have the following meaning: $\D_3$ is defined by
\beqa
\label{degree 3 pol}
\D_3&=&b_2 a_0^2-b_1a_1a_0+b_0a_1^2
\eeqa
which is of course the following resultant
(cf.~app.~\ref{sym})
\beqa
\label{D3 matrix}
\D_3 := Res(B_2, A_1)  = 
\det\, \left( {\begin{array}{ccc}
b_2 & b_1  & b_0  \\
a_1 & a_0  & 0   \\
0   & a_1  & a_0 
\end{array}} \right) 
\eeqa

{\em $Q_{11}$: The Giant}

The structure of the second factor is more involved, it has the 
following 132 terms
\beqa
Q_{11}&=&-2 a_1^6 b_0 b_1^3 c_0 + 2 a_0 a_1^5 b_1^4 c_0 
+ 4 a_1^6 b_0^2 b_1 b_2 c_0 + 
 4 a_0 a_1^5 b_0 b_1^2 b_2 c_0 - 10 a_0^2 a_1^4 b_1^3 b_2 c_0
\nonumber\\
&&- 8 a_0 a_1^5 b_0^2 b_2^2 c_0 + 20 a_0^3 a_1^3 b_1^2 b_2^2 c_0 - 
 20 a_0^4 a_1^2 b_1 b_2^3 c_0 + 8 a_0^5 a_1 b_2^4 c_0 + 48 a_1^8 b_1 c_0^2
\nonumber
\eeqa
\newpage
\beqa
&&
- 96 a_0 a_1^7 b_2 c_0^2 + 2 a_1^6 b_0^2 b_1^2 c_1 - 2 a_0 a_1^5 b_0 b_1^3 c_1 
- 2 a_1^6 b_0^3 b_2 c_1 - 6 a_0 a_1^5 b_0^2 b_1 b_2 c_1 
\nonumber\\ 
&&
+ 10 a_0^2 a_1^4 b_0 b_1^2 b_2 c_1 + 10 a_0^2 a_1^4 b_0^2 b_2^2 c_1 - 
 20 a_0^3 a_1^3 b_0 b_1 b_2^2 c_1 + 10 a_0^4 a_1^2 b_0 b_2^3 c_1 + 
 2 a_0^5 a_1 b_1 b_2^3 c_1
\nonumber\\ 
&& 
- 2 a_0^6 b_2^4 c_1 - 24 a_1^8 b_0 c_0 c_1 - 
 72 a_0 a_1^7 b_1 c_0 c_1 + 168 a_0^2 a_1^6 b_2 c_0 c_1 
+ 24 a_0 a_1^7 b_0 c_1^2 
\nonumber\\ 
&&
+24 a_0^2 a_1^6 b_1 c_1^2 - 72 a_0^3 a_1^5 b_2 c_1^2 - 2 a_1^6 b_0^3 b_1 c_2 + 
 2 a_0 a_1^5 b_0^2 b_1^2 c_2 + 8 a_0 a_1^5 b_0^3 b_2 c_2 
\nonumber\\ 
&&
-10 a_0^2 a_1^4 b_0^2 b_1 b_2 c_2 + 10 a_0^4 a_1^2 b_0 b_1 b_2^2 c_2 - 
 2 a_0^5 a_1 b_1^2 b_2^2 c_2 - 8 a_0^5 a_1 b_0 b_2^3 c_2 
+ 2 a_0^6 b_1 b_2^3 c_2 
\nonumber\\ 
&&
+48 a_0 a_1^7 b_0 c_0 c_2 + 48 a_0^2 a_1^6 b_1 c_0 c_2 - 
 144 a_0^3 a_1^5 b_2 c_0 c_2 - 72 a_0^2 a_1^6 b_0 c_1 c_2 - 
 24 a_0^3 a_1^5 b_1 c_1 c_2 
\nonumber\\ 
&&
+ 120 a_0^4 a_1^4 b_2 c_1 c_2 + 
 48 a_0^3 a_1^5 b_0 c_2^2 - 48 a_0^5 a_1^3 b_2 c_2^2 + 2 a_1^6 b_0^4 c_3 - 
 2 a_0 a_1^5 b_0^3 b_1 c_3 
\nonumber\\ 
&&
- 10 a_0^2 a_1^4 b_0^3 b_2 c_3 + 
 20 a_0^3 a_1^3 b_0^2 b_1 b_2 c_3 - 10 a_0^4 a_1^2 b_0 b_1^2 b_2 c_3 + 
 2 a_0^5 a_1 b_1^3 b_2 c_3 - 10 a_0^4 a_1^2 b_0^2 b_2^2 c_3 
\nonumber\\ 
&&
+6 a_0^5 a_1 b_0 b_1 b_2^2 c_3 - 2 a_0^6 b_1^2 b_2^2 c_3 
+ 2 a_0^6 b_0 b_2^3 c_3 - 
 72 a_0^2 a_1^6 b_0 c_0 c_3 - 24 a_0^3 a_1^5 b_1 c_0 c_3 
\nonumber\\ 
&&
+120 a_0^4 a_1^4 b_2 c_0 c_3 + 96 a_0^3 a_1^5 b_0 c_1 c_3 - 
 96 a_0^5 a_1^3 b_2 c_1 c_3 - 120 a_0^4 a_1^4 b_0 c_2 c_3 + 
 24 a_0^5 a_1^3 b_1 c_2 c_3 
\nonumber\\ 
&&
+ 72 a_0^6 a_1^2 b_2 c_2 c_3 + 
 72 a_0^5 a_1^3 b_0 c_3^2 - 24 a_0^6 a_1^2 b_1 c_3^2 - 24 a_0^7 a_1 b2 c_3^2 - 
 8 a_0 a_1^5 b_0^4 c_4 
\nonumber\\ 
&&
+ 20 a_0^2 a_1^4 b_0^3 b_1 c_4 - 
 20 a_0^3 a_1^3 b_0^2 b_1^2 c_4 + 10 a_0^4 a_1^2 b_0 b_1^3 c_4 - 
 2 a_0^5 a_1 b_1^4 c_4 - 4 a_0^5 a_1 b_0 b_1^2 b_2 c_4
\nonumber\\ 
&&
+ 2 a_0^6 b_1^3 b_2 c_4 + 
 8 a_0^5 a_1 b_0^2 b_2^2 c_4 - 4 a_0^6 b_0 b_1 b_2^2 c_4 + 
 96 a_0^3 a_1^5 b_0 c_0 c_4 - 96 a_0^5 a_1^3 b_2 c_0 c_4 
\nonumber\\ 
&&
-120 a_0^4 a_1^4 b_0 c_1 c_4 + 24 a_0^5 a_1^3 b_1 c_1 c_4 + 
 72 a_0^6 a_1^2 b_2 c_1 c_4 + 144 a_0^5 a_1^3 b_0 c_2 c_4 - 
 48 a_0^6 a_1^2 b_1 c_2 c_4 
\nonumber\\ 
&&
- 48 a_0^7 a_1 b_2 c_2 c_4 - 
 168 a_0^6 a_1^2 b_0 c_3 c_4 + 72 a_0^7 a_1 b_1 c_3 c_4 
+ 24 a_0^8 b_2 c_3 c_4 + 96 a_0^7 a_1 b_0 c_4^2 
\nonumber\\ 
&&
- 48 a_0^8 b_1 c_4^2 - 2 a_1^8 b_0^2 b_1 g_0 + 
 4 a_0 a_1^7 b_0 b_1^2 g_0 - 2 a_0^2 a_1^6 b_1^3 g_0 
+ 4 a_0 a_1^7 b_0^2 b_2 g_0 
\nonumber\\ 
&&
-12 a_0^2 a_1^6 b_0 b_1 b_2 g_0 + 8 a_0^3 a_1^5 b_1^2 b_2 g_0 + 
 8 a_0^3 a_1^5 b_0 b_2^2 g_0 - 10 a_0^4 a_1^4 b_1 b_2^2 g_0 + 
 4 a_0^5 a_1^3 b_2^3 g_0 
\nonumber\\ 
&&
+ a_1^8 b_0^3 g_1 - a_0 a_1^7 b_0^2 b_1 g_1 - 
 a_0^2 a_1^6 b_0 b_1^2 g_1 + a_0^3 a_1^5 b_1^3 g_1 
- a_0^2 a_1^6 b_0^2 b_2 g_1 
\nonumber\\ 
&&
+6 a_0^3 a_1^5 b_0 b_1 b_2 g_1 - 5 a_0^4 a_1^4 b_1^2 b_2 g_1 - 
 5 a_0^4 a_1^4 b_0 b_2^2 g_1 + 7 a_0^5 a_1^3 b_1 b_2^2 g_1 - 
 3 a_0^6 a_1^2 b_2^3 g_1 
\nonumber\\ 
&&
- 2 a_0 a_1^7 b_0^3 g_2 + 4 a_0^2 a_1^6 b_0^2 b_1 g_2 - 
 2 a_0^3 a_1^5 b_0 b_1^2 g_2 - 2 a_0^3 a_1^5 b_0^2 b_2 g_2 + 
 2 a_0^5 a_1^3 b_1^2 b_2 g_2 
\nonumber\\ 
&&
+ 2 a_0^5 a_1^3 b_0 b_2^2 g_2 - 
 4 a_0^6 a_1^2 b_1 b_2^2 g_2 + 2 a_0^7 a_1 b_2^3 g_2 
+ 3 a_0^2 a_1^6 b_0^3 g_3 - 
 7 a_0^3 a_1^5 b_0^2 b_1 g_3 
\nonumber\\ 
&&
+ 5 a_0^4 a_1^4 b_0 b_1^2 g_3 - 
 a_0^5 a_1^3 b_1^3 g_3 + 5 a_0^4 a_1^4 b_0^2 b_2 g_3 - 
 6 a_0^5 a_1^3 b_0 b_1 b_2 g_3 + a_0^6 a_1^2 b_1^2 b_2 g_3 
\nonumber\\ 
&&
+a_0^6 a_1^2 b_0 b_2^2 g_3 + a_0^7 a_1 b_1 b_2^2 g_3 - a_0^8 b_2^3 g_3 - 
 4 a_0^3 a_1^5 b_0^3 g_4 + 10 a_0^4 a_1^4 b_0^2 b_1 g_4 
\nonumber\\ 
&&
-8 a_0^5 a_1^3 b_0 b_1^2 g_4 + 2 a_0^6 a_1^2 b_1^3 g_4 - 
 8 a_0^5 a_1^3 b_0^2 b_2 g_4 + 12 a_0^6 a_1^2 b_0 b_1 b_2 g_4 - 
 4 a_0^7 a_1 b_1^2 b_2 g_4 
\nonumber\\ 
&&
- 4 a_0^7 a_1 b_0 b_2^2 g_4 + 2 a_0^8 b_1 b_2^2 g_4
\eeqa

{\em The Giant is friendly}

The expression for $Q_{11}$ looks unwieldily.
However meditation reveals\footnote{as a minor difference
we find in contrast to [\ref{BDO}] a minus-sign in front of $D_1$ 
(although not an overall sign this is tunable by the sign of $G_4$) 
and get (\ref{exp result Example 1}) (with a prefactor $4^3$)
with (\ref{Q11 decomposition})
by using elements $z^2x, z^2y, zx^2, zxy, x^3, x^2y, xy^2, y^3$
for the $H^1(\E, \cO(9s-F))$ decomposition, 
cf.~also app.~\ref{concrete matrix representations}; note that
in line with the treatment for the $f$-factor would actually 
be a representation as {\em one} determinant} 
\beqa
\label{Q11 decomposition}
Q_{11}&=&-D_1+D_2+D_3
\eeqa
(cf.~app.~\ref{Giant decomposition})
where one has the individual terms
(with $zy^2=4x^3-G_4xz^2-G_6z^3$ the elliptic fibre over $b$; thus here 
a dependence on the complex structure moduli of $X$ enters)
\beqa
\mbox{type } "(ga)b^3a^7"\;\;\;\;\;\;
D_1&=&\D_3^2\cdot \frac{R(G_4, B_2^2, A_1^2)}{\D_3}\\
\mbox{type } "c^2ba^8"\;\;\;\;\;\;
D_2&=&24\, \D_5 \cdot \frac{R(C_4, B_2^2, A_1^2)}{\D_3}\\
\label{D3 equation}
\mbox{type } "cb^4a^6"\;\;\;\;\;\;
D_3&=&2\, \D_3\cdot R(C_4, A_1^4, B_2)
\eeqa
Here the meaning of the factors is the following:
$\D_5$ is defined by
\beqa
\label{D5 explicit}
\D_5&=&\sum_{i=0}^4(-1)^i c_i a_0^i a_1^{4-i}
\eeqa
which is of course the following resultant
\beqa
\label{D5 matrix}
\D_5 := Res(C_4,A_1)=
\det\, \left( {\begin{array}{ccccc}
c_4  & c_3   & c_2 & c_1  & c_0  \\
a_1 & a_0  &   0 &   0  &   0  \\
0   & a_1  & a_0 &   0  &   0  \\
0   &   0  & a_1 & a_0  &   0  \\
0   &   0  &   0 & a_1  & a_0
\end{array}} \right) 
\eeqa
The other terms are the following 'multi-resultants'
\beqa
R(C_4, B_2^2, A_1^2)&=&
\det\, \left( {\begin{array}{ccccc}
c_4  & c_3   & c_2 & c_1  & c_0  \\
b_2^2 & 2b_2 b_1  &   2b_2 b_0 + b_1^2 &   2b_1 b_0  &   b_0^2  \\
a_1^2   & 2a_1 a_0  & a_0^2 &   0  &   0  \\
0   &   a_1^2  & 2a_1 a_0& a_0^2  &   0  \\
0   &   0  &   a_1^2 & 2a_1 a_0  & a_0^2
\end{array}} \right) 
\eeqa
and correspondingly for $G_4$; realise that
this contains actually a $\D_3$-factor !
Finally
\beqa
R(C_4, A_1^4, B_2)&=&
- \det\, \left( {\begin{array}{ccccc}
c_4   & c_3        & c_2          & c_1        & c_0  \\
b_2   & b_1        & b_0          &   0        &   0  \\
0     & b_2        & b_1          & b_0        &   0  \\
0     &   0        & b_2          & b_1        & b_0  \\
a_1^4 & 4a_1^3 a_0 & 6a_1^2 a_0^2 & 4a_1 a_0^3 & a_0^4
\end{array}} \right) 
\eeqa

\subsection{Detailed consideration of Example 2}

The spectral curve equation is
$w=C_5z+B_3x+A_2y=0$ and the experimental result is
\beqa
Pfaff &=& (\D_5^{II})^4
\eeqa
where 
\beqa
\D_5^{II}&=&a_2^3 b_0^2 - a_1 a_2^2 b_0 b_1 + a_0 a_2^2 b_1^2 
+ a_1^2 a_2 b_0 b_2 -  2 a_0 a_2^2 b_0 b_2 
 - a_0 a_1 a_2 b_1 b_2 + a_0^2 a_2 b_2^2 \nonumber\\
&&- a_1^3 b_0 b_3 + 
 3 a_0 a_1 a_2 b_0 b_3 + a_0 a_1^2 b_1 b_3 
- 2 a_0^2 a_2 b_1 b_3 - a_0^2 a_1 b_2 b_3 +  a_0^3 b_3^2
\eeqa
which is the following resultant
\beqa
\D_5^{II}&=&R(B_3, A_2)=\det \left( {\begin{array}{ccccc}
b_3 & b_2  & b_1 & b_0 & 0  \\
0   & b_3  & b_2 & b_1 & b_0\\
a_2 & a_1  & a_0 & 0   & 0  \\
0   & a_2  & a_1 & a_0 & 0  \\
0   & 0    & a_2 & a_1 & a_0
\end{array}} \right) 
\eeqa

\subsection{Detailed consideration of Example 4}

The spectral curve equation is 
$w=C_2z+B_2x+A_2y=0$ and the experimental result is
\beqa
Pfaff &=& (\D_3^{IV})^4
\eeqa
where 
\beqa
\D_3^{IV}&=&c_2b_1a_0-c_2b_0a_1+c_1b_0a_2-c_1b_2a_0+c_0b_2a_1-c_0b_1a_2
\eeqa
which is the following 'multi-resultant'
\beqa
\D_3^{IV}&=&R(C_2, B_2, A_2)=\det\left( {\begin{array}{ccc}
c_2 & c_1  & c_0 \\
b_2  & b_1 & b_0 \\
a_2 & a_1  & a_0 
\end{array}} \right) 
\eeqa

\newpage

\subsection{\label{Giant decomposition}
The decomposition of the Giant factor $Q_{11}$}

To follow in greater detail the process how 
the decomposition of $Q_{11}$ arises note that in the expression 
for $Q_{11}$ three types of terms occur:
first those with $g_i$ (and with no $c_i$), second those with 
$c_ic_j$ and third those with $c_i$.
More precisely the first group of 46 terms has the {\bf type $"(ga)b^3a^7"$}
\beqa
D_1&=&
g_0 a_0\Bigg(10 a_1^2 b_1^2 b_2 + 
    10 b_2^2 (a_1^2 b_0 - 2 a_0 a_1 b_1 + a_0^2 b_2)\Bigg) a_0^2 a_1^3  
\nonumber\\
&&
+ (g_0 a_1+g_1 a_0) \Bigg(-2 a_1^5 b_0^2 b_1 - 2 a_0^2 a_1^3 b_1^3 
- 12 a_0^2 a_1^3 b_0 b_1 b_2+ 
    10 a_0^4 a_1 b_1 b_2^2 - 6 a_0^5 b_2^3 
\nonumber\\
&&
\;\;\;\;\;\;\;\;\;\;\;\;\;\;\;\;\;\;\;\;\;\;\;\;\;\;
+ 4 a_0 a_1^4 b_0(b_1^2 + b_0 b_2)- 
    2 a_0^3 a_1^2 b_2 (b_1^2 + b_0 b_2)\Bigg)a_1^2 
\nonumber\\
&&
+ (g_2 a_0+g_1 a_1) 
\Bigg(a_1^6 b_0^3 + a_0 a_1^5 b_0^2 b_1 + 3 a_0^3 a_1^3 b_1^3 + 
    18 a_0^3 a_1^3 b_0 b_1 b_2 - 3 a_0^5 a_1 b_1 b_2^2 
\nonumber\\
&&
\;\;\;\;\;\;\;\; \;\;\;\;\;\;\;\;\;\;\;\;\;\;\;\;\;\;
+ 3 a_0^6 b_2^3 - 5 a_0^2 a_1^4 b_0 (b_1^2 + b_0 b_2) - 
    3 a_0^4 a_1^2 b_2 (b_1^2 + b_0 b_2)\Bigg) a_1 
\nonumber\\
&&
+ (g_3 a_0+g_2 a_1) 
\Bigg(-3 a_1^6 b_0^3 + 3 a_0 a_1^5 b_0^2 b_1 - 3 a_0^3 a_1^3 b_1^3 - 
    18 a_0^3 a_1^3 b_0 b_1 b_2 - a_0^5 a_1 b_1 b_2^2 
\nonumber\\
&&
\;\;\;\;\;\;\;\; \;\;\;\;\;\;\;\;\;\;\;\;\;\;\;\;\;\;
-a_0^6 b_2^3 + (3 a_0^2 a_1^4 b_0 + 5 a_0^4 a_1^2 b_2) (b_1^2 + b_0 b_2)\Bigg) 
a_0 
\nonumber\\
&&
+(g_4 a_0+g_3 a_1) 
\Bigg(6 a_1^5 b_0^3 - 10 a_0 a_1^4 b_0^2 b_1 + 2 a_0^3 a_1^2 b_1^3 + 
    12 a_0^3 a_1^2 b_0 b_1 b_2 + 2 a_0^5 b_1 b_2^2 
\nonumber\\
&&
\;\;\;\;\;\;\;\;\;\;\;\;\;\;\;\;\;\;\;\;\;\;\;\;\;\;
+2 a_0^2 a_1^3 b_0 (b_1^2 + b_0 b_2) - 4 a_0^4 a_1 b_2 (b_1^2 + b_0 b_2)\Bigg) 
a_0^2 
\nonumber\\
&&
+ g_4 a_1\Bigg(-10 a_0^2 b_0 b_1^2 - 
    10 b_0^2 (a_1^2 b_0 - 2 a_0 a_1 b_1 + a_0^2 b_2)\Bigg) a_0^3 a_1^2 
\eeqa
(with terms already regrouped according to
the coefficients not of $G_4$ but of $G_4A_1$).

The second group of 40  terms has the {\bf type $"c^2ba^8"$}
\beqa
D_2&=&
24 \Bigg[c_0^2 (-2 a_1^2 b_1 + 4 a_0 a_1 b_2) a_1^6 + 
   c_0 c_1 (a_1^2 b_0 + 3 a_0 a_1 b_1 - 7 a_0^2 b_2) a_1^6 
\nonumber\\
&&
+ (c_1^2 + 2 c_0 c_2) (-a_1^2 b_0 - a_0 a_1 b_1 + 3 a_0^2 b_2) a_0 a_1^5 + 
   (c_1 c_2 + c_0 c_3)(3 a_1^2 b_0 + a_0 a_1 b_1 - 5 a_0^2 b_2) a_0^2 a_1^4  
\nonumber\\
&&
+ (c_2^2 + 2 c_1 c_3 + 2 c_0 c_4) (-2 a_1^2 b_0 + 2 a_0^2 b_2) a_0^3 a_1^3 + 
   (c_2 c_3 + c_1 c_4)(5 a_1^2 b_0 - a_0 a_1 b_1 - 3 a_0^2 b_2) a_0^4 a_1^2  
\nonumber\\
&&
+ (c_3^2 + 2 c_2 c_4)(-3 a_1^2 b_0 + a_0 a_1 b_1 + a_0^2 b_2) a_0^5 a_1 
\nonumber\\
&&
+ c_3 c_4 (7 a_1^2 b_0 - 3 a_0 a_1 b_1 - a_0^2 b_2) a_0^6 + 
   c_4^2 (-4 a_0 a_1 b_0 + 2 a_0^2 b_1) a_0^6 \Bigg]
\eeqa
Finally the last group of 46 terms has the {\bf type $"cb^4a^6"$}
\beqa
D_3&=&c_0\Bigg(2 a_1^6 b_0 b_1^3 - 2 a_0 a_1^5 b_1^4 - 4 a_1^6 b_0^2 b_1 b_2 - 
    4 a_0 a_1^5 b_0 b_1^2 b_2 + 10 a_0^2 a_1^4 b_1^3 b_2 \nonumber\\
&&\;\;\;\;\;\;\; + 8 a_0 a_1^5 b_0^2 b_2^2 - 20 a_0^3 a_1^3 b_1^2 b_2^2 + 
    20 a_0^4 a_1^2 b_1 b_2^3 - 8 a_0^5 a_1 b_2^4\Bigg) 
\nonumber\\
&&
+c_1 \Bigg(-2 a_1^6 b_0^2 b_1^2 + 
    2 a_0 a_1^5 b_0 b_1^3 + 2 a_1^6 b_0^3 b_2 + 6 a_0 a_1^5 b_0^2 b_1 b_2 - 
    10 a_0^2 a_1^4 b_0 b_1^2 b_2 \nonumber\\
&&\;\;\;\;\;\;\;\;\; - 10 a_0^2 a_1^4 b_0^2 b_2^2 + 
    20 a_0^3 a_1^3 b_0 b_1 b_2^2 - 10 a_0^4 a_1^2 b_0 b_2^3 - 
    2 a_0^5 a_1 b_1 b_2^3 + 2 a_0^6 b_2^4\Bigg)  \nonumber\\
&&
+ c_2 \Bigg(2 a_1^6 b_0^3 b_1 - 
    2 a_0 a_1^5 b_0^2 b_1^2 - 8 a_0 a_1^5 b_0^3 b_2 + 
    10 a_0^2 a_1^4 b_0^2 b_1 b_2 - 10 a_0^4 a_1^2 b_0 b_1 b_2^2 
\nonumber\\
&&\;\;\;\;\;\;\;\;\; +  2 a_0^5 a_1 b_1^2 b_2^2 + 8 a_0^5 a_1 b_0 b_2^3 - 
    2 a_0^6 b_1 b_2^3\Bigg)
\nonumber\\
&&
+ c_3 \Bigg(-2 a_1^6 b_0^4 + 2 a_0 a_1^5 b_0^3 b1 + 
    10 a_0^2 a_1^4 b_0^3 b_2 - 20 a_0^3 a_1^3 b_0^2 b_1 b_2 + 
    10 a_0^4 a_1^2 b_0 b_1^2 b_2 \nonumber\\
&&\;\;\;\;\;\;\;\;\; - 2 a_0^5 a_1 b_1^3 b_2 + 
    10 a_0^4 a_1^2 b_0^2 b_2^2 - 6 a_0^5 a_1 b_0 b_1 b_2^2 + 
    2 a_0^6 b_1^2 b_2^2 - 2 a_0^6 b_0 b_2^3\Bigg)  \nonumber\\
&&
+c_4 \Bigg(8 a_0 a_1^5 b_0^4 - 
    20 a_0^2 a_1^4 b_0^3 b_1 + 20 a_0^3 a_1^3 b_0^2 b_1^2 - 
    10 a_0^4 a_1^2 b_0 b_1^3 + 2 a_0^5 a_1 b_1^4 
\nonumber\\
&&\;\;\;\;\;\;\;\;\; + 4 a_0^5 a_1 b_0 b_1^2 b_2 - 
    2 a_0^6 b_1^3 b_2 - 8 a_0^5 a_1 b_0^2 b_2^2 + 4 a_0^6 b_0 b_1 b_2^2\Bigg) 
\eeqa

From $D_1$ and $D_2$ one can split off a factor $\D_3^2$ and $24 \D_5$, 
respectively, and gets (for $D_2$, say)
a complicated polynomial
(with the coefficients $c_i$ of $C_4$ replaced by the coefficients of $G_4$
for $D_1$)
\beqa
P&=&
a_1^4(-2 b_1 c_0 + b_0 c_1) - a_1^3 a_0(-4 b_2 c_0 - b_1 c_1 + 2 b_0 c_2) + 
 a_1^2 a_0^2 (-3 b_2 c_1 + 3 b_0 c_3) \nonumber\\
&&- a_1 a_0^3(-2 b_2 c_2 + b_1 c_3 + 4 b_0 c_4) + a_0^4(-b_2 c_3 + 2 b_1 c_4)
\eeqa
The structural meaning of this expression is revealed by
the following identity 
\beqa
\det\, \left( {\begin{array}{ccccc}
c_4  & c_3   & c_2 & c_1  & c_0  \\
b_2^2 & 2b_2 b_1  &   2b_2 b_0 + b_1^2 &   2b_1 b_0  &   b_0^2  \\
a_1^2   & 2a_1 a_0  & a_0^2 &   0  &   0  \\
0   &   a_1^2  & 2a_1 a_0& a_0^2  &   0  \\
0   &   0  &   a_1^2 & 2a_1 a_0  & a_0^2
\end{array}} \right) 
&=& P \cdot 
\det\, \left( {\begin{array}{ccc}
b_2 & b_1  & b_0  \\
a_1 & a_0  & 0   \\
0   & a_1  & a_0 
\end{array}} \right) 
\eeqa
(cf.~sect.~\ref{intro Ex 1}).
Similarly $D_3$ is $2$ times the product of two determinants,
cf.~(\ref{D3 equation}).


\section{\label{rational curves}Rational Curves $P$ in $X$}
\resetcounter

Which (smooth) rational curves $P$, suitable as support for the 
world-sheet instanton, exist in $X$ ? 
As we want to bring to bear the elliptically fibered
structure of $X$ and the spectral nature of $V$ we concentrate on 
horizontal curves: if $p \si + a_P F$ denotes the cohomology class 
then $P$ is said to lie 'horizontally' (embedded in $B$ via $\si$)
if $a_P =0$. We first search for such {\em rational} base curves;
then we treat the question of isolatedness.

\subsection{The different rational base surfaces $B$}

The Calabi-Yau threefold $X$ has
the following possible rational surfaces as bases $B$:
a Hirzebruch surface ${\bf F_k}$ with $k=0,1,2$ (or blow-ups of it); 
or $B$ is ${\bf P^2}$
or blow-ups of it, i.e. one of the del Pezzo surfaces $dP_k$ with $k\leq 8$;
finally the Enriques surface is possible.

The surface ${\bf F_k}$ is a ${\bf P^1}$-fibration over
a base ${\bf P_1}$ denoted by $b$ (the fibre is denoted by $f$;
if no confusion arises $b$ and $f$ will denote also the 
cohomology classes). One finds $c_1({\bf F_k})=2b+(2+k)f$.
$b$ of $b^2=-k$ is a section of the fibration; 
there is another section ("at infinity") of class $b_{\infty}=b+kf$
of self-intersection $+k$.
The Kaehler cone (the very ample classes) equals the positive 
(ample) classes and is given (cf.~footn.~\ref{Hartshorne}) by the 
numerically effective classes $xb+yf$ with $x>0, y>kx$. An irreducible
non-singular curve exists in a class $xb+yf$ exactly if the class lies in the
mentioned cone or is one of the elements $b, f$ 
or $ab_{\infty}$ (with $a>0$) on the boundary of the cone; 
these classes
together with their positive linear combinations span the effective cone 
($x,y\geq 0$). 
$c_1$ is positive for ${\bf F_0}$ and ${\bf F_1}$, for 
${\bf F_2}$ it lies on the boundary of the positive cone.

We will concentrate on the case $B={\bf F_k}$ for illustration;
in ${\bf P^2}$ one has (non-isolated)
rational curves given by a line or a quadric
(of classes $l$ and $2l$); on a $dP_k$ one has among
various further curves the exceptional ${\bf P^1}$'s (from the blow-up) 
of self-intersection $-1$.

\subsection{Horizontal rational curves}

For $B$ a Hirzebruch surfaces ${\bf F_k}$ ($k=0,1,2$)
let us find the possible rational instanton curves on $B$
besides the three immediate candidates $P=b, b_{\infty}$ (of class $b+kf$) 
and $f$.

The cohomology class $P=xb+yf$ is represented by an irreducible smooth curve 
for\footnote{\label{Hartshorne}Cf.~R. Hartshorne, {\em Algebraic Geometry}, 
Springer Verlag (1977).} $P=b,f, a\, b_{\infty}$ 
(with $a>0$ and $k>0$) or $P>0$, i.e. $P$ ample (positive) 
which comes down to $P\cdot f = x >0$ and 
$P\cdot b = y - kx >0$. So, except for $f$ and $b$,
one has $x>0, y>0$ which we will assume from now on. Now
the numerical rationality condition
\beqa
\label{Euler number}
2 &\stackrel{\mbox{\normalsize{!}}}{=}&
c_1\cdot P - P^2 = 2(x+y)-kx+kx^2-2xy 
\eeqa
leads to the following possibilities: \\
${\bf F_0}:\;$ $P= b+yf$ or $P=xb+f$\\
${\bf F_1}:\;$ $P= b+yf$ or $P=2(y-1)b+yf$\\
${\bf F_2}:\;$ $P= b+yf$ or $P=(y-1)b+yf$.\\
Combining this finding with the requirement that 
$P$ (if not equal to $b$ or $f$) is either of the form $a(b+kf)$ or $P>0$ 
the following possibilities remain in total
\beqa
\label{remaining curves}
P = b\; , \;\;\; f \; , \;\;\; b+yf\;\; (y\geq k)
\eeqa
together with the mirrored case $xb+f$ on ${\bf F_0}$ and the 
exceptional $P=2b+2f$ on ${\bf F_1}$.

\subsection{The question of isolatedness}

Now let us study which of the rational curves found so far
are furthermore isolated. To make the discussion transparent we recall
first the relevant facts in general (cf.~also [\ref{AChor}]). 
For this we decompose the problem in 
steps: we first study in the next two subsections
the rigidity question for a base curve $P$ 
with respect to the two surfaces in which $P$ is contained, that is
for $B$ and $\E=\pi^{-1}(P)$. Finally we point to the decomposed nature
of the problem. The upshot is that the base curve $b$ in ${\bf F_1}$
remains (as the corresponding other rational blow-up curves of 
self-intersection $(-1)$ in a $dP_k$ base).

\subsubsection{\label{horizontal defs}
The deformations of $P$ in the rational base surface $B$}

We have a 'local' information ${\rm def}_B^{loc}(P):=h^0(P,N_B P)$ 
about deformations of $P$ in $B$
as well as a 'global' one 
${\rm def}_B^{glob}(P):={\rm def}_{B}(P):=h^0(B,{\cal O}(P))-1$.
Using Riemann-Roch
\beqa
\label{ro}
\sum_{i=0}^{2}(-1)^ih^i(B,{\cal O}_B(P))=h^0(P,N_{B}P)-h^1(P,N_{B}P)+1
\eeqa
(with $\chi (B,{\cal O})=1$ from Noether's theorem for our rational $B$)
we get 
\beqa
\label{defs}
{\rm def}_{B}(P)=h^0(P,N_{B}P)-h^1(P,N_{B}P)+s-h^2(B,{\cal O}_B(P))
\eeqa
(with the {\it superabundance} $s=h^1(B,{\cal O}_B(P))$)
with the local terms
\beqa
\label{deor}
h^0(P,N_B P)-h^1(P,N_B P)=\frac{1}{2}\chi(P)+{\rm deg} N_B P=\frac{Pc_1+P^2}{2}
\eeqa
Let us investigate the two higher cohomological corrections $s$ and
$h^2(B, {\cal O}_B(P))$ in (\ref{defs}).
For any curve $P$ on a rational surface $B$ 
(like a Hirzebruch surface ${\bf F_n}$ or a del Pezzo surface $dP_k$)
one has $h^2(B, {\cal O}_B(P))=0$ 
which can be seen from the exact sequence
$0\rightarrow {\cal O}_B\rightarrow {\cal O}_B(P)\rightarrow 
{\cal O}_P(P)\rightarrow 0$
with associated long exact sequence 
(where $B$ being rational one has $p_g(B)=h^2(B, {\cal O}_B)=0$ 
and  $q(B)=h^1(B, {\cal O}_B)=0$)
\beqa
\label{lesq}
0& \rightarrow &h^0(B,{\cal O}_B)\rightarrow h^0(B,{\cal O}_B(P))
\rightarrow h^0(P, {\cal O}_P(P)) \nonumber\\
 & \rightarrow &h^1(B,{\cal O}_B)\rightarrow h^1(B,{\cal O}_B(P))
\rightarrow h^1(P, {\cal O}_P(P)) \nonumber\\
 & \rightarrow &h^2(B,{\cal O}_B)\rightarrow h^2(B,{\cal O}_B(P))
\rightarrow 0
\eeqa
so $h^2(B, {\cal O}_B(P))=0$ and 
$h^1(B,{\cal O}_B(P))=h^1(P, {\cal O}_P(P))$ in (\ref{defs}), giving
\beqa
B\; {\rm rational} & \Lra & {\rm def}_B^{glob} (P)
= {\rm def}_B^{loc} (P)
\eeqa
The mentioned middle terms in
(\ref{defs}) not only cancel but vanish individually if $h^0(P,K_P-N_B P)=0$ 
which happens if $0>{\rm deg}( K_P-N_B P)=P(K_B+P)-P^2=PK_B$
which is guaranteed if $-K_B$ is ample,
as for $B={\bf F_n}$ with\footnote{For ${\bf F_2}$ of $c_1=2b_{\infty}$
 the Kodaira vanishing theorem 
gives $s=h^1(B,{\cal O}(P))=h^1(B,{\cal O}(K-P))=0$ 
if $-K+P=(x+2)b+(y+4)f$ is ample, i.e. for $y>2x, x>-2$, so clearly 
for all ample $P=xb+yf$ (where even $x>0$) and for $f$;
finally
$kb_+\cdot c_1({\bf F_2})=4k>0$ making (\ref{resBdef}) again applicable.} 
$n=0,1$ 
and $dP_k$ ($k\neq 9$); in general 
\beqa
\label{resBdef}
P\cdot c_1>0 & \Lra & {\rm def}_{B}(P)=h^0(P,N_{B}P)=\frac{Pc_1+P^2}{2}
\eeqa

{\em Isolated rational curves in ${\bf F_k}$}

We restrict ourselves to {\em isolated} instantons; so for
$p=b$, say, we restrict us to ${\bf F_1}, {\bf F_2}$
where $b^2<0$ enforces 
isolatedness\footnote{Having negative self-intersection $b$ cannot move;
more formally ${\rm def}_B(b)=0$ on ${\bf F_1}$ by (\ref{resBdef}), and
${\rm def}_B(P)=
\frac{Pc_1+P^2}{2}+h^1(P, N_B P)=-1+h^0(P, K_P-N_B P)=0$ on ${\bf F_2}$ 
from (\ref{deor}) and $N_B b=K_b$.}
in ${\bf F_k}$.\footnote{The same follows in $X$ for $k\neq 2$ 
(for $k=2$ one has $\E_b=b\x F$ showing a deformation), 
cf.~sect.~\ref{vertical defs}.}
Let us complete the discussion with the case $P=b+yf$ where $y> k$:
then the number of deformations of the very ample $P$ is
\beqa
def_{{\bf F_k}}(P)&=&2y-k+1
\eeqa
(here was $y>k$ but $P=b_{\infty}$ of $y=k$ is also covered:
$h^0(b_{\infty}, N_{\bf F_k} b_{\infty})
=h^0\big(b_{\infty}, {\cal O}_{b_{\infty}}(k)\big)=k+1$).
For $f$ obviously ${\rm def}_{F_n}f=1$ (also from (\ref{resBdef})).
Finally (cf.~(\ref{remaining curves})) $P=2b+2f$ on ${\bf F_1}$ has
$def_B(P)=5$ by (\ref{resBdef}). 
Thus only $b$ in ${\bf F_1}$ remains (similarly the blow-up curves
of self-intersection $(-1)$ in $dP_k$).

\subsubsection{\label{vertical defs}The deformations of $P$ in the vertical 
elliptic surface ${\cal E}$}

The Kodaira formula identifies the canonical bundle of ${\cal E}$ 
as a pull-back class 
$K_{{\cal E}}=\pi_{{\cal E}}^*K_P+\chi({\cal E},{\cal O}_{{\cal E}})F$.
So $\chi({\cal E},{\cal O}_{\cal E})={1\over 12}e({\cal E})$. 
But $e({\cal E})=12c_1\cdot P$ as the elliptic fibration 
has discriminant $12c_1$ so that
$\chi({\cal E},{\cal O}_{\cal E})=c_1\cdot P$.
Alternatively one can see from adjunction 
\beqa
\label{adju}
c({\cal E})=
c(P)
\frac{(1+r)(1+r+2c_1)(1+r+3c_1)}{1+3r+6c_1}
\eeqa
that
\beqa
\label{ok}
c_1({\cal E})= \Big(e(P)-c_1\cdot P\Big)F \; , \;\;\;\;\;\;\;\;\;\;
e({\cal E})= 12 c_1\cdot P
\eeqa
Note that the number $c_1\cdot P$ has the following important 
interpretation:
as $c_1({\cal E})$ is a pull-back class, i.e. a number of fibers,
one has from (\ref{ok}) that $P_{{\cal E}}\cdot c_1({\cal E})=e_P-c_1\cdot P_B$
(we write $P_{\E}$ when we wish to emphasize that $P$ 
is considered as a curve in $\E$);
so with adjunction $-e_P=P^2_{{\cal E}}-P_{{\cal E}}\cdot c_1({\cal E})$ 
inside ${\cal E}$ one gets that the self-intersection of $P_{\E}$ in $\E$ is 
$P^2_{{\cal E}}= -c_1\cdot P_B=-\chi({\cal E},{\cal O}_{\cal E})$.
So with ${\rm deg}(N_{\E}\, P_{\E})=P^2_{{\cal E}}$ one has the criterion
\beqa
\label{defE}
P\cdot c_1>0 & \Longrightarrow & def_{{\cal E}}^{loc}\, P_{{\cal E}}=0
\eeqa
So, assuming $P\cdot c_1>0$ as in (\ref{resBdef}), we have no 
further deformations in the vertical direction except for the case 
$P=b$ of $b\cdot c_1=0$ in $B={\bf F_2}$:
there ${\cal E}=b\times F$ shows 
a deformation in $X$ (while one has no deformation in the base $B$,
cf.~sect.~\ref{horizontal defs}).

\noindent
{\it Examples}

The cases $c_1\cdot P=0,1,2$
lead to $\E=b\times F, dP_9, K3$ of $e({\cal E})=0,12,24$.
These $\E$ occur over $P=b$ in $B={\bf F_k}$ 
with $k=2,1,0$ (cf.~sect.~\ref{chi subsection}): 
first, the ruled surface $b\times F$ over $F$, being one of
the two exceptional divisors (the other is the base ${\bf F_2}$ itself)
of the $STU$ Calabi-Yau ${\bf P}_{1,1,2,8,12}(24)$; 
secondly, the rational elliptic $dP_9$ surface which occurs also
over any exceptional curve ($b^2=-1, b\cdot c_1=1$; rationality
and the second property imply the first) in a $dP_k$ base; finally
for ${\bf F_0}$ one gets the well-known $K3$ fibers, which occur
also over each fiber of a ${\bf F_k}$ base ($c_1\cdot P=2$ by adjunction). 

\subsubsection{The deformations of $P$ in $X$}

The total deformation space can be considered fibered together out of
the pieces investigated so far. Concerning $def_X^{loc} \, P$ consider 
\beqa
\label{seqn}
0\rightarrow N_{B}\, P\rightarrow N_{X}\, P\rightarrow 
N_{X}\, B|_P\rightarrow 0
\eeqa
the last term being $N_{{\cal E}}\, P$.
To get $def_X^{loc}\, P=def_B^{loc}\, P$ one has to show that 
$H^0(P, N_{X}\, B|_P)=0$, 
i.e.~that there are no further deformations of $P$ in ${\cal E}$. This 
will hold if ${\rm deg} \, N_{{\cal E}}\, P=P^2_{{\cal E}}=-\chi<0$, 
cf.~(\ref{defE}).


\section{\label{Lemmata}Some Lemmata}

\resetcounter

\subsection{Lemma 1}

We follow the notation in section \ref{WS instantons} where one 
finds that $b$ contributes iff $V|_b$ is trivial
\beqa
W_b\neq 0 \Longleftrightarrow V|_b = \bigoplus_{i=1}^n\, \cO_b
\eeqa
A corresponding framing would give $n$ linearly independent 
global sections such that
\beqa
W_b\neq 0 \Longrightarrow h^0(b, V|_b)=n
\eeqa
This is also directly a consequence of 
$W_b\neq 0 \Llra h^0\Big(b, V|_b(-1)\Big)=0$ in (\ref{main criterion}).
For this note that the short exact sequence
\beqa
0 \lra l(-F)|_c \lra l|_c \lra {\bf C^n} \lra 0
\eeqa
and its associated long exact cohomology sequence
\beqa
0 \lra H^0(c, l(-F)|_c ) \lra H^0(c, l|_c ) \lra  {\bf C^n} \lra
H^1(c, l(-F)|_c ) \lra H^1(c, l|_c ) \lra 0\;\;\;
\eeqa
show that $h^0\Big(b, V|_b(-1)\Big)= h^0\Big(c, l(-F)|_c \Big)=0
\Longrightarrow h^0(b, V|_b)=  h^0(c, l|_c)=n$ 
as one has
\beqa
\label{index}
h^0\Big( c, l(-F)|_c \Big)=h^1\Big( c, l(-F)|_c \Big)
\eeqa
This follows either, arguing downstairs on $b$,  using
$h^i\Big( c, l(-F)|_c \Big)=h^i\Big(b, V|_b(-1)\Big)$ from
\beqa
h^0\Big(b, V|_b(-1)\Big)-h^1\Big(b, V|_b(-1)\Big)=
\int_b c_1\Big(V|_b(-1)\Big)+\frac{c_1(b)}{2}=\int_b c_1(V)|_b=0
\eeqa
or, with (\ref{degree equalities}), also directly upstairs on $c$ as one has
with (\ref{degree equalities})
\beqa
h^0\Big( c, l(-F)|_c \Big)-h^1\Big( c, l(-F)|_c \Big)=
\frac{1}{2}\, \mbox{deg}\, K_c + \frac{1}{2}\, \mbox{deg}\, K_c^{-1}=0
\eeqa

\subsection{\label{Lemma 2}Lemma 2}

As $\pi_{\E}$ is a projection one has $\pi_{\E}^{-1}(b)=\E$ and so
$H^0\Big(\E, {\cal O}_{\E}(\al s + \be F)\Big)=
H^0\Big(b, \pi_{\E *}{\cal O}_{\E}(\al s + \beta F)\Big)$.
From the Leray spectral sequence for the elliptically fibered surface $\E$ 
one has
\beqa
\label{Leray ss}
0 \; \lra \; H^1\Big( b, \, \pi_* \cO_{\E}(\al s)\Big) \; \lra 
H^1\Big( \E, \, \cO_{\E}(\al s)\Big) \; \lra 
H^0\Big( b, \, R^1\pi_* \cO_{\E}(\al s)\Big)&& \\
\lra \;\;\;\;\;\;\;\;\;\;\;\;\;\;\; 0 \;\;\;\;\;\;\;\;\;\;\; \lra 
H^2\Big( \E, \, \cO_{\E}(\al s)\Big) \; \lra 
H^1\Big( b, \, R^1\pi_* \cO_{\E}(\al s)\Big) \; &\lra &\; 0\nonumber 
\eeqa
\un{{\em The case $\al >0$}}

For $\al > 0$ one has (recall that $\cL|_b=K_B^{-1}|_b=\cO_b(\chi)$)
\beqa
\label{direct image}
\;\; \pi_* \cO_{\E}(\al s+ \beta F) &=& {\cal O}_b(\beta)\; \oplus \; 
\bigoplus_{i=2}^{\al}{\cal O}_b\Big(\beta - i\chi\Big)\\
\label{vanishing R^1}
R^1  \pi_* \cO_{\E}(\al s +\beta F) &=& 0
\eeqa 
and the Leray spectral sequence gives 
$H^2\Big(\E, {\cal O}_{\E}(\al s + \be F)\Big)=0$ and
\beqa
\label{reduction}
H^1\Big(\E, {\cal O}_{\E}(\al s + \be F)\Big)=
H^1\Big(b, \pi_{\E *}{\cal O}_{\E}(\al s + \beta F)\Big)
\eeqa
One finds that $h^0\Big(\E, {\cal O}_{\E}(\al s + \be F)\Big)$ 
vanishes, if $\al > 0$, just for negative $\be$ 
\beqa
\label{lemma 2}
\underline{\al >0:}\;\;\;\;\;\;\;\;
h^0\Big(\E, {\cal O}_{\E}(\al s + \be F)\Big)=
h^0\Big(b, \pi_{\E *}{\cal O}_{\E}(\al s + \beta F)\Big)
=0 \;\; \Llra \;\; \beta < 0 \;\;\;\;\;
\eeqa
More precisely note that (where $\{ m \}: = m+1=h^0(b, \cO_b(m))$ 
for $m\geq 0$ or else zero)
\beqa
h^0\Big(\E, {\cal O}_{\E}(\al s + \be F)\Big)
\!\!\!\!&=&\!\!\!\!\{ \beta \} + \sum_{i=2}^{\al} \{ \beta - i\chi \}
\ra \al(\be +1)-\Big(\frac{\al(\al+1)}{2}-1\Big)\chi
\;\;\;\;\;\;\;\;\; \\
\label{h^1 reduction}
h^1\Big(\E, {\cal O}_{\E}(\al s + \be F)\Big)
\!\!\!\!&=&\!\!\!\!\{ -\beta -2\} + \sum_{i=2}^{\al} \{ -\beta -2 + i\chi \}
\ra  -\al(\be +1)+\Big(\frac{\al(\al+1)}{2}-1\Big)\chi\;\;\;\;\;\;\;\;\;\;
\eeqa
Here the final evaluations for $\be$ or $-\be$ sufficiently big,
i.e. $\beta - \al \chi\geq 0$ or $-\beta-2\geq 0$; actually 
(\ref{h^1 reduction}) still holds for $\beta=-1$. 

\un{{\em The case $\al =0$}}

Next, in the case $\al =0$ one has $\pi_* \cO_{\E}=\cO_b$ and
$R^1\pi_* \cO_{\E}=K_B|_b=\cO_b(-\chi)$. One finds
\beqa
H^1(b, R^1\pi_*\cO_{\E}(0s+\beta F))
\; & \cong& \; H^1(b, \cO_b(\beta-\chi))\cong H^0(b, \cO_b(\chi-\beta-2))^*\\
H^0(b, R^1\pi_* \cO_{\E}(0 s+\beta F))\; & \cong & \; H^0(b, \cO_b(\beta-\chi))
\eeqa

\un{{\em The case $\al <0$}}

Finally in the case $\al <0$ one has 
\beqa
\label{neg al case}
\pi_*\cO_{\E}(\al s)\; &=&\; 0\\
\Big(R^1\pi_*\cO_{\E}(\al s)\Big)^*\; &=&
\; \pi_*\Big( \cO_{\E}(-\al s)\ox (K_{\E}\ox K_b^{-1})\Big)
\;=\;\pi_*\cO_{\E}(-\al s)\ox K_B^{-1}|_b\;\;\;
\eeqa 
and finds 
\beqa
H^1\Big(b, R^1\pi_*\cO_{\E}(\al s+\beta F)\Big)\; 
& =& \; H^0\Big(b, (\cO_b\oplus 
\bigoplus_{i=2}^{-\al}\cO_b(-i\chi))\ox \cO_b(\chi-\beta-2)\Big)^*\;\;\;\;\\
H^0\Big(b, R^1\pi_*\cO_{\E}(\al s+\beta F)\Big)\; & =& \; H^0\Big(b, 
(\cO_b\oplus \bigoplus_{i=2}^{-\al}\cO_b(i\chi))\ox \cO_b(\beta -\chi)\Big)
\eeqa

To summarize: we get that $H^1\big(b, R^1\pi_*\cO_{\E}(\al s+\beta F)\big)$ 
vanishes\footnote{such that then 
one will have, from (\ref{Leray ss}), still $H^2(\E, \cO_{\E}(\al s))=0$,
leaving in the long exact sequence (\ref{long exact sequence})
again only the three $H^0$- and the three $H^1$-terms} and that
$H^0(b, R^1\pi_* \cO_{\E}(\al s+\beta F))$ is 
vanishing\footnote{such that (\ref{reduction}) holds}
in the following cases
\beqa
\label{H^2 vanishing conditions}
 H^2(\E, \cO_{\E}(\al s+\beta F))=0 &\Lla &\left\{ \begin{array}{lll}
\beta \; \mbox{arbitrary} & \;\;\;\mbox{for} & \al >0\\
\beta>\chi-2              & \;\;\;\mbox{for} & \al \leq 0
\end{array} \right. \;\;\;\;\;\;\;\;\;\\
\label{H^1 reduction conditions}
H^1(\E, \cO_{\E}(\al s+\beta F))\cong H^1(b, \pi_*\cO_{\E}(\al s+\beta F)) 
&\Lla & \left\{ \begin{array}{lll}
\beta \; \mbox{arbitrary} & \mbox{for} & \al >0\\
\beta <(\al +1) \chi      & \mbox{for} & \al \leq 0
\end{array} \right. \;\;\;\;\;\;\;\;\;
\eeqa
\underline{{\em Example:}} For $\la=1/2$ one has also to consider 
the case $p=n(\la - \frac{1}{2})=\al-n=0$ 
by (\ref{the critical map}), so $\chi-2<q=\beta -r <\chi$ 
from the conditions
(\ref{H^2 vanishing conditions}), (\ref{H^1 reduction conditions})
for $\cO_{\E}(ps+qF)$;
but the resulting consequence $\beta -r = -r+\frac{1+n}{2}\chi-1\geq \chi-1$ 
(from the lower bound), i.e.~$r-\frac{n-1}{2}\chi-1<0$
contradicts (\ref{r chi relation}) or (\ref{canonical degree});
so then $H^2(\E, \cO_{\E}(\al s+\beta F))\neq 0$.

\subsection{Lemma 3 (cf. sect. \ref{Criteria Evaluation}, 
equ. (\ref{dim equality}))}

A sufficient criterion for the condition in (\ref{criterion}) to hold
(necessary for $W_b\neq 0$) is $\beta < 0$
\beqa
\label{remarkably}
h^0\Big(\E, \, l(-F-c)\Big) \; 
= h^0\Big(\E, \, l(-F)\Big)
\;\;\;\; \stackrel{\mbox{\normalsize{$\Lra$}}}{\Lla} \;\;\;\;
\beta < 0 
\eeqa
as then both dimensions vanish by (\ref{vanishing crit}).
The converse holds also: the dimensions can be equal only if both vanish.
So assume the lhs of (\ref{remarkably}) and $\beta \geq 0$ such that
\beqa
h^0\Big( \E, l(-F)\Big) &=& \beta + 1 + 
\sum_{i=2}^{\al} \{ \beta - i\chi \}\nonumber\\
h^0\Big( \E, l(-F-c)\Big) &=& \{ \beta -r \} + 
\sum_{i=2}^{\al -n} \{ \beta -r - i\chi \}\; ,
\eeqa
from $l(-F)=\cO_{\E}(\al s+ \beta F)$ with $\al:=n(\la + \frac{1}{2})$.
Then one gets 
\beqa
\label{difference}
\sum_{i=2}^{\al - n} 
\Bigg( \{ \beta - i\chi \} - \{ \beta -r - i\chi \}\Bigg)
+ \sum_{i=\al - n + 1}^{\al} \{ \beta - i\chi \}=\{\beta -r\}-(\beta+1)
\eeqa
(note that $\la > 1/2$ by our standing technical assumption 
(\ref{technical assumption}), so $\al > n$).
This leads to a contradiction:
the lhs of (\ref{difference}) is always $\geq 0$
as $r\geq 0$ by (\ref{r chi relation}); 
so $\beta \geq r$ (otherwise the rhs would be $<0$) 
and the rhs is $-r \leq 0$, such that
$r=0$. But then (\ref{r chi relation})
gives $\chi =0$ (i.e. $B={\bf F_2}$) 
and therefore from (\ref{beta eval}) the contradiction $\beta = -1$.
Therefore actually $\beta <0$ and both $H^0$-terms 
on the lhs of (\ref{remarkably}) are zero by (\ref{lemma 2}).

So, indeed, the necessary criterion (for $W_b\neq 0$) 
that the equality on the lhs of (\ref{remarkably}) holds 
is equivalent to $\beta$ being negative, which also means that 
the two $H^0$-dimensions were actually zero. 


\section{\label{augmentation}An alternative method}

\resetcounter

For another method to get a 4-term exact sequence 
with $H^0(c, l(-F)|_c)$ as kernel of a map $\rho$
between spaces of equal dimension (cf.~after (\ref{criterion 2}))
choose, besides the varying curve $c\subset \E$,
a second fixed effective divisor $c'\subset \E$ which may be
reducible (some fibers, say) or even non-reduced (some fibers coalescing); 
assume $\deg D'> \deg K_c^{1/2}$ where $D':=c'|_c$. 

\noindent
From $\deg \cL = \deg K_c^{1/2}$ 
with $\cL=l(-F)=\cO_{\E}(\alpha s + \beta F)$
one has ($\deg (K_c - \cL|_c(D'))<0$)
\beqa
\label{the other sequence}
0 \lra H^0(c, \cL|_c)\lra H^0(c, \cL|_c(D'))\buildrel \rho \over \lra
H^0(D', \cL|_c(D')|_{D'})\lra H^1(c, \cL|_c)\lra 0\;\;\;\;\;\;
\eeqa
(middle terms have
equal dimension: $h^0(c, \cL|_c(D'))
=\deg D' = c' \cdot c = h^0(D', \cL|_c(D')|_{D'})$).

Let $c'$ be a set of
$m\geq r-\frac{n-1}{2}\chi $ fibers.
For $m>-\beta + r - 2 +(\alpha-3)\chi$ one has
\beqa
\label{2. term from global sections}
0 \lra H^0(\E, \cL(mF-c))\lra H^0(\E, \cL(mF))\lra H^0(c, \cL(mF)|_c)\lra 0
\eeqa
Similarly here also the third term in 
(\ref{the other sequence}) can be represented as
\beqa
0 
\lra \oplus_{i=1}^m H^0(F, \cO_{F_i}(\alpha -n))
\lra \oplus_{i=1}^m H^0(F, \cO_{F_i}(\alpha )) 
\lra \oplus_{i=1}^m H^0(F_i|_c, \cO_{F_i}(\alpha )|_c)
\lra 0\nonumber
\eeqa

All of these sequences are interwoven with our original sequence after
(\ref{criterion 2})
\beqa
\begin{array}{ccccccccc}
&&&&&& 0 && \\
&&&&&& \da &&\\
 & & 0 & & 0\;\; & \ra & H^0(c, \cL|_c) & \buildrel \delta \over \ra & \\
 & & \da & & \da \;\; & & \da & & \\
0 & \ra & H^0(\E, \cL(mF-c))& \ra & H^0(\E, \cL(mF)) & \ra & 
H^0(c, \cL(mF)|_c)& \ra & 0 \\
 & & \da & & \;\;\;\; \da r_m& & \;\;\;\;\;\; \da \rho_m & & \\
0 &\ra & \displaystyle \bigoplus_{i=1}^{m} 
H^0(F_i, \cO_{F_i}(\alpha -n))
&\ra & \displaystyle \bigoplus_{i=1}^m H^0(F_i, \cO_{F_i}(\alpha )) 
&\ra & \displaystyle \bigoplus_{i=1}^m H^0(F_i|_c, \cO_{F_i}(\alpha )|_c)
& \ra & 0\\
 & & \da & & \da & & \da & & \\
 & \buildrel \delta \over \ra & H^1(\E, \cL(-c)) & \ra & 
H^1(\E, \cL) & \ra & H^1(c, \cL|_c) & \ra & 0\\
 & & \da & & \da & & \da & & \\
 & & 0 & & 0 & & 0 & & 
\end{array}\nonumber
\eeqa
Here one gets
again\footnote{The map $\rho_m$ over $c$ comes from the moduli-independent
restriction map $r_m$. 
The moduli-dependence can be understood from the necessity to select 
representatives inside $H^0(\E, \cL(mF))$
of a set of basis elements in $H^0(c, \cL(mF)|_c)$; for this one has
to take into account the equivalences arising from embedding
$H^0(\E, \cL(mF-c))$ into $H^0(\E, \cL(mF))$ via multiplication with
the defining polynomial $w_c$ of $c$.} 
that $Pfaff(t)=0 \Llra \det \rho_{c'}(t) =0$.
As an example take $m\geq 3$ in Example 1: among 
the $\cO(9s+yF)$ with $y\geq 2$ those of $y\geq 9$ have 
all global sections arising from (restriction from) $\E$, 
cf.~(\ref{2. term from global sections}).


\section{\label{negative lambda}The cases $\la < - 1/2$}
\resetcounter

To treat also $\la < - \frac{1}{2}$, 
outside the range (\ref{technical assumption}),
let $\widetilde{\la} := - \la > \frac{1}{2}$ and
note that with
\beqa
\label{normal expression}
l(-F) & \, = & \cO_{\E}\Big( n (\la+\frac{1}{2})s+\beta F\Big)\\
\label{tilde expression}
\widetilde{l}(-F) & := & 
\cO_{\E}\Big( n (\widetilde{\la}+\frac{1}{2})s+\widetilde{\beta}F\Big)
\eeqa
(where $\beta=\beta(\la), 
\widetilde{\beta}=\beta(\widetilde{\lambda})$)
one gets with (\ref{index}) and (\ref{degree equalities}) that
\beqa
\label{h^0 equality}
h^0\Big(c, l(-F)|_c\Big)&=&h^1\Big(c, l(-F)|_c\Big)
=h^0\Big(c, \Big( K_c^{1/2}\ox \cF_{\lambda}\Big)^* \ox K_c \Big)
=h^0\Big(c,K_c^{1/2}\ox \cF_{-\lambda}\Big) \nonumber\\ 
&=&h^0\Big(c, \widetilde{l}(-F)|_c\Big)
\eeqa
So, by (\ref{prec criterion}) and (\ref{h^0 equality}), the question of 
contribution (or not) is independent of the sign of $\lambda$.

\noindent
So, following (\ref{prec criterion}), one can work equally well with the new 
$h^0$-expression for $\widetilde{l}$, 
i.e., when one wants to consider (\ref{normal expression}) 
with $\la <-\frac{1}{2}$ one applies the same arguments as before to 
(\ref{tilde expression}) assuming the necessary condition (\ref{criterion}), 
i.e. $\beta<0$, what leads again to a map (\ref{the critical map}), now
for $\widetilde{l}$ with $\widetilde{\la}>\frac{1}{2}$.
The map $H^1(\E, l(-F-c))\lra H^1(\E, l(-F))$
becomes with Serre duality 
$H^1(\E, l(-F-c)^*\ox K_{\E})^*\lra 
H^1(\E, l(-F)^*\ox K_{\E})^*$ which is dual to 
\beqa
H^1(\E, l(-F)^*\ox K_{\E})& \lra & 
H^1(\E, l(-F-c)^*\ox K_{\E})
\eeqa
This amounts to the original map (\ref{the critical map}), 
now for $\widetilde{l}$ which has $\widetilde{\la}>1/2$,
\beqa
\label{new map}
H^1(\E, \widetilde{l}(-F-c))&\lra &H^1(\E, \widetilde{l}(-F))
\eeqa 
as
$l(-F)^*\ox \cO_{\E}(c)\ox K_{\E}=\widetilde{l}(-F)$
because $l(-F)=(K_{\E}\ox \cO_{\E}(c))^{1/2}\ox \un{\G}_{\la}|_{\E}$ 
by (\ref{l def}) and
$\widetilde{l}(-F)$ arises from $l(-F)$ by going from $\la$ to 
$\widetilde{\la}=-\la$ (the argument we gave in 
(\ref{h^0 equality}) on $c$).
So, as necessary condition for contribution to the superpotential 
one has to consider $\beta(\widetilde{\lambda})<0$ also 
in the case $\la < - \frac{1}{2}$.

{\em Example 1: the case with $\la =-5/2$}

Consider $\la = -5/2$ where $\beta=3r+\frac{1-5n}{2}\chi-1$ 
and $l(-F)=\cO_{\E}(-2ns+\beta F)$ and
apply the usual reasoning to $\widetilde{l}(-F)$ 
in (\ref{tilde expression})
where $\widetilde{\la}=+5/2$ and $\widetilde{\beta}=-\beta +r-2+\chi $.
Assuming the necessary condition (\ref{criterion}), i.e.~$\widetilde{\beta}<0$,
consider the map (\ref{new map})
\beqa
I_r\, : \;\; H^1\Big( \E, \cO_{\E}\big( 2ns+(\widetilde{\beta}-r)F \big) \Big)
\lra
H^1\Big( \E, \cO_{\E}\big( 3ns + \widetilde{\beta}F\big) \Big)
\eeqa
Concretely, consider an $SU(3)$ bundle over $B={\bf F_1}$ 
and take $r=4$ such that
$\beta =4, \widetilde{\beta}=-1$ and $l(-F)=\cO_{\E}(-6s+4F),
\widetilde{l}(-F)=\cO_{\E}\big( 9s-F\big)$. 
One gets a map\footnote{which is 
dual to the map arising via Serre duality from the map
$H^1(\E, \cO_{\E}(-9s+0F))\lra H^1(\E, \cO_{\E}(-6s+4F))$}
(\ref{ex 1tilde map})
between $44$-dimensional spaces (by (\ref{h^1 reduction})).


\section{\label{Reduction cases}Reduction cases with equality condition}

\resetcounter

\subsection{\label{cases al > n}The case $\bar{\al}>n$}

Note that $\bar{\al}\leq \frac{1}{p}\al=\frac{n}{p}(\la + \frac{1}{2})$
gives here $\la > p - \frac{1}{2}$.
Then $\bar{\beta} \leq \frac{1}{p}\beta$ gives with (\ref{betabar fixing})
\beqa
\bar{\al}&\geq & \frac{1}{p}\al 
+ \frac{p-1}{p} \frac{n}{r-n\chi}\Big( r - \frac{n-1}{2}\chi -1\Big)
\eeqa
such that
\beqa
r - \frac{n-1}{2}\chi -1 & \leq & 0
\eeqa
With (\ref{r chi relation sharp}) one finds that $\chi =0, r=1$ such that
one gets with (\ref{betabar fixing})
(for $\la+\frac{1}{2}\in p{\bf Z}^{>1}$)
\beqa
l(-F) & = & \cO_{\E}\Big(n(\la + \frac{1}{2}) s - (\la + \frac{1}{2})F\Big)\\
\bar{l}(-F)&=&
\cO_{\E}\Big(\frac{n(\la + \frac{1}{2})}{p}s -\frac{\la + \frac{1}{2}}{p}F\Big)
\eeqa

\subsection{\label{list subsection}The case $\bar{\al}=n$}

Note that $h^0(c, \bar{l}(-F)|_c)>0$ needs $\deg \bar{l}(-F)|_c\geq 0$; 
but in the case (\ref{bar l(-F)}) one has
$\deg \bar{l}(-F)|_c=n(\frac{n-2}{n-1}r-(\frac{n}{2}-1)\chi-1)$ 
so we do not have to consider $SU(2)$ bundles here.

Here $\bar{\al}\leq \frac{1}{p}\al=\frac{n}{p}(\la + \frac{1}{2})$
gives $\la \geq p - \frac{1}{2}$ and the $\bar{\beta}$-bound gives
with (\ref{last expression})
\beqa
\label{first bound}
\Bigg(\la - ( \frac{1}{2}+\frac{p}{n-1})\Bigg) (r -n\chi)
- \frac{1}{2}\Bigg(\frac{2p}{n-1}- n(p-1)+1\Bigg)\chi& \leq & p-1
\eeqa
Thereby one derives (after checking the sign of the prefactor) 
with (\ref{sharper bound}) that one has
$
\Big[ \Big(\la - ( \frac{1}{2}+\frac{p}{n-1})\Big) 
\frac{\frac{n+1}{2}}{\la - \frac{1}{2}}
- \frac{1}{2}\Big(\frac{2p}{n-1}- n(p-1)+1\Big)\Big] \chi  \leq  p-1
$
or
\beqa
\label{third bound}
\Big(n-2-\frac{1}{\la - \frac{1}{2}}\Big) \chi & \leq &
2 \, \frac{p-1}{p}\, \frac{n-1}{n+1}
\eeqa
For $p=1$ this can be fulfilled by $\chi=0$ 
where then (\ref{first bound}) gives $\la \leq \frac{1}{2}\, \frac{n+1}{n-1}$,
allowing just $\la = \frac{3}{2}$ for $n=2$, or, for even $r$, also $\la =1$;
but we need $n>2$ so let us now assume that $\chi \geq 1$;
then, for $n>2$, one can have $SU(3)$ bundles with  
$\la = \frac{3}{2}$ where one gets $2\chi \leq r-3\chi\leq 2\chi$
from (\ref{first bound}) and (\ref{sharper bound made explicit})
such that $r=5\chi$ (case $\sharp$ 2 there); 
or $SU(4)$ bundles with $\la =1$ 
where one gets $5\chi \leq r-4\chi \leq 5\chi$ such that $r=9\chi$
(case $\sharp$ 1 there).

So let now $p\geq 2$. One derives then
\beqa
\label{fourth bound}
(n-3)\chi \; \leq \; (n-2-\frac{1}{p-1})\chi & \leq &
2 \, \frac{n-1}{n+1}
\eeqa

{\em $SU(3)$ bundles}

Here, where $\la \in \frac{1}{2}+{\bf Z}$, 
one gets from (\ref{fourth bound}) that $\frac{p-2}{p-1}\chi \leq 1$; 
we will discuss the case $p=2$ separately. 
So let $p>2$ such that\footnote{the case $p=3, \chi=2$ with $\la = 3/2$ 
from (\ref{fifth bound}) contradicts $\la \geq p - \frac{1}{2}$} 
$\chi=0$ or $1$;
more precisely one has from (\ref{third bound}) 
\beqa
\label{fifth bound}
\frac{\la - \frac{3}{2}}{\la - \frac{1}{2}}\chi \leq \frac{p-1}{p}
\eeqa
For $\chi=0$, where $r$ must be even by (\ref{last expression}), 
one gets from (\ref{first bound}) that 
$(\frac{p}{2}-1)r \leq (\la - \frac{p+1}{2})r\leq p-1$;
so $r=2$ or $r=4, p=3$; here $r=2$ (case $\sharp$ 1) leads to
$\la \leq p$, so $\la = p - \frac{1}{2}$ and
\beqa
\;\;\;\mbox{case}\; \sharp 1\;\;\;\;\;\;\;\;\;\;\;\;\;\;\;\;\;\;\;
l(-F)&=&\cO_{\E}\Big( 3(\la+\frac{1}{2})s-2\la F\Big)\\
\bar{l}(-F)&=&\cO_{\E}\Big( 3s-2F\Big)
\eeqa
whereas the $r=4, p=3$ case (case $\sharp$ 4)
leads to $\la \leq \frac{5}{2}$ such that  
$\la = p - \frac{1}{2}=\frac{5}{2}$ and
\beqa
\mbox{case}\; \sharp 4\;\;\;\;\;\;\;\;\;\;\;\;\;\;
\;\;\;\;\;\;\;\;\;\;\;\;\;\;\;\;
l(-F)&=&\cO_{\E}\Big( 9s -9F\Big)\\
\bar{l}(-F)&=&\cO_{\E}\Big( 3s-3F\Big)
\eeqa
For $\chi=1$ one gets $p-\frac{1}{2}\leq \la \leq p+\frac{1}{2}$ 
from (\ref{fifth bound}), 
giving $r\geq 3+\frac{2}{p-1}$ for $\la = p-\frac{1}{2}$ 
and $r\geq 3+ \frac{2}{p}$ for $\la = p + \frac{1}{2}$ 
by (\ref{sharper bound}); here
$\frac{1}{2}(r-3)=(\la - \frac{p+1}{2})(r-3)\leq 1$
from (\ref{first bound}), leaving only the case 
$p=3, \la = p - \frac{1}{2}=\frac{5}{2}$ and $r=5$ (case $\sharp$ 5)
as $r$ must be odd by (\ref{last expression}) 
\beqa
\mbox{case}\; \sharp 5\;
\;\;\;\;\;\;\;\;\;\;\;\;\;\;\;\;\;\;\;\;\;\;\;\;\;\;\;\;
l(-F)&=&\cO_{\E}\Big( 9s -3F\Big)\\
\bar{l}(-F)&=&\cO_{\E}\Big( 3s-F\Big)
\eeqa
Finally, for $p=2$ now (\ref{first bound})
says that $(\la - \frac{3}{2})(r-3\chi)\leq 1$, 
so $\la = 3/2$ (case $\sharp$ 3) is a solution 
(the only one\footnote{For, if $\la> 3/2$, 
note that 
$\frac{2}{\la - \frac{1}{2}}\chi \leq r-3\chi \leq \frac{1}{\la-\frac{3}{2}}$ 
by (\ref{sharper bound made explicit}) such that 
$2\frac{\la-\frac{3}{2}}{\la - \frac{1}{2}}\chi\leq 1$ which implies
$\la = 5/2$ (as for $\la \geq 7/2$ the coefficient of $\chi$ 
becomes $>1$ which enforces $\chi =0$ where (\ref{first bound}) 
gives the contradiction $(\la - \frac{3}{2})r\leq 1$ as $r>0$).
$\la = 5/2$ is excluded as one gets $r-3\chi \leq 1$ while 
$\chi\leq r-3\chi$ by (\ref{sharper bound made explicit}): 
$\chi=0$ giving $r=1$ and $\chi=1$ giving $r=4$ are both
excluded as (\ref{last expression}) must be integral.}), so one has
for $r\geq 5\chi$ and $r\equiv \chi\, (\mbox{mod}\, 2)$
(with $\deg \det \bar{\iota}_1=\frac{3r-5\chi}{2}$)
\beqa
\;\;\;\;\;\mbox{case}\; \sharp 3\;\;\;\;\;\;\;\;\;\;\;\;\;\;\;\;\;\;
l(-F)&=&\cO_{\E}(6s-(r-5\chi+1)F)\\
\bar{l}(-F)&=&\cO_{\E}(3s-(\frac{r-5\chi}{2}+1)F)
\eeqa 
Let us also recall the earlier mentioned $p=1$ case
\beqa
\mbox{case}\; \sharp 2\;\;\;\;\;\;\;\;\;\;\;\;\;\;\;\;\;\;\;\;\;\;\;\;\;\;\;
l(-F)&=&\cO_{\E}\Big(6s-F\Big)\\
\bar{l}(-F)&=&\cO_{\E}\Big(3s-F\Big)
\eeqa

{\em $SU(4)$ bundles}

Here one gets from (\ref{fourth bound}) that $\chi=0$ or $1$.
Actually\footnote{For $\chi=1$ one gets from (\ref{first bound}) that
$(\frac{2}{3}p-1)(r-4) \leq
(\la - \frac{1}{2}-\frac{p}{3})(r-4)\; \leq \; -\frac{2}{3}\, p+\frac{3}{2}$
or $r\leq 3+ \frac{3/2}{2p-3}$, i.e. $r=3$ 
as $3|r$ by (\ref{last expression}) (also here $\la \in {\bf Z}$
and $p=2$ by (\ref{fourth bound})); so (\ref{r chi relation}) is violated.} 
$\chi=0$ and one gets, with $\la - \frac{1}{2}\geq p-1$ 
from the remark before (\ref{first bound}), that
 $(\frac{2}{3}p-1)r\leq p-1$ or
$r\leq 1+\frac{p}{2p-3}$, such that $p=2, r=3$ (case $\sharp$ 2)
as $3|r$ by (\ref{last expression}); 
one gets from (\ref{first bound}) that $\la = \frac{3}{2}$
and one has (with $\deg \det\bar{\iota}_1=4$ and $\deg \det \iota_1=24$)
\beqa
\mbox{case}\; \sharp 2\;\;\;\;\;\;\;\;\;\;\;\;\;\;\;\;\;\;\;\;\;\;\;\;\;\;\;
l(-F)&=&\cO_{\E}\Big(8s-4F\Big)\\
\bar{l}(-F)&=&\cO_{\E}\Big(4s-2F\Big)
\eeqa
Let us also recall the earlier mentioned $p=1$ case
\beqa
\mbox{case}\; \sharp 1\;\;\;\;\;\;\;\;\;\;\;\;\;\;\;\;\;\;\;\;\;\;\;\;\;\;\;
l(-F)&=&\cO_{\E}\Big(6s-F\Big)\\
\bar{l}(-F)&=&\cO_{\E}\Big(4s-F\Big)
\eeqa
{\em $SU(n)$ bundles with $n\geq 5$}

Here one gets $\chi=0$ such that
$(p-1-\frac{p}{n-1})r \, \leq \, 
(\la - \frac{1}{2}-\frac{p}{n-1})r  \leq  p-1$
or $(1-\frac{2}{n-1})r\leq (1-\frac{1}{n-1}\frac{p}{p-1})r\leq 1$; 
this has no solutions as $(n-1)|r$ by (\ref{last expression}).

So in total one gets the following lists

\underline{$SU(3)$ bundles}
\begin{center}
\begin{tabular}{|c||c|c|c||c|}
\hline
$\sharp$ &$\chi$ & $r$ & $\la $ & $p$ \\
\hline
\hline
1 &
$0$ & $2$& $p-\frac{1}{2}$ & $>2$\\
\hline
2 & $\geq 1$ & $5\chi$ & $\frac{3}{2}$ & $1$\\
\hline
3 & $\chi$ & $\geq 5\chi, \equiv \chi(2)$ & $\frac{3}{2}$ & $2$\\
\hline
4 & $0$ & $4$ & $\frac{5}{2}$ & $3$\\
\hline
5 & $1$ & $5$ & $\frac{5}{2}$ & $3$\\
\hline
\end{tabular}
\end{center}
Here the case 1 for $p=2$ is the case 3 for $\chi=0, r=2$.
Note that in case 3 one has also the assumption $r\geq 5\chi$.
\vspace{.5cm}

\underline{$SU(4)$ bundles}

\begin{center}
\begin{tabular}{|c||c|c|c||c|}
\hline
$\sharp$ & $\chi$ & $r$ & $\la $ & $p$ \\
\hline
\hline
1 & $\geq 1$ & $9\chi$ & $1$ & $1$\\
\hline
2 & $0$ & $3$ & $\frac{3}{2}$ & $2$\\
\hline
\end{tabular}
\end{center}
\vspace{.5cm}

\subsection{\label{vanishing n=3 cases}$SU(3)$ and $SU(4)$
bundles with $Pfaff\equiv 0$}

\subsubsection{$SU(3)$ bundles}

According to sect.~\ref{The vanishing condition} one uses here besides the
strong reduction condition $p\bar{\beta}\leq \beta$ the condition that 
(\ref{positivity condition}) is positive
\beqa
\label{aux SU(3)}
-\frac{1}{p}\beta \;\; \leq \;\; -\bar{\beta} \;\; < \;\; \frac{r-5\chi}{2}+1
\eeqa
Furthermore the other part $p\bar{\al}\leq \al$ of the strong reduction 
condition gives (with $\bar{\al}\geq 3$) that 
$p\leq \la + \frac{1}{2}$ such that one gets
\beqa
2(\la - \frac{1}{2})r-2(3\la + \frac{1}{2})\chi+2\;\; <
(\la + \frac{1}{2})r-5(\la + \frac{1}{2})\chi+2(\la + \frac{1}{2})
\eeqa
or equivalently
\beqa
\label{r condition}
(\la - \frac{3}{2})(r-\chi)& < & 2(\la - \frac{1}{2})
\eeqa

Let us first assume that $\la = 3/2$. Then one gets 
$\frac{1}{p}(r-5\chi+1)\leq -\bar{\beta}<\frac{r-5\chi}{2}+1$
from (\ref{aux SU(3)});
as $p\leq 2$ and $r-5\chi \geq 0$ by (\ref{sharper bound})
one gets $p=2$ with $\frac{r-5\chi}{2}+\frac{1}{2}\leq -\bar{\beta}
<\frac{r-5\chi}{2}+1$; the two ensuing cases $r\equiv\chi(2)$,
where no (integral !) solution for $\bar{\beta}$ exists, 
and $r\not\equiv\chi(2)$ where $\bar{\beta}=-(\frac{r-5\chi+1}{2})$
are discussed further in sect.~\ref{The vanishing condition}.
We show now that no further cases exist.

Let us therefore assume $\la \not = 3/2$; as $n=3$ one has 
$\la = \frac{1}{2}+m$ with $m\in {\bf Z^{\geq 1}}$, so let now $m\not =1$.
From (\ref{sharper bound}) one gets 
\beqa
\label{aux2 SU(3)}
r-\chi \geq 2 \, \frac{m+1}{m}\chi
\eeqa
such that one gets from (\ref{r condition})
that $\chi \leq \frac{m^2}{m^2-1}$ and therefore $\chi =0$ or $1$.

For $\chi=0$ one has $r<2\, \frac{m}{m-1}$ from (\ref{r condition})
such that $m=2$ with $r=1,2,3$
or $m\geq 3$ with $r=1,2$; further $p\leq m+1$.
The $Pfaff\equiv 0 $ case $\chi=0, r=1$ was already covered 
in sect.~\ref{red cases for bar al > n}.
From $\frac{1}{p}(mr+1)\leq -\bar{\beta}<\frac{r}{2}+1$
one gets for $r=2$ that $-\bar{\beta}=1$ (recall $\bar{\beta}<0$)
such that one gets
the contradiction $2m+1\leq p$; for $r=3, m=2$ one gets the contradiction
$\frac{1}{p}7\leq -\bar{\beta}<\frac{5}{2}$ 
(allowing no integral $\bar{\beta}$ solution for $p=3$)
where $p\leq 3$.

For $\chi=1$ one gets $r-1<2\frac{m}{m-1}$ from (\ref{r condition})
contradicting $r-1\geq 2\frac{m+1}{m-1}$ from (\ref{aux2 SU(3)}).

\subsubsection{$SU(4)$ bundles}

Following the same procedure one finds that here no further
solutions exist besides the already covered case $\chi=0, r=1$.
For $\la\in \frac{1}{2}+{\bf Z}$ one finds $\chi=0$ and 
a contradiction for $r\not =1$; for $\la \in {\bf Z}$ one has the same
for $\la \not =1$ and gets for $\la=1$ a contradiction to 
(\ref{sharper bound}).


\section{\label{concrete matrix representations}
Explicit matrix representations for $n=3, \la = 3/2$}

\resetcounter

Making the map $\iota_1$ 
in sect.~\ref{Exa 2} explicit via canonical bases one gets
($\beta = -r+5\chi -1$)
\beqa
\bigoplus_{w\in StB(\Si)}w\, H^0\Big(b, 
\cO_b(r -\be + [w] \, \chi-2)\Big)^*  \stackrel{\iota_1}{\lra}
\bigoplus_{w_2\in StB(S^2 \Si)}
w_2\, H^0\Big(b, \cO_b(-\be + [w_2] \, \chi -2)\Big)^* \;\;\;
\eeqa
(using (\ref{the critical map}))
or, with the notation $V:=H^0(b,\cO_b(1))$ and $S:=Sym$
(cf.~sect.~\ref{sym}),
\beqa
\label{map}
M_r: \;\;\;\;\;\;\;\;\;
\widetilde{\Sigma}_w \odot S^{2r- 5\chi-1}\, V^* 
\lra
(\widetilde{S^2\, \Sigma})_{w_2=w\, w'} \odot S^{r- 5\chi-1}\, V^* \;
\eeqa
Here we use the symmetrized tensor product 
$\odot$ (cf.~sect.~\ref{sym}) and elements 
\beqa
w\in \{z,x,y\}=StB(\Si)\;\;\; , \;\;\; \;\;\;\;\;
w_2 \in \{z^2, zx, zy, x^2, xy, y^2\}= StB(S^2 \Si)
\eeqa
of standard bases ($StB$) of
$\Si=z{\bf C}\oplus x{\bf C}\oplus +y{\bf C}$ and $S^2 \Si=Sym^2 \Si$\, :
\beqa
\label{standard base elements}
w\in H^0\Big(\E, \cO_{\E}(3s + [w] \, \chi F)\Big)\;\;\;, \;\;\;\;\;\;\;\;
w_2 \in H^0\Big(\E, \cO_{\E}(6s + [w_2] \, \chi F)\Big)\;\;\;\;\;\;
\eeqa
Furthermore\footnote{By $\beta<0\Leftrightarrow r\geq 5\chi$
only the space belonging to $z^2$ can be zero-dimensional (for $r=5\chi$;
the other case $r=0$ over ${\bf F_2}$ leads to a map between two 
zero-dimensional spaces where (\ref{criterion 2}) is trivially fulfilled).}
we made use of the notion of degree where $[w]:= 2q+3r$ 
for $w=z^p\, x^q \, y^r$
\beqa
\widetilde{\Sigma}_w  &:= &
\bigoplus_{w\in StB(\Si)}w\, S^{[w] \chi}V^*
=z\, {\bf C}\oplus x\, S^{2\chi}\, V^* \oplus y\, S^{3\chi}\, V^*\\
(\widetilde{S^2\, \Sigma})_{w_2} & := &
\bigoplus_{w_2\in StB(S^2 \Si)}w_2\, S^{[w_2] \chi}V^*\;\;
\eeqa
One finds in  (\ref{map}) that
$\mbox{dim } lhs = 3(2r-5\chi)+5\chi = 6r-10\chi = 6(r-5\chi)+20\chi 
= \mbox{dim } rhs$.
In this representation $M_r$ 
(cf.~sect.~\ref{Exa 2}) 
is multiplication with an element 
\beqa
\label{tildeiota}
\tilde{\iota} \; = \; Cz+Bx+Ay 
& \in & H^0\Big( \E, \cO_{\E}(3s+rF)\Big)
=\bigoplus_{w'\in StB(\Si)}\, w' \, H^0\Big(b, \cO_b(r- [w'] \, \chi)\Big)
\nonumber\\
&=&\bigoplus_{w'\in StB(\Si)}\, w'\, S^{r- [w'] \, \chi}\, V
= \widetilde{\Sigma}_w  \, \odot \, S^r\, V 
\eeqa
with the accompanying coefficients
\beqa
C\in S^r\, V, \;\; \;\; \;\; 
B\in S^{r-2\chi}\, V, \;\; \;\; \;\; A\in S^{r-3\chi}\, V
\eeqa
For the mentioned bases one gets a block matrix 
(with first column {\em not} $(C, 0, B, A, 0, 0)^t$)
\beqa
\label{matrix}
M_r\, : \;\;\;\;\;\;\;\;\;\;\;\;
\begin{array}{cc}
 & \begin{array}{ccc} z & x & y \end{array}\\
\begin{array}{c} z^2 \\ zx \\zy \\x^2 \\xy \\y^2 \end{array} &
\left( \begin{array}{ccc}
C & 0 & 0 \\
B & C & 0 \\
A & 0 & C \\
0 & B & 0 \\
0 & A & B \\
0 & 0 & A
\end{array} \right)
\end{array}
\eeqa
This is a square matrix of size $\big(6r-10\chi\big)\x \big(6r-10\chi\big)$.
If, say, (cf.~sect.\ref{sym} for the notation)
\beqa
C=\sum_{i=0}^r c_i u^{r-i}v^i\in S^r\, V = Hom ( S^r\, V^*, {\bf C} )
\eeqa
then we consider in (\ref{matrix}) actually induced maps
(or $(k+1)\x (k+r+1)$ - matrices)
\beqa
C\; \underline{\odot}\;  S^k\, V^*: S^{k+r}\, V^*\ra S^k\, V^*
\eeqa
So, in (\ref{matrix}),
where we indicated the expansion coefficients in $x,y,z$ accompanying
the respective spaces in (\ref{map}), each entry has to be suitable extended:
a $D$ in the line of $w_2$ stands actually for
$D \; \underline{\odot}\; 
S^{r+\big([w_2]-5\big)\chi-1}\, V^*$ 
(where $D\in \{A, B, C\}$; cf.~(\ref{C matrix example})).
This is for $D\in S^{r-[w']\chi}\, V$ in the $w$ - column a matrix of size
$(r+\big([w_2]-5\big)\chi)\x (2r+\big( [w]-5\big) \chi )$.

It will be shown that
half of the determinants of all these matrices vanish, 
cf.~sect.~\ref{The vanishing condition},
the other ones (for $r\equiv \chi\, (2) $) have as factor
the determinant of the following square matrix $m_r$ of size 
$\frac{1}{4}\big(6r-10\chi\big)\x \frac{1}{4}\big(6r-10\chi\big)$
(for $r=5\chi$ the first line is absent)
\beqa
\label{m_r matrix}
m_r\, : \;\;\;\;\;\;\;\;\;\;\;\;\;\;
\left( \begin{array}{c}
\, C \odot S^{\frac{r-5\chi}{2}-1}\;\;\;\;\;\, V^*    \\
B \odot S^{\frac{r-5\chi}{2}-1+2\chi}\, V^*     \\
A \odot S^{\frac{r-5\chi}{2}-1+3\chi}\, V^*     
\end{array} \right)
\;\;\; \;\;\; \;\;\; (\; \mbox{where} \;\;\; 
\begin{array}{c}
\, C\in S^r\, \;\;\;\;\; V \\ 
B\in S^{r-2\chi}\, V \\
A\in S^{r-3\chi}\, V
\end{array} )
\eeqa
$m_r$ is mediated by multiplication with the element (\ref{tildeiota})).
For $\chi=1, r=5$ one has
\beqa
C=c_5u^5+c_4u^4v+c_3u^3v^2+c_2u^2v^3+c_1uv^4+c_0 v^5
&\in &S^r\, V \;\;\;\;\, = S^5\, V = Hom(S^5\, V^*, {\bf C})\;\;\;
\;\;\;\;\;\;\nonumber\\
B=b_3u^3+b_2u^2v+b_1uv^2 + b_0v^3\;\;\;\;\;\;\;\;\;\;\;
&\in &S^{r-2\chi}\, V 
= S^3\, V = Hom(S^3\, V^*, {\bf C})\;\;\;\;\;\;\;\;\;\;\;\nonumber\\ 
A=a_2u^2+a_1uv+a_0v^2 \;\;\;\;\;\;\;\;\;\;\;\;\;\;\;\;
& \in &S^{r-3\chi}\, V=  S^2\, V= Hom(S^2\, V^*, {\bf C})
\;\;\;\;\;\;\;\;\;\;\nonumber
\eeqa
and gets for the map (\ref{m_r matrix})
\beqa
\label{D_5}
m_5\, =\D_5: \;\;\;\;\;\;\;\;\;\;\;\;\;\;
\left( {\begin{array}{ccccc}
b_3 & b_2  & b_1 & b_0 & 0  \\
0   & b_3  & b_2 & b_1 & b_0\\
a_2 & a_1  & a_0 & 0   & 0  \\
0   & a_2  & a_1 & a_0 & 0  \\
0   & 0    & a_2 & a_1 & a_0
\end{array}} \right) 
\eeqa

For the case $\chi=0, r=2$ one gets for $C=C_2$ (and the same for 
$B=B_2$ and $A=A_2$)
\beqa
C=c_2u^2+c_1uv+c_0v^2&\in & S^2\, V = Hom(S^2\, V^*, {\bf C})
\eeqa
and gets for the map (\ref{m_r matrix})
\beqa
\label{D_3 chi=0}
m_3\, =\D_3^{\chi=0}: \;\;\;\;\;\;\;\;\;\;\;\;\;\;
\left( {\begin{array}{ccc}
c_2 & c_1  & c_0 \\
b_2  & b_1 & b_0 \\
a_2 & a_1  & a_0 
\end{array}} \right) 
\eeqa


\section{\label{chi = 0 app} Bundles of $\la=3/2$ over a $\chi=0$ curve}

\resetcounter

Over ${\bf F_2}$ the curve $b$ is movable along $F$ in $\E=b\x F$ 
(i.e.~the naive moduli space of motions, $F$, is of Euler number zero). 
Besides the issue of an integral over a moduli space the whole 
physical interpretation changes. 
As nevertheless some simplifications occur in this case we illustrate 
the general procedure (on a formal, i.e.~purely mathematical level)
also with examples from this case.

Note first that one has by (\ref{direct image}) 
(with $-\be -2 \geq 0$ 
by $\beta=-(\lambda-\frac{1}{2})r-1$ and (\ref{r chi relation sharp}))
\beqa
\pi_{\E *}{\cal O}_{\E}(\al s + \beta F)
=\bigoplus_{i=1}^{\al}{\cal O}_b(\beta)
\eeqa
As we will have cause to consider the higher cohomology groups in
(\ref{long exact sequence}) we note also
\beqa
\label{general H1 computation}
H^1\Big(\E, \cO_{\E}(\al s + \be F \Big)&=&
H^1\Big(b, \; \bigoplus_{i=1}^{\al} {\cal O}_b(\be)\Big)=
\bigoplus_{i=1}^{\al} H^0\Big(b, \, {\cal O}_b(-\be -2)\Big)^*\\
&=&H^0\Big(F, \cO_F(\al)\Big) \ox H^0\Big(b, \, {\cal O}_b(-\be -2)\Big)^*
={\bf C_F^{\al}} \ox Sym^{-\be -2}V^*\nonumber
\eeqa
where $V:=H^0\Big(b, \, {\cal O}_b(1)\Big)\cong {\bf C^2}$, generated by the 
first order monomials $u$ and $v$ 
(or linear polynomials in $t=u/v$, including a constant term). 
We use the identifications 
\beqa
H^0\Big(b, \cO_b(p)\Big)&=&{\bf C}[t]_{\leq p}=Sym^p \, V={\bf C}^{p+1}\\
H^0\Big(F, \cO_F(q)\Big)&=&({\bf C[x, y]}/rel)_{\leq q}={\bf C_F^q}
\eeqa
($p\geq 0, q\geq 1$).
The subscripts $|_{\leq p}$ and $|_{\leq q}$ indicate bounds
on the degree (with deg $t = 1$ and deg $x =2$, deg $y=3$).  
We have divided out the relation $rel: \, y^2=4x^3-g_2 x -g_3$ (we also 
apply the corresponding homogeneous version including $z$ with deg $z=0$). 

With $\beta =-(\la - \frac{1}{2})r-1<0$ here we have, 
according to (\ref{the critical map}), to consider the map
\beqa
\label{F_2 critical map}
H^1\Big(\E, l(-F-c)\Big) \stackrel{\iota_1}{\lra} H^1\Big(\E, l(-F)\Big)
\eeqa
or, explicitly with (\ref{general H1 computation}) 
and (\ref{the critical map}),
\beqa
H^0\Big( F, n (\la -\frac{1}{2})\Big) \ox 
H^0\Big( b, (\la + \frac{1}{2})r-1\Big)^* \stackrel{\un{\iota}}{\lra}
H^0\Big( F, n (\la +\frac{1}{2})\Big) \ox 
H^0\Big( b, (\la - \frac{1}{2})r-1\Big)^*\nonumber
\eeqa
which is a map between equidimensional spaces 
(by (\ref{dimensions}) where lhs $=0$ by $\be <0$)
\beqa
Sym^{(\la +\frac{1}{2})r-1}\, V^* \ox {\bf C_F^{n (\la -\frac{1}{2})}}
\stackrel{\un{\iota}}{\lra}
Sym^{(\la -\frac{1}{2})r-1}\, V^* \ox {\bf C_F^{n (\la +\frac{1}{2})}}
\eeqa
making manifest the equal dimension $n(\la^2-\frac{1}{4})r$.
The map $\un{\iota}$ is induced by multiplication with an element
\beqa
\tilde{\iota}\in H^0\Big(\E, \cO_{\E}(c)\Big)\cong H^0(b, r)\ox H^0(F, n)\cong
Sym^r\, V \ox {\bf C_F^n}
\eeqa

\subsection{Example 3: the non-contributing case $r=1$}

For $r=1$ we have
$\la \in \frac{1}{2}+{\bf Z^{(>0)}}$ as $\la \in {\bf Z}$
needs $n$ and $r$ even. So take first $\la= 3/2$. 
The identifications (\ref{general H1 computation})
\beqa
H^1\Big(\E, l(-F-c)\Big)
&=&H^1\Big(\E, {\cal O}_{\E}(ns-(2r+1)F)\Big)
\cong H^1\Big(b, \; \oplus_{i=1}^n {\cal O}_b(-2r-1)\Big)\nonumber\\
&\cong &\oplus_{i=1}^n H^0\Big(b, \, {\cal O}_b(2r-1)\Big)^*
\nonumber\\
H^1\Big(\E, l(-F)\Big)
&=&H^1\Big(\E, {\cal O}_{\E}(2ns-(r+1)F)\Big)
\cong H^1\Big(b, \; \oplus_{i=1}^{2n} {\cal O}_b(-r-1)\Big)\nonumber\\
&\cong &\oplus_{i=1}^{2n} H^0\Big(b, \, {\cal O}_b(r-1)\Big)^*
\nonumber
\eeqa
show that the relevant map in (\ref{criterion 2}) is given by
the map between $2nr$-dimensional spaces
\beqa
Sym^{2r-1}\, V^*\ox {\bf C_F^n} \; \buildrel \un{\iota} \over \lra \; 
Sym^{r-1}\, V ^*\ox {\bf C_F^{2n}}
\eeqa
with $\un{\iota}$ multiplication by 
$\ti{\iota} \in Sym^r\, V\ox {\bf C_F^n}$.
For $r=1$ the map
$\un{\iota}:V^*\ox {\bf C_F^n} \ra {\bf C_F^{2n}}$ has non-trivial kernel as 
multiplication with $\ti{\iota}=\sum_{i=1}^n 
p_i \ox x_i \in V \ox {\bf C_F^n}$ 
maps\footnote{Here the evaluation 
$< \; , \; >: V \x V^* \ra {\bf C}$ is, via the canonical scalar product
$<p,q>=p^{(1)}q^{(1)}+ p^{(2)}q^{(2)}$, 
understood as a map $V\ox V \lra {\bf C}$ 
(thus reinterpreting $V^*$ as $V$, 
i.e.~we take $p=p^{(1)}u+p^{(2)}v$ and $q=q^{(1)}u^*+q^{(2)}v^*$);
furthermore, by combination with the map $V\ni q=(q^{(1)}, q^{(2)})
\lra q^{\perp}=(-q^{(2)}, q^{(1)})\in V$, this can be understood as the map
$V \wedge V \ra \Lambda^2 V \cong {\bf C}$ as one has
$p\wedge q^{\perp}=p^{(1)}( q^{\perp})^{(2)}-p^{(2)}( q^{\perp})^{(1)}
=p^{(1)}q^{(1)}+ p^{(2)}q^{(2)}=<p,q>$.}
(for $y_j=x_j$)
\beqa
V^* \ox {\bf C_F^n} \ni \sum q_j \ox y_j & \lra &
\sum_{i,j} <p_i, p_j^{\perp}> (x_i\cdot x_j)
=\sum_{i,j}p_i\wedge p_j \, (x_i\cdot x_j)= 0\;\;\;
\eeqa
for $q_j=p_j^{\perp}$ (reinterpreting $V^*$ as $V$).
So $W_b=0$ as (\ref{criterion 2}) is violated; 
cf.~sect.~\ref{red cases for bar al > n}.

This reasoning explains algebraically the case $n=3$, found experimentally
as Example 3 in [\ref{BDO}], 
also a special case of sect.~\ref{The vanishing condition};
the argument extends to arbitrary $n$. 
In sect.~\ref{The vanishing condition} the same conclusion was argued 
even for any $\la = l+1/2$ with $l \in {\bf Z}^{>0}$ where $\beta =-lr-1$; 
one can consider the map 
(using the symmetrized tensor product $\odot$; cf.~sect.~\ref{sym})
\beqa
Sym^{(l+1)r-1}\, V^* \ox 
\bigodot_1^{l}{\bf C_F^n} \lra
Sym^{lr-1}\, V^* \ox 
\bigodot_1^{l+1}{\bf C_F^n} 
\eeqa
mediated by multiplication with
$\tilde{\iota}=\sum   p_i\ox x_i\in Sym^r\, V\ox {\bf C_F^n}$ 
and may contemplate a reasoning similar to above. 

\noindent
{\em \underline{Remark:}}
There are some further obvious exceptional cases which do not contribute.
For this note that for $\un{\iota}$ to be an isomorphism the map 
$H^0\Big(F, n (\la -\frac{1}{2})\Big)\ox H^0(F, n)\lra 
H^0\Big(F, n (\la +\frac{1}{2})\Big)$
must be surjective;
this excludes the cases $n (\la -\frac{1}{2})=1$, i.e. $n=2, \la =1$
as $H^0(F, 3)$ is not generated by $H^0(F, 1)$ and $H^0(F, 2)$, and
$n=2, \la =3/2$ as $H^0(F, 4)$ is not generated by  $H^0(F, 2)$ 
and $H^0(F, 2)$ (in both cases one can not get $y$ from $x$).


\section{\label{sym}The symmetrized tensor product}
\resetcounter

Let us consider symmetrized tensor powers of the vector spaces
\beqa
\Sigma & = & z\, {\bf C}\oplus x\, {\bf C}\oplus y\, {\bf C}\\
V      & = & H^0\Big(b, \cO(1)\Big) = u\, {\bf C}\oplus v\, {\bf C}
\eeqa
Symmetrized tensor powers of degree $n$ mean that one considers 
expressions of order $n$ 
(linear combinations of monomials which themselves consist of
$n$ factors of basis elements)
within the polynomial algebra on the basis elements of the vector space.
Thus one has for example (with $S^k:=Sym^k$ and the dual basis elements 
$u^*, v^*$ of $V^*$)
\beqa
S^2 \, \Si \; & = & z^2\, {\bf C}\oplus zx\, {\bf C}\oplus zy\, {\bf C}
\oplus x^2\, {\bf C}\oplus xy\, {\bf C}\oplus y^2\, {\bf C}\\
S^3\, V^* & = & (u^*)^3 \, {\bf C}\oplus (u^*)^2 v^* \, {\bf C}\oplus 
u^* (v^*)^2 \, {\bf C}\oplus (v^*)^3\, {\bf C}\\
S^2\, V \; & = & u^2 \, {\bf C}\oplus uv \, {\bf C}\oplus v^2 \, {\bf C}
\eeqa
where we have used the polynomial notation for the composed elements.
In general we denote by $\odot$ the symmetric tensor product. So one has
(with $\cdot$ for evaluation; let $n\geq m$) 
\beqa
S^a\, V \odot S^b\, V & = & S^{a+b}\, V\\
S^n\, V^* \cdot S^m\, V & = & S^{n-m}\, V^*
\eeqa
For example one has
\beqa
u^{*3}v^{*2}\cdot(au^2+buv+cv^2)&=&(cu^{*2}+bu^*v^*+av^{*2})u^*
\eeqa

We will apply the prescription $\cdot \lra \cdot \odot \, S^k \, V^*$
(i.e. symmetric product with $S^k \, V^*$), not only
to spaces, but also, functorially, to maps $f:A\lra B$.
Then we use the notation
\beqa
f\, \underline{\odot} \, S^k \, V^*: \;\;\;
A\odot S^k \, V^*\lra B\odot S^k \, V^*
\eeqa

If, to give an example, $C=c_0u^3+c_1u^2v+c_2uv^2+c_3v^3\in S^3\, V = 
Hom(S^3\, V^*, {\bf C})$, then one gets for 
$C\; \underline{\odot}\; S^2\, V^*: \; S^5\, V^*\lra  S^2\, V^*$ the matrix
representation
\beqa
\label{C matrix example}
C\underline{\odot} S^2\, V^*=
\left( {\begin{array}{cccccc}
c_0 & c_1 & c_2 & c_3 &   0 & 0   \\
0   & c_0 & c_1 & c_2 & c_3 & 0   \\
0   & 0   & c_0 & c_1 & c_2 & c_3 
\end{array}} \right)
\eeqa

\subsection{\label{resultant}Interpretation of the resultant criterion}

Let $f=\sum_{i=0}^m a_i z^i$ and $g=\sum_{i=0}^n b_i z^i$ 
polynomials in the complex variable $z$ (with $a_m\neq 0, b_n\neq 0$;
homogeneous polynomials are treated analogously). 
Euler and Sylvester remarked that $f$ and $g$ have a common zero $z_*$ 
precisely if a certain determinant vanishes, the resultant $Res(f,g)$
of $f$ and $g$, a polynomial of degree $m+n$ in
the coefficients $a_i$ and $b_i$. If we take 
for example $m=5$ and $n=2$ the determinant in question is that of
\beqa
\label{multmatrix}
\left( {\begin{array}{ccccccc}
a_5 & 0   & b_2 & 0   &   0 & 0   & 0   \\
a_4 & a_5 & b_1 & b_2 &   0 & 0   & 0   \\
a_3 & a_4 & b_0 & b_1 & b_2 & 0   & 0   \\
a_2 & a_3 &   0 & b_0 & b_1 & b_2 & 0   \\
a_1 & a_2 &   0 &   0 & b_0 & b_1 & b_2 \\
a_0 & a_1 &   0 &   0 &   0 & b_0 & b_1 \\
  0 & a_0 &   0 &   0 &   0 &   0 & b_0 \\
\end{array}} \right)
\eeqa
(or of its transpose).
For the assertion 
$f=(z-z_*) \tilde{f}, \; g=(z-z_*) \tilde{g}$ is equivalent to have
\beqa
\label{fg relation}
f\tilde{g}=g\tilde{f}
\eeqa 
for some polynomials $\tilde{f}, \tilde{g}$
of degree $m-1$ and $n-1$ ($\tilde{f}, \tilde{g}\neq 0$). 
For the equivalence note that
clearly not all linear factors of $f$ 
can come from $\tilde{f}$; the other direction is obvious.

From the equality (\ref{fg relation}) of polynomials of degree $m+n-1$ 
the classical reasoning proceeds by comparison of their $m+n$ coefficents 
to the indicated matrix of a system of $m+n$ linear equations. 
More in the spirit of our investigation is to argue as follows. 
For $f\in S^m\, V\cong Hom( {\bf C}, S^m\, V)$ one has the map
given by multiplication with $f$
\beqa
f\; \un{\odot}\, S^k\, V \in Hom(S^k\, V, S^{k+m}\, V)
\eeqa
(correspondingly for $g$). With this definition consider the following map
\beqa
\un{m}:   S^{n-1}\, V \oplus S^{m-1}\, V \ni (p,q)\lra 
(f \un{\odot}\, S^{n-1}\, V )\, p +  (g \un{\odot}\, S^{m-1}\, V )\,  q  
= fp+gq \in S^{m+n-1}\, V
\nonumber
\eeqa
Now (\ref{fg relation}) just expresses the fact 
$\un{m}(\, (\tilde{g}, -\tilde{f}  ) \, )=0$, 
i.e. that the map $\un{m}$ is not injective. However $\un{m}$ 
has just (\ref{multmatrix}) as matrix and $Res(f,g)=\det \, \un{m}$.

\noindent
Example $1$ has
$Res(B_2, A_1)=\det (B \, \oplus \, A\, \underline{\odot}\, V^*)$
with $B \, \oplus \, A\, \underline{\odot}\, V^*
\in \, Hom(S^2\, V^*, {\bf C} \oplus V^*)$, cf.~(\ref{D3 matrix}),
and $Res(C_4, A_1)=\det (C \, \oplus \, A\, \underline{\odot}\, S^3\,V^*)$,
cf.~(\ref{D5 explicit}), (\ref{D5 matrix}).

\subsection{\label{more than one root}
Polynomials having more that one root in common}

The classical case gives a determinantal criterion 
for two polynomials to have one root in common.
We will also be interested in the case of having more roots in common. 

Let us assume that 
(where $\deg \tilde{f}=m-2, \deg \tilde{g}=n-2$)
\beqa
f&=&(z-z_1)(z-z_2)\tilde{f}\\
g&=&(z-z_1)(z-z_2)\tilde{g}
\eeqa
As before an precise criterion for this case is a relation
\beqa
\label{crit mult}
fp&=&gq
\eeqa
where $\deg \tilde{q}=m-2, \deg \tilde{p}=n-2$. Equivalently
consider the map
\beqa
\underline{m}:\; S^{n-2}V\oplus S^{m-2}V \ni (p ,q) &\lra &fp+gq \in S^{m+n-2}V
\eeqa
By (\ref{crit mult})
we search a criterion for $\ker \underline{m}\neq 0$.
As $\dim \, \mbox{source} \, \un{m}=m+n-2$ 
and $\dim \, \mbox{target} \, \un{m}=m+n-1$
we should consider $\un{m}$ restricted (in its target) to its image.

To describe this in greater detail we focus on the example $m=2, n=3$ 
which is of importance in sect.~\ref{Exa 2} (cf.~Example $2$).
So let $f=\sum_{i=0}^2f_iz^i, g=\sum_{i=0}^3g_iz^i$ and
\beqa 
\un{m}:\; V\oplus {\bf C} \ni (p ,q) &\lra &fp+gq \in S^{3}V
\eeqa
(where $p=p_1z+p_0, q=q_0$)
of matrix $Mat[g,f]^{tr}$ (cf.~(\ref{matrix hgf}) without first row). 
Now
\beqa
\label{image criterion}
h=\sum_{i=0}^3h_iz^i
\in im \; \un{m}&\Llra & \det Mat[h, g, f ]=-\sum_{i=0}^3 (-1)^ih_i M_i =0
\eeqa
where the matrix of the map 
$h\oplus g\oplus f \un{\circ}V^*:\, 
S^3 V^* \lra {\bf C}\oplus {\bf C}\oplus V^*$ occurs
\beqa
\label{matrix hgf}
Mat[h, g, f ]&=&\left( {\begin{array}{cccc}
h_3 & h_2   & h_1 & h_0   \\
g_3 & g_2   & g_1 & g_0   \\
f_2 & f_1   & f_0 & 0      \\
  0 & f_2   & f_1 & f_0 
\end{array}} \right)
\eeqa
with minors $M_i$ associated to the development w.r.t.~the first row.
To see (\ref{image criterion}) one may eliminate $p_1, p_0, q_0$ 
from the coefficents of a typical image element 
to get directly $\sum_{i=0}^3 (-1)^ih_i M_i =0$.
Note that one gets from taking $h=f$ or $fz$ the relations
$f_2M_2=-f_0M_0+f_1M_1$ and $f_2M_3=-f_0M_1+f_1M_2$ between the minors.
A basis for the three-dimensional image of $\un{m}$ 
is given by $fz, f$ and $g$ 
(which explains (\ref{image criterion}) in an elementary way).
For these elements to become linearly dependent one finds by direct inspection
\beqa
ker \; \un{m} \neq 0 &\Llra & M_0=0=M_1
\eeqa
A common root of $M_0$ and $M_1$ implies of course a second
order zero of the resultant as
\beqa
Res(g, f) \;=\;-\frac{1}{f_1}\det \left( {\begin{array}{cc}
M_3 & M_2 \\
M_1 & M_0
\end{array}} \right)
\;=\;\frac{1}{f_2^2}\det
\left( {\begin{array}{ccc}
f_0 & f_1 & f_2 \\
M_1 & M_0 &  0  \\
 0  & M_1 & M_0
\end{array}} \right)
\eeqa

\section*{References}
\begin{enumerate}

\item
\label{Wittenhet}
E. Witten, {\em World-Sheet Corrections Via D-Instantons},
hep-th/9907041, JHEP {\bf 0002} (2000) 030.

\item
\label{FMW}
R. Friedman, J. Morgan and E. Witten, {\em Vector Bundles and F-Theory},
hep-th/9701162, Comm. Math. Phys. {\bf 187} (1997) 679.

\item
\label{FMWIII}
R. Friedman, J.W. Morgan and E. Witten,
{\em Vector Bundles over Elliptic Fibrations}, alg-geom/9709029,
Jour. Alg. Geom. {\bf 8} (1999) 279.

\item
\label{C}
G. Curio, {\em Chiral matter and transitions in heterotic string models},
hep-th/9803224, Phys.Lett. {\bf B435} (1998) 39. 

\item
\label{AChor}
B. Andreas and G. Curio,
{\em Horizontal and Vertical Five-Branes in Heterotic/F-Theory Duality},
hep-th/9912025, JHEP {\bf 0001} (2000) 013.

\item
\label{CD}
G. Curio and R. Y. Donagi, {\em Moduli in N=1 Heterotic/F-Theory 
Duality}, hep-th/9801057, Nucl.Phys. {\bf B518} (1998) 603.

\item
\label{irred}
B.A. Ovrut, T. Pantev and J. Park,
{\em Small Instanton Transitions in Heterotic M-Theory},
hep-th/0001133, JHEP 0005 (2000) 045.

\item
\label{BDO}
E.I. Buchbinder, R. Donagi and B.A. Ovrut,
{\em Vector Bundle Moduli Superpotentials in Heterotic Superstrings 
and M-Theory}, hep-th/0206203, JHEP {\bf 0207} (2002) 066.

\end{enumerate}
\end{document}